\documentclass[11pt,a4paper]{article}
\pdfoutput=1
\usepackage{jstyle}
\usepackage{amsmath, amsfonts, amssymb}
\usepackage{graphicx, subfig, hyperref, color, verbatim}

\definecolor{cardinal}{rgb}{0.6,0,0}

\title{More on holographic correlators: \\Twisted and dimensionally reduced structures}
\author[a]{Connor Behan,}
\author[a]{Pietro Ferrero,}
\author[b]{Xinan Zhou}
\affiliation[a]{Mathematical Institute, University of Oxford, Andrew Wiles Building, Radcliffe Observatory Quarter, Woodstock Road, Oxford, OX2 6GG, U.K.}
\affiliation[b]{Princeton Center for Theoretical Science, Princeton University, Princeton, NJ 08544, USA}
\emailAdd{connor.behan@maths.ox.ac.uk, \\ \hskip 42pt pietro.ferrero@maths.ox.ac.uk, \\ \hskip 42pt xinanz@princeton.edu}
\abstract{
Recently four-point holographic correlators with arbitrary external BPS operators were constructively derived  in \cite{az20a,az20} at tree-level for maximally superconformal theories. 
In this paper, we capitalize on these theoretical data, and perform a detailed study of their analytic properties. We point out that these maximally supersymmetric holographic correlators exhibit a hidden dimensional reduction structure {\it \`a la} Parisi and Sourlas. This emergent structure allows the correlators to be compactly expressed in terms of only scalar exchange diagrams in a dimensionally reduced spacetime, where formally both the AdS and the sphere factors have four dimensions less. We also demonstrate the superconformal properties of holographic correlators under the chiral algebra and topological twistings. For $AdS_5\times S^5$ and $AdS_7\times S^4$, we obtain closed form expressions for the meromorphic twisted correlators from the maximally R-symmetry violating limit of the holographic correlators. The results are compared with  independent field theory computations in 4d $\mathcal{N}=4$ SYM and the 6d $(2,0)$ theory, finding perfect agreement. For $AdS_4\times S^7$, we focus on an infinite family of near-extremal four-point correlators, and extract various protected OPE coefficients from supergravity. These OPE coefficients provide new holographic predictions to be matched by future supersymmetric localization calculations. In deriving these results, we also develop many technical tools which should have broader applicability beyond studying holographic correlators. 
}

\begin{document}
\maketitle
\tableofcontents

\section{Introduction}
\label{sec:intro}
Historically, the holographic computation of superconformal correlators using AdS supergravity has been an extremely difficult task, even just at tree-level and for four half-BPS operators (see, {\it e.g.}, \cite{DHoker:1999pj,Arutyunov:2000py,Arutyunov:2002ff,Arutyunov:2003ae,Arutyunov:2002fh} for early progress). For example, the complete quartic vertices  for $AdS_5\times S^5$ IIB supergravity were worked out in \cite{Arutyunov:1999fb} and occupied 15 pages. The sheer complexity of the vertices presents a daunting challenge for the standard diagrammatic expansion method, rendering it practically useless for generic correlators. However, an important breakthrough was made in \cite{rz16,rz17}, where a different line of attack was introduced by incorporating bootstrap ideas. It was argued that holographic correlators have very rigid structures, and can be therefore completely determined by using only supersymmetry and consistency conditions. This led to several efficient new methods to compute holographic correlators, which avoid inputting the details of the complicated supergravity effective actions altogether.\footnote{For an overview of these bootstrap methods, see section 2 of \cite{az20}.} Most spectacularly, the bootstrap methods gave a complete solution to  {\it all} tree-level four-point functions of half-BPS operators for $AdS_5\times S^5$ IIB supergravity \cite{rz16,rz17}, without computing a single Witten diagram. On the other hand, the success of these methods in other maximally supersymmetric backgrounds ($AdS_4\times S^7$ and $AdS_7\times S^4$) was much more modest. While many new results, unattainable by brute force, were generated by using these methods \cite{rz17b,z17,Zhou:2018ofp}, the algebraic bootstrap problems were too difficult to solve in general. It was only recently that a different constructive method was developed \cite{az20a,az20} which completed the program for tree-level half-BPS four-point functions in maximally supersymmetric theories. The constructive method starts with a special R-symmetry polarization configuration, dubbed ``maximally R-symmetry violating'' in \cite{az20a,az20}, where correlators drastically simplify and can be easily computed. The full correlators are then reconstructed from this limit by using symmetries. This method applies to any spacetime dimension, and leads to a closed form formula for all tree-level four-point functions in all maximally supersymmetric theories \cite{az20a,az20}. 

The solution of general tree-level four-point functions generates  a wealth of new theoretical data. The purpose of the current paper is to capitalize on these data and extract useful physical information. Our analysis will include two complementary aspects. The first is to find new hidden structures in the correlators. Identifying these hidden structures not only leads to more compact expressions, but also suggests new symmetry properties of the bulk theory. In particular, we will show that the correlators exhibit an emergent dimensional reduction structure, which is closely related to the Parisi-Sourlas supersymmetry \cite{Parisi:1979ka,Kaviraj:2019tbg}. The second aspect concerns certain protected subsectors of the superconformal field theories. Their information can be cleanly isolated by focusing on special twisted configurations of the supergravity correlators. We also perform independent boundary calculations, which perfectly match the bulk predictions. This provides non-trivial checks for several conjectures in the protected sector. Below we provide a more detailed summary of our main results, embedded in brief reviews of related backgrounds.
\newpage
\noindent{\bf Emergent dimensional reduction}
\vspace{0.3cm}

\noindent It is well known that scattering amplitudes in flat space often contain surprising structures which reveal unexpected   symmetries hidden from the Lagrangian description. Viewed as scattering amplitudes in anti de Sitter space, it is not surprising that holographic correlators can similarly exhibit hidden structures which cannot be seen from the diagrammatic expansion. A beautiful example is the hidden ten dimensional conformal symmetry observed in $AdS_5\times S^5$ \cite{ct18}, which organizes all tree-level four-point functions into a generating function. The generating function is nothing but the stress tensor four-point function after replacing four dimensional distances with ten dimensional distances. A similar six dimensional version was also discovered for $AdS_3\times S^3\times K3$ correlators \cite{rrz19,grtw20}. However, it is generally believed that the hidden conformal symmetry is non-existent for backgrounds which are not conformally flat (such as $AdS_4\times S^7$ and $AdS_7\times S^4$). 
 
 In this paper, we will point out a different hidden structure in tree-level four-point functions that involves dimensional reduction, and is present in {\it all} maximally superconformal theories. We recall from \cite{az20a,az20} that four-point functions in $AdS_{d+1}\times S^{\mathtt{d}-1}$ are sums of exchange amplitudes over finitely many supergravity multiplets, with no additional contact interactions.  We will show that the exchange amplitude of a multiplet $p$ can be simplified as a differential operator acting on only three {\it scalar} exchange amplitudes. Schematically, we have
 \begin{equation}\label{introeq1}
{\rm Diff}\circ (\mathcal{M}_{\epsilon p}^{(d)}+\#_1 \mathcal{M}_{\epsilon p+2}^{(d)}+\#_2 \mathcal{M}_{\epsilon p+4}^{(d)})
\end{equation}
where $\epsilon=(d-2)/2$, and $\mathcal{M}_{\Delta}^{(d)}$ is the scalar exchange amplitude in $AdS_{d+1}$ with internal dimension $\Delta$. Note that in addition to superconformal primary of the multiplet with $\Delta=\epsilon p$, there are also new scalars with shifted conformal dimensions. The spinning field contributions in the multiplet are generated by the differential operator action. Quite magically, the combination of diagrams in (\ref{introeq1}) turns out to be just {\it a single} scalar exchange diagram in a {\it lower} dimensional $AdS_{d-3}$ space
\begin{equation}\label{introeq2}
\mathcal{M}_{\epsilon p}^{(d)}+\#_1 \mathcal{M}_{\epsilon p+2}^{(d)}+\#_2 \mathcal{M}_{\epsilon p+4}^{(d)}=\mathcal{M}_{\epsilon p}^{(d-4)}\;.
\end{equation}
A similar pattern can be found for the R-symmetry part, which leads to a dimensionally reduced internal space $S^{\mathtt{d}-5}$. These observations allow us to repackage the full multiplet amplitude into the following simple form 
\begin{equation}
{\rm Diff}' \circ (\mathcal{Y}_{p,0}^{\mathtt{d}-4} \mathcal{M}_{\epsilon p}^{(d-4)})
\end{equation}
where $\mathcal{Y}_{p,0}^{\mathtt{d}-4}$ is the R-symmetry polynomial of the rank-$p$ symmetric traceless representation of the reduced R-symmetry group $SO(\mathtt{d}-4)$ associated with $S^{\mathtt{d}-5}$. Consequently, the correlators in the original theory are now fully captured by the correlators of a simple scalar theory in the lower dimensional spacetime $AdS_{d-3}\times S^{\mathtt{d}-5}$!

The appearance of dimensionally reduced spacetimes is quite curious, and we do not have a good understanding of its physical origin. Nevertheless, we find strong evidence indicating that this reduction phenomenon is intimately related to the Parisi-Sourlas dimensional reduction, which relates a $d$ dimensional theory with Parisi-Sourlas supersymmetry and a $d-2$ dimensional non-supersymmetric theory. The connection comes from a dimensional reduction formula for exchange Witten diagrams in $AdS_{d+1}$ and $AdS_{d-1}$, which was shown in \cite{Zhou:2020ptb} as the consequence of the holographically realized Parisi-Sourlas supersymmetry. It turns out that using the dimensional reduction formula twice gives precisely (\ref{introeq2}).

\vspace{0.7cm}
\noindent{\bf Twisted correlators and protected sectors}
\vspace{0.3cm}

\noindent Common to the maximally superconformal theories considered in this paper is the existence of a protected subsector of operators.  These operators  are formed by restricting a certain class of operators (half-BPS operators are such examples) to a two dimensional plane or a one dimensional line, and ``twisting'' the operators by giving them special R-symmetry polarizations specified by the locations of the insertions \cite{bllprv13,brv14,clpy14,bpr16}. Crucially, the subsector involves only members of short multiplets, closes on itself under OPE, and does not depend on marginal deformations (if there are any) \cite{bllprv13}. Therefore this construction isolates a fully protected subsector. The protected subsector has the form of a unitary chiral algebra in six dimensions \cite{brv14}, a non-unitary chiral algebra in four dimensions \cite{bllprv13} and topological quantum mechanics in three dimensions \cite{clpy14,bpr16}. The presence of the protected subsector immediately imposes strong constraints on the correlation functions, as they must become meromorphic or topological in the twisted configurations. It was also quickly recognized that non-trivial unitarity bounds are encoded in these rigid structures analytically \cite{bllprv13,lrs15,ll15,b18}. Along the same lines, one can use chiral algebra or topologically protected data to greatly increase the power of the numerical conformal bootstrap \cite{brv13,clpy14,bllrv14,blrv15,llmm16,brv16,acp18,acp19,gl20,bmv20,bcjp20}. When combined with supersymmetric localization, these techniques can also be used to predict new holographic dualities \cite{br16,mpw17,f20}.

While the existence of a 6d/2d or 4d/2d correspondence, in the sense described above, is a proven fact, a precise description of the lower dimensional theory is not always readily available. In well studied cases, \textit{e.g.}, theories of Argyres-Douglas type or class $\mathcal{S}$, much of the recent progress in understanding their chiral algebras has been guided by a set of compelling conjectures \cite{bllprv13,brv14,bprv14,lp14,bmr18}. The 3d/1d correspondence is on a somewhat different footing due to the explicit Lagrangians derived in \cite{dpy16,dfpy17,dfpy18,Panerai:2020boq}.\footnote{See however also \cite{Pan:2017zie,Pan:2019bor} for 4d/2d examples.} Nevertheless, applying these results to ABJM theory requires the use of conjectural dualities \cite{bk10,bk11} and even then the calculations can be very involved.

In this work, we further strengthen the evidence for these conjectures in the case of holographic CFTs with maximal supersymmetry. We perform independent calculations from both the field theory side and the supergravity side, finding perfect agreement. For 6d and 4d, where the twisting leads to meromorphic functions, we compute four-point chiral correlators on the field theory side by improving the holomorphic bootstrap method of \cite{hmpz15}. The basic idea is that meromorphic functions are determined by singularities which are dictated by the OPE of the chiral algebra. However, the improved algorithm allows us to write down closed form expressions for all chiral four-point functions, in a form that foreshadows the structures anticipated from the supergravity side. We then reproduce these results in Mellin space by taking residues, from the MRV limit of the supergravity amplitudes \cite{az20a,az20}. For the 3d case, the topological nature of the twisted correlators makes the situation inevitably more involved. Therefore, we will content ourselves with  mostly focusing on the next-to-next-to-extremal correlators of the form $\langle\mathcal{O}_2\mathcal{O}_2\mathcal{O}_k\mathcal{O}_k\rangle$. Moreover, instead of attempting to find closed form expressions for the topological correlators, we will focus on identifying the finitely many operators which contribute to a topological correlator, and computing their OPE coefficients in a $1 / c_T$ expansion. Similarly to the 6d and 4d cases, we will approach the 3d problem both from the boundary TQFT description, and from the bulk side using the Mellin space results. To summarize, the calculations outlined above provide non-trivial checks for the conjectures in the protected sector. Given that many aspects of the protected subsectors appear to be within reach of a formal proof, it might also be helpful to turn the logic around and view our results as a tree-level check of AdS/CFT. In performing these checks, we have developed many technical tools for studying holographic four-point functions. We expect these tools will have broader applicability. \\

The rest of the paper is organized as follows. We set the stage in section \ref{sec:setup} by reviewing some basic superconformal kinematics.  In section \ref{sec:ps}, we show that the holographic correlators have an emergent dimensional reduction structure, and explain the connection with the Parisi-Sourlas supersymmetry. Discussions of the protected subsectors begin in section \ref{sec:6d}, where we show the agreement between the field theory and supergravity calculations for 6d. In section \ref{sec:4d}, we present a parallel check for the 4d case. The 3d case is discussed in section \ref{sec:topological} to \ref{sec:ope-coeffs}. We explain in section \ref{sec:topological} the reason why the topological property makes the calculations different. The field theory computations are carried out in section \ref{sec:matrix-model}, and the supergravity analysis is done in section \ref{sec:ope-coeffs}. We conclude in section \ref{sec:conc} with a brief discussion of future directions. Various technical details are relegated to the two appendices.

\section{Superconformal kinematics}
\label{sec:setup}
To start, we will use this section to fix our notation and review some basic superconformal kinematics of four-point correlation functions. In particular, we will review a powerful kinematic constraint, namely the superconformal Ward identities in diverse dimensions \cite{no04,Dolan:2004mu}, which served as a precursor to the SCFT/chiral algebra correspondence.  

\subsection{Four-point functions in $d>2$}
We focus on the supergravity backgrounds $AdS_4 \times S^7$, $AdS_5 \times S^5$ and $AdS_7 \times S^4$, which have the maximal amount of superconformal symmetry. In these backgrounds all supergravity single-particle states are dual to components of the half-BPS supermultiplets. The super primaries of these half-BPS multiplets are scalar operators of the form $\mathcal{O}_k^{I_1 \dots I_k}(x)$, which transform in rank-$k$ symmetric traceless representations of the R-symmetry group. The scaling dimensions of the half-BPS operators are fixed by the R-symmetry charges
\begin{equation}
\Delta = \epsilon k, \hspace{1cm} \epsilon \equiv \frac{d - 2}{2}\label{eps-def}
\end{equation}
where $k \in \{ 2\,,3\,,\ldots \}$. We will saturate the R-symmetry indices of these operators with polarization vectors $t_I$ that are null in order to respect tracelessness
\begin{equation}
\mathcal{O}_k(x, t) = \mathcal{O}_k^{I_1 \dots I_k}(x) \; t_{I_1} \dots t_{I_k}, \hspace{1cm} t \cdot t = 0\;. \label{polarization1}
\end{equation}
The central objects of this paper are the four-point correlation functions of these operators
\begin{equation}
\left < \mathcal{O}_1(x_1,t_1) \mathcal{O}_2(x_2,t_2) \mathcal{O}_3(x_3,t_3) \mathcal{O}_4(x_4,t_4) \right >\;.
\end{equation}
The extremality for a correlator is defined as
\begin{equation}
\mathcal{E} = \begin{cases}
\frac{k_{\bar{1}} + k_{\bar{2}} + k_{\bar{3}} - k_{\bar{4}}}{2}, & k_{\bar{1}} + k_{\bar{4}} \geq k_{\bar{2}} + k_{\bar{3}} \\
k_{\bar{1}}, & k_{\bar{1}} + k_{\bar{4}} < k_{\bar{2}} + k_{\bar{3}}
\end{cases} \label{extremality}
\end{equation}
where $k_{\bar{\imath}}$ is the $i$-th smallest element of $\{k_1, k_2, k_3, k_4\}$. The standard conformal and R-symmetry cross-ratios are
\begin{equation}
U = \frac{x_{12}^2 x_{34}^2}{x_{13}^2 x_{24}^2}, \;\; V = \frac{x_{14}^2 x_{23}^2}{x_{13}^2 x_{24}^2}, \;\; \sigma = \frac{t_{13} t_{24}}{t_{12} t_{34}}, \;\; \tau = \frac{t_{14} t_{23}}{t_{12} t_{34}}\;, \label{cross-ratio1}
\end{equation}
with $x_{ij} \equiv x_i - x_j$ and $t_{ij} \equiv t_i \cdot t_j$. We will exploit conformal symmetry and R-symmetry to extract a kinematic factor such that the correlator becomes a function of the cross-ratios, and depends on $\sigma$ and $\tau$ as a polynomial of total degree $\mathcal{E}$. If the weights are ordered as $k_1 \leq k_2 \leq k_3 \leq k_4$, a convention that accomplishes this is
\begin{align}
& \left < \mathcal{O}_1(x_1,t_1) \mathcal{O}_2(x_2,t_2) \mathcal{O}_3(x_3,t_3) \mathcal{O}_4(x_4,t_4) \right > = \left ( \frac{t_{34}}{x_{34}^{2\epsilon}} \right )^{\frac{1}{2}(k_{3} + k_{4} - k_{1} - k_{2})} \left ( \frac{t_{24}}{x_{24}^{2\epsilon}} \right )^{\frac{1}{2}(k_{2} + k_{4} - k_{1} - k_{3})} \nonumber \\
& \hspace{2.2cm} \left ( \frac{t_{23}}{x_{23}^{2\epsilon}} \right )^{\frac{1}{2}(k_{1} + k_{2} + k_{3} - k_{4}) - \mathcal{E}} \left ( \frac{t_{14}}{x_{14}^{2\epsilon}} \right )^{k_{1} - \mathcal{E}} \left ( \frac{t_{12} t_{34}}{x_{12}^{2\epsilon} x_{34}^{2\epsilon}} \right )^{\mathcal{E}} \mathcal{G}_{1234}(U, V; \sigma, \tau)\;. \label{unbarred-correlator}
\end{align}
However, we should stress that when working with various correlators, we will specify the ordering and often it will be different from $k_1 \leq k_2 \leq k_3 \leq k_4$. The formulas we introduce in this section and section \ref{sec:ps}, which form the basis for later results, will be valid for arbitrary ordering of $k_i$ as long as the $\sigma$ and $\tau$ dependence remains polynomial with degree $\mathcal{E}$. In practice, we implement this by interchanging the four labels in \eqref{unbarred-correlator} according to the permutation which takes us from $k_1 \leq k_2 \leq k_3 \leq k_4$ to whatever the new ordering is.

The dynamical function $\mathcal{G}_{1234}(U, V; \sigma, \tau)$ will have two important pieces in this work -- the disconnected part and the tree-level part. We are allowed to use these terms because holographic CFTs admit a ``large $N$ limit'', and disconnected and tree-level correspond to the first two orders in the $1/N$ expansion. In AdS/CFT, $N$ is typically the number of branes that give rise to the near-horizon geometry that has an $AdS_{d + 1}$ factor. Since tree-level corrections scale with a different power of $N$ in each dimension, namely $N^{-1-\epsilon}$, it is useful to quote results in terms of the central charge.\footnote{CFTs in even dimension have an additional notion of central charge because they exhibit a Weyl anomaly when placed on a curved manifold \cite{d93}. A universal prefactor, computed in \cite{brv14,bllprv13}, relates the coefficient of the Euler density to the corresponding Virasoro central charge which we will call $c_{2d}$ in later sections.}
\begin{align}
c_T^{(6\mathrm{D})} &= 70(4N^3 - 3N - 1) \label{ct-6d} \\
c_T^{(4\mathrm{D})} &= 30(N^2 - 1) \label{ct-4d} 
\end{align}
\begin{align}
c_T^{(3\mathrm{D})} &\approx 1 - \frac{112}{3\pi^2} - (8N + 9) \left ( \frac{\pi^2}{2} \right )^{-\frac{2}{3}} \frac{8 \mathrm{Ai}^\prime \left [ \left ( N - \frac{3}{8} \right ) \left ( \frac{\pi^2}{2} \right )^{\frac{1}{3}} \right ]}{3 \mathrm{Ai} \left [ \left ( N - \frac{3}{8} \right ) \left ( \frac{\pi^2}{2} \right )^{\frac{1}{3}} \right ]} \label{ct-3d} \\
&\approx \frac{64\sqrt{2}}{3\pi} N^{\frac{3}{2}} \nonumber
\end{align}
The 3d expression, which includes all perturbative terms, can be computed using the techniques of \cite{mp11,h15}. Note that we are using conventions such that
\begin{align}
\left < T_{\mu\nu}(x) T_{\rho\sigma}(0) \right > &= \frac{d}{d - 1} \frac{\Gamma \left ( \frac{d}{2} \right )^2}{4\pi^d} \left [ \frac{1}{2} I_{\mu\sigma} I_{\nu\rho} + \frac{1}{2} I_{\mu\rho} I_{\nu\sigma} - \frac{1}{d} \delta_{\mu\nu} \delta_{\sigma\rho} \right ] \frac{c_T}{x^{2d}} \label{ct-def} \\
I_{\mu\nu} &\equiv \delta_{\mu\nu} - 2 \frac{x_\mu x_\nu}{x^2} \nonumber
\end{align}
which means that a single free boson has $c_T = 1$ \cite{op93}.

\subsection{Superconformal Ward identities}
The four-point functions \eqref{unbarred-correlator}, by construction, obey all of the Ward identities from ordinary conformal symmetry and R-symmetry. In addition to this, we must consider superconformal Ward identities associated with the fermionic generators. The action of the fermionic generators generically relate correlation functions of superconformal primaries to those of their super descendants -- a complicated bootstrap problem in general (see, {\it e.g.}, \cite{cls17}). The situation becomes nicer when the external operators are the super primaries of the half-BPS multiplets. In this case, Ward identities from superconformal symmetry lead to constraints which only involve the original four-point function. More precisely, this constraint takes the universal form \cite{Dolan:2004mu}
\begin{equation}
\left [ \chi^\prime \frac{\partial}{\partial \chi^\prime} - \epsilon \alpha^\prime \frac{\partial}{\partial \alpha^\prime} \right ] \mathcal{G}_{1234}(\chi, \chi^\prime; \alpha, \alpha^\prime) \biggl |_{\alpha^\prime = \frac{1}{\chi^\prime}} = 0 \label{scwi}
\end{equation}
which is often called the \textit{superconformal Ward identity}.
Here we have introduced the change of variables
\begin{equation}
U = \chi \chi^\prime, \;\; V = (1 - \chi)(1 - \chi^\prime), \;\; \sigma = \alpha \alpha^\prime, \;\; \tau = (1 - \alpha)(1 - \alpha^\prime). \label{cross-ratio2}
\end{equation}
These new variables will be identified with cross-ratios in one dimension ({\it i.e.}, for $\mathfrak{sl}(2)$ and $\mathfrak{su}(2)$), as we will be more specific about in the next subsection.

A direct consequence of \eqref{scwi} is that there are certain loci in cross-ratio space, on which the four-point functions are topological along certain directions. We will call these special choices of cross-ratio configurations \textit{twisted configurations}. These configurations exist in $d = 3, 4, 6$, with the prescription being slightly different in each case. We will be interested in seeing how this structure is captured in the results of $AdS_4 \times S^7$, $AdS_5 \times S^5$ and $AdS_7 \times S^4$ supergravity respectively.

Starting with $d = 6$, \eqref{scwi} implies
\begin{equation}
\epsilon = 2: \hspace{2cm} \frac{\partial}{\partial \chi^\prime} \mathcal{G}_{1234} \left ( \chi, \chi^\prime; \frac{1}{\chi^\prime}, \frac{1}{\chi^\prime} \right ) = 0. \label{scwi-6d}
\end{equation}
This is the statement that the twisted correlator is a holomorphic function. It is now understood that, far from being an arbitrary holomorphic function, the holomorphic function is really a four-point function of an auxiliary chiral algebra in two dimensions \cite{brv14}. Since this conclusion required us to have two independent R-symmetry cross-ratios, it is easy to see that it only applies to the 6d superconformal algebra $\mathfrak{osp}(8^* | 4)$.

Chiral symmetry appears again in $d = 4$ according to
\begin{equation}
\epsilon = 1: \hspace{2cm} \frac{\partial}{\partial \chi^\prime} \mathcal{G}_{1234} \left ( \chi, \chi^\prime; \alpha, \frac{1}{\chi^\prime} \right ) = 0. \label{scwi-4d}
\end{equation}
This time, the structure is somewhat richer since we have a holomorphic dependence on $\alpha$ as well. This means there is still a notion of R-symmetry for the twisted correlator. As explained in \cite{bllprv13}, this is a consequence of the chiral algebra including super-Virasoro generators whenever the 4d theory has $\mathfrak{psu}(2, 2 | 4)$ or $\mathfrak{su}(2, 2 | 3)$ symmetry. The chiral algebra is non-supersymmetric for $\mathfrak{su}(2, 2 | 2)$, in agreement with the fact that $\alpha$ is not an independent cross-ratio in these theories.

Finally, for $d = 3$ we have
\begin{equation}
\epsilon = \frac{1}{2}: \hspace{2cm} \frac{\partial}{\partial \chi^\prime} \mathcal{G}_{1234} \left ( \chi^\prime, \chi^\prime; \alpha, \frac{1}{\chi^\prime} \right ) = 0. \label{scwi-3d}
\end{equation}
This indicates that, if we take a diagonal limit first, the remaining spatial cross-ratio drops out after the twist. More precisely, the correlator only depends on space-time positions through the ordering since the superconformal Ward identity is a local statement. This is what gives the one dimensional protected subsector the structure of topological quantum mechanics \cite{clpy14,bpr16}. To still interpret this object as a CFT in one dimension, it would have to consist solely of dimension zero operators which can be thought of as conserved currents for a higher-spin algebra \cite{mpw17}. This 3d/1d correspondence applies to $\mathfrak{osp}(\mathcal{N} | 4)$ with $\mathcal{N} \geq 4$ which means that the cross-ratio $\alpha$ is always present.

Although \eqref{scwi} is a statement about half-BPS four-point functions, the results of \cite{brv14,bllprv13,clpy14,bpr16} apply to a larger family of operators in short multiplets, and also to higher-point functions. From a holographic point of view, the easiest way to study correlators of these operators would be to perform the OPE on half-BPS correlators. However, this will not be pursued here. It is also worth noting that we will encounter the $d = 2$ superconformal Ward identity in section \ref{sec:4d} when we construct $\mathfrak{psu}(1, 1 | 2)$ blocks. The implications of this for protected quantities were discussed in \cite{rrz19}.

\subsection{Four-point functions in lower dimensions}
Correlators of operators in lower dimensions emerge from correlators in higher dimensional theories after performing superconformal twisting, as we have seen from the discussion on superconformal Ward identities in the last subsection. In this subsection, we give a summary of the kinematics of correlators of local operators on the plane (or restricted to a line). 

Let us consider an operator with conformal weight $h$ and $\mathfrak{su}(2)$ spin $j$. The $\mathfrak{su}(2)$ refers to the possible residual R-symmetry after twisting. We can denote the operator as
\begin{equation}
\mathcal{O}_{h, j}(z, y) = \mathcal{O}_{h, j | a_1 \dots a_{2j}}(z) \; y^{a_1} \dots y^{a_{2j}} \label{polarization2}
\end{equation}
where we have similarly used $\mathfrak{su}(2)$ spinors $y^a$ to keep track of the R-symmetry indices. For $\mathfrak{su}(2)$ and $\mathfrak{sl}(2)$, only one independent cross-ratio can be formed from four points. These lower dimensional cross-ratios are defined in terms of the coordinates as
\begin{equation}
\chi = \frac{z_{12} z_{34}}{z_{13} z_{24}}, \hspace{1cm} \alpha = \frac{y_{13} y_{24}}{y_{12} y_{34}}\;, \label{cross-ratio3}
\end{equation}
and can be identified with the symbols in \eqref{cross-ratio2}. Note that $z_{ij} \equiv z_i - z_j$ and $y_{ij} \equiv y_i^a y_j^b \epsilon_{ab}$.

Covariance under $\mathfrak{sl}(2)$ and $\mathfrak{su}(2)$ constrains the possible terms in the OPE of the operators \eqref{polarization2}. To discuss these constraints individually, let us pretend for a moment that the $y$ and $z$ dependence can be separated.\footnote{Clearly, a pure $\mathfrak{su}(2)$ representation -- one with $j$ as the only non-zero quantum number -- cannot satisfy the unitarity bounds. We will also not see operators like this in the non-unitary chiral algebra discussed in section \ref{sec:4d}. Instead, to apply \eqref{z-ope} and \eqref{y-ope} to a reasonable theory, they should be combined into a single OPE. Note however that non-unitary multiplets, even when absent from the theory, play a role in determining the superconformal blocks \cite{oy16,sy18}.}
\begin{subequations}
\begin{align}
\mathcal{O}_{h_1}(z_1) \mathcal{O}_{h_2}(z_2) &= \sum_{\mathcal{O}} \lambda_{12 \mathcal{O}} \sum_{m = 0}^\infty \frac{(h_{12} + h)_m}{m! (2h)_m} \frac{\partial^m \mathcal{O}_h(z_2)}{z_{12}^{h_1 + h_2 - h - m}} \label{z-ope} \\
\mathcal{O}_{j_1}(y_1) \mathcal{O}_{j_2}(y_2) &= \sum_{\mathcal{O}} \lambda_{12 \mathcal{O}} \frac{1}{2j!} \sum_{\sigma \in S_{2j}} \mathcal{O}_{j | a_{\sigma(1)} \dots a_{\sigma(2j)}} y_1^{a_1} \dots y_1^{a_{j + j_{12}}} y_2^{a_{j + j_{12} + 1}} \dots y_2^{a_{2j}} \label{y-ope}
\end{align}
\end{subequations}
In both cases, the outer sum runs over (quasi)primaries while the inner sum covers the states other than the highest weight. The chiral OPE \eqref{z-ope}, first applied to the search for $\mathcal{W}$-algebras in \cite{b91}, leads to the familiar $s$-channel expansion of the four-point function.
\begin{align}
\left < \mathcal{O}_{h_1}(z_1) \mathcal{O}_{h_2}(z_2) \mathcal{O}_{h_3}(z_3) \mathcal{O}_{h_4}(z_4) \right > &= \left ( \frac{z_{24}}{z_{14}} \right )^{h_{12}} \left ( \frac{z_{14}}{z_{13}} \right )^{h_{34}} \frac{\mathcal{F}_{1234}(\chi)}{z_{12}^{h_1 + h_2} z_{34}^{h_3 + h_4}} \label{z-4pt} \\
\mathcal{F}_{1234}(\chi) &= \sum_{\mathcal{O}} \lambda_{12 \mathcal{O}} \lambda_{34 \mathcal{O}} \; g_h^{h_{12}, h_{34}}(\chi) \nonumber
\end{align}
where the $\mathfrak{sl}(2)$ blocks are given by
\begin{equation}
g_h^{h_{12}, h_{34}}(\chi) = \chi^h \; {}_2F_1(h - h_{12}, h + h_{34}; 2h; \chi)\;. \label{sl2-block}
\end{equation}
In fact, the coefficients in \eqref{z-ope} can be fixed by demanding that \eqref{z-4pt} is reproduced. A convenient result, which can be shown with embedding space methods \cite{cppr11a,cppr11b}, is that the second OPE \eqref{y-ope} leads to an entirely analogous expression with $h \mapsto -j$ everywhere.\footnote{This is far from the only reason to consider the OPE directly in embedding space \cite{fs19}.}
\begin{align}
\left < \mathcal{O}_{j_1}(y_1) \mathcal{O}_{j_2}(y_2) \mathcal{O}_{j_3}(y_3) \mathcal{O}_{j_4}(y_4) \right > &= \left ( \frac{y_{14}}{y_{24}} \right )^{j_{12}} \left ( \frac{y_{13}}{y_{14}} \right )^{j_{34}} y_{12}^{j_1 + j_2} y_{34}^{j_3 + j_4} \; \mathcal{F}_{1234}(\alpha) \label{y-4pt} \\
\mathcal{F}_{1234}(\alpha) &= \sum_{\mathcal{O}} \lambda_{12 \mathcal{O}} \lambda_{34 \mathcal{O}} \; g_{-j}^{j_{21}, j_{43}} \left ( \frac{1}{\alpha} \right ) \nonumber \\
&= \sum_{\mathcal{O}} \lambda_{12 \mathcal{O}} \lambda_{34 \mathcal{O}} \frac{(-1)^j (j + j_{34})! P_{j + j_{34}}^{j_{21} - j_{34}, j_{12} - j_{34}}(1 - 2\alpha)}{(-\alpha)^{j_{34}} (j - j_{34} + 1)_{j + j_{34}}} \nonumber
\end{align}
We have used a standard hypergeometric function identity in the last line, relating $\mathfrak{sl}(2)$ blocks with half-integer weights and Jacobi polynomials, in order to make contact with another convention in the literature. Note also that the kinematic $y_{ij}$ factors we extracted here are different from the ones in \eqref{unbarred-correlator}.

\section{Emergent Parisi-Sourlas supersymmetry}
\label{sec:ps}
In this section, we point out interesting hidden structures in tree-level maximally supersymmetric four-point Mellin amplitudes. We start by reviewing the results of \cite{az20a,az20} in section \ref{sec:3.1}. We then show in section \ref{sec:3.2} that correlators can be expressed as finite linear combinations of scalar exchange Witten diagrams, acted on by differential operators of cross-ratios. The combination coefficients  are highly special and exhibit an emergent Parisi-Sourlas symmetry. This allows the sum of Witten diagrams in each multiplet to be written in terms of  a single scalar exchange Witten diagram in an AdS space with four dimensions fewer. An analogous structure exists for the R-symmetry dependence, where the dimension of the internal manifold is similarly reduced by four. In section \ref{sec:3.3} we use these observations to provide a  compact way to rewrite the holographic correlators, which manifests supergraph-like structures. The form of the result also suggests a lower dimensional scalar seed theory which encodes all the essential data. Correlators of the original theory can be obtained by dressing the seed theory correlators with differential operators.

\subsection{Holographic Mellin amplitudes}\label{sec:3.1}
The natural language for holographic correlators is the Mellin representation \cite{m09,p10,fkprv11} which makes the analytic structure manifest. In the four-point case, the Mellin amplitude involves Mandelstam-like variables satisfying $s + t + u = \sum_{i = 1}^4 \Delta_i$. Tree-level Mellin amplitudes for a CFT with a weakly coupled gravity dual take the form
\begin{equation}
\mathcal{M}(s, t) = \mathcal{M}^{(s)}(s, t) + \mathcal{M}^{(t)}(s, t) + \mathcal{M}^{(u)}(s, t) + \mathcal{M}_{\mathrm{contact}}(s, t)\;. \label{m-is-stu}
\end{equation}
Here $\mathcal{M}^{(s)}$, $\mathcal{M}^{(t)}$, $\mathcal{M}^{(u)}$ correspond to exchange contributions, and have simple poles in the Mandelstam variable of the respective channel. $\mathcal{M}_{\mathrm{contact}}$ is regular and accounts for additional contact interactions.  Upon transforming back to position space, the poles in the Mellin amplitude -- from exchange Witten diagrams -- produce conformal blocks for an internal single-particle operator. Poles corresponding to double-particle operators are instead contained in the measure for the inverse Mellin transformation which is ideally suited to theories at large $N$.\footnote{One often hears these types of operators referred to as ``single-trace'' and ``double-trace'' respectively. Strictly speaking, this is only correct in the strict $N \rightarrow \infty$ limit. Even if one is not concerned with any microscopic Lagrangian, there is a well defined basis of double-particle operators which have correction terms in addition to normal ordered squares of single-particle operators \cite{Aprile:2019rep,az19}.}

Restoring the dependence on $\sigma$ and $\tau$, the inverse Mellin transformation is given by the following contour integral\footnote{The $4\pi i$, which might look strange, becomes $2\pi i$ again after mapping $s$, $t$ and $u$ to the variables which generalize more easily to arbitrary $n$-point correlators.}
\begin{align}
\mathcal{G}^{\mathrm{conn}}_{1234}(U, V; \sigma, \tau) = \int_{-i\infty}^{i\infty} & \frac{\textup{d}s \textup{d}t}{(4\pi i)^2} U^{\frac{s}{2} - a_s} V^{\frac{t}{2} - a_t} \mathcal{M}_{1234}(s, t; \sigma, \tau) \label{inverse-mellin} \\
&\times \Gamma \left [ \frac{\Delta_1 + \Delta_2 - s}{2} \right ] \Gamma \left [ \frac{\Delta_1 + \Delta_4 - t}{2} \right ] \Gamma \left [ \frac{\Delta_1 + \Delta_3 - u}{2} \right ] \nonumber \\
&\times \Gamma \left [ \frac{\Delta_3 + \Delta_4 - s}{2} \right ] \Gamma \left [ \frac{\Delta_2 + \Delta_3 - t}{2} \right ] \Gamma \left [ \frac{\Delta_2 + \Delta_4 - u}{2} \right ] \nonumber
\end{align}
where 
\begin{equation}\label{asat}
a_s=\frac{\epsilon}{2}(k_1+k_2)-\epsilon\mathcal{E}\;,\quad a_t=\epsilon\mathcal{E}-\frac{\epsilon}{2}(k_1-k_4)\;,
\end{equation}
and we recall that $\Delta_i=\epsilon k_i$. To make sense of this, we need to realize that the shape of the contour is different from the schematic one above. The well known prescription is to choose a contour which makes the real part positive in the arguments of all six gamma functions. This is always possible for a $G_{\mathrm{conn}}$ which does not require regularization. This restriction on the contour, based on double-particle poles, is not sufficient because the critical strip described above usually contains some of the single-particle poles of the Mellin amplitude itself. To resolve this ambiguity, we must put single-particle and double-particle exchanges on the same footing, \textit{i.e.}, by choosing a contour such that the poles of $\mathcal{M}^{(s)}$ and $\mathcal{M}^{(t)}$ lie to the right while those of $\mathcal{M}^{(u)}$ lie to the left. In other words, whether we are talking about the gamma functions or Mellin amplitudes, we keep semi-infinite sequences of poles that increase the exponents but not semi-infinite sequences of poles that decrease them.\footnote{\label{overlapping} In the case of $AdS_5 \times S^5$ and $AdS_7 \times S^4$, the spectrum is such that the single-particle sequences are in fact finite \cite{dfr99}. The truncation occurs as a consistency condition of the $1/N$ expansion \cite{rz17}. The residues of $\mathcal{M}(s, t; \sigma, \tau)$ must conspire to stop the single-particle and double-particle poles from overlapping which would lead to higher order singularities that are unphysical at tree-level. This phenomenon was historically very important in the holographic correlator program before methods were developed in \cite{z17} to crack the $AdS_4 \times S^7$ case as well.} Clearly, these towers have integer spacing because an exchanged primary appears together with its descendants.\footnote{We should note that this logic only defines a correlator when there is some notion of which poles belong together. In the non-perturbative context \cite{psz19}, a Mellin amplitude is only useful if the appropriate contour is specified along with it.}

The Mellin amplitudes of interest to us were constructively derived for all external half-BPS operators in \cite{az20a,az20} for all three  maximally supersymmetric backgrounds. The derivation exploited a special R-symmetry configuration, dubbed maximally R-symmetry violating (MRV) in  \cite{az20a,az20}, where major simplifications occur. The full amplitudes were then obtained from the MRV amplitudes by using symmetries. A remarkable feature in these results is that all the contact terms in the amplitudes vanish, after using a natural prescription to symmetrize the exchange amplitudes recovered from the MRV limit. Therefore the full amplitudes can be written as the sum over only exchange amplitudes in three channels. Specifically, \cite{az20} gives a formula for the Mellin amplitudes as a sum over simple poles
\begin{equation}
\mathcal{M}_{1234}^{(s)}(s, t; \sigma, \tau) = \sum_{\substack{p = \mathrm{max}\{|k_{12}|, |k_{34}|\} + 2 \\ \mathrm{step} \; 2}}^{\mathrm{max}\{|k_{12}|, |k_{34}|\} + 2\mathcal{E} - 2} \sum_{m = 0}^\infty \sum_{0 \leq i + j \leq \mathcal{E}} \sigma^i \tau^j \frac{\mathcal{R}^{p, m; i, j}_{1234}(t, u)}{s - 2m - \epsilon p} \label{s-channel}
\end{equation}
with
\begin{eqnarray}
\mathcal{M}_{1234}^{(t)}(s, t; \sigma, \tau) &=& \tau^{\mathcal{E}} \mathcal{M}_{3214}^{(s)} \left ( t, s; \frac{\sigma}{\tau}, \frac{1}{\tau} \right ) \nonumber \\
\mathcal{M}_{1234}^{(u)}(s, t; \sigma, \tau) &=& \sigma^{\mathcal{E}} \mathcal{M}_{4231}^{(s)} \left (u, t; \frac{1}{\sigma}, \frac{\tau}{\sigma} \right ) \label{tu-crossing}
\end{eqnarray}
determining the other channels.\footnote{Identities for the $s$-channel amplitude itself, which might be useful to keep in mind, are
\begin{equation}
\mathcal{M}^{(s)}_{1234}(s, t; \sigma, \tau) = \mathcal{M}^{(s)}_{3412}(s, t; \sigma, \tau) = \mathcal{M}^{(s)}_{2134}(s, u; \tau, \sigma). \label{s-identities}
\end{equation}
}
The integer $p$ labels the exchanged supergravity multiplets whose super primaries are dual to the half-BPS operators $\mathcal{O}_p$. The R-symmetry selection rule, and the requirement of the effective action being finite, restrict the range of $p$ to the finite set as indicated in the above sum. 

For the reader's convenience, we now give explicit expressions for the residues. In \cite{az20a,az20}, they took the form
\begin{equation}\label{residueRa}
\mathcal{R}^{i,j}_{p,m}(t, u)= K^{i,j}_{p}(t,u)\, L^{i,j}_{p,m}\, N^{i,j}_{p}\;,
\end{equation}
where we have dropped the position labels as we will often do when they are clear from the context. The factor of \eqref{residueRa} which encodes all the $t$ and $u$ dependence is
\begin{eqnarray}
\nonumber K^{i,j}_{p}&=& 2i(2i+\kappa_u)t^-t^++2j(2j+\kappa_t)u^-u^+-2j(\tfrac{2}{\epsilon}-2+\kappa_u)t^+u^--2i(\tfrac{2}{\epsilon}-2+\kappa_t)u^+t^-\\
\nonumber &+& \frac{1}{4}(2p-\kappa_t-\kappa_u)(2p+\tfrac{4}{\epsilon}-4+\kappa_t+\kappa_u)(u^- t^-+4\epsilon^2 ij)\\
\nonumber &+&\tfrac{\epsilon}{2}(\kappa_u+\kappa_t-2p)(\kappa_u+\kappa_t+2p+\tfrac{4}{\epsilon}-4)(i t^-+ju^-)\\
\label{k-expression} &+&4\epsilon ij(t^+(\kappa_u+\tfrac{2}{\epsilon}-2)+u^+(\kappa_t+\tfrac{2}{\epsilon}-2))-8ijt^+u^+\;,
 \end{eqnarray}
 where we have defined
 \begin{equation}
\kappa_s\equiv|k_3+k_4-k_1-k_2|\;,\;\; \kappa_t\equiv|k_1+k_4-k_2-k_3|\;,\;\; \kappa_u\equiv|k_2+k_4-k_1-k_3|\;,
\end{equation}
and introduced the shorthand notation
\begin{equation}
u^\pm=u\pm \frac{\epsilon}{2}\kappa_u-\frac{\epsilon}{2}\Sigma\;,\quad\;\; t^\pm=t\pm \frac{\epsilon}{2}\kappa_t-\frac{\epsilon}{2}\Sigma
\end{equation}
with $\Sigma = k_1 + k_2 + k_3 + k_4$.

For the other two factors in \eqref{residueRa}, we will find it convenient to rewrite them in a way which is manifestly proportional to the OPE coefficients of the exchanged super primaries $\mathcal{O}_p$. In the resulting expression
\begin{equation}\label{residueR}
\mathcal{R}^{i,j}_{p,m}(t, u)= C_{k_1k_2p}\, C_{k_3k_4p}\, K^{i,j}_{p}(t,u)\, H^{i,j}_{p,m}\;,
\end{equation}
we have
\begin{align}
H^{i,j}_{p,m} &= \frac{2^{-\tfrac{1}{3} \left ( 2\epsilon - \tfrac{2}{\epsilon} + 9 \right )} \Gamma[\epsilon p]\Gamma[\epsilon (p + \tfrac{1}{\epsilon} - 1)] \Gamma[\frac{p + k_{12}}{2}] \Gamma[\frac{p - k_{12}}{2}] \Gamma[\frac{p + k_{34}}{2}] \Gamma[\frac{p - k_{34}}{2}]}{\Gamma[p]\Gamma[p + \tfrac{1}{\epsilon} - 1] \Gamma[\frac{\epsilon}{2}(p + k_{12})] \Gamma[\frac{\epsilon}{2}(p - k_{12})] \Gamma[\frac{\epsilon}{2}(p + k_{34})] \Gamma[\frac{\epsilon}{2}(p - k_{34})]} \\
&\times \frac{(-1)^{i+j+\frac{2p-\kappa_t-\kappa_u}{4}} \Gamma[\frac{\kappa_u+2+2i}{2}]^{-1} \Gamma[\frac{1}{4}(\tfrac{4}{\epsilon}-4+2p+\Sigma-\kappa_s-4l)] \Gamma[\frac{\kappa_t+2+2j}{2}]^{-1}}{i!j!m!\Gamma[2-\epsilon+m+\epsilon p]\Gamma[\frac{\epsilon(k_1+k_2-p)}{2} - m]\Gamma[\frac{2(p+2)-\Sigma+\kappa_s+4l}{4}]\Gamma[\frac{\epsilon(k_3+k_4-p)}{2} - m]}. \nonumber
\end{align}
and $l = \mathcal{E} - i - j$, which is the analogue of $u$ for the R-symmetry. The OPE coefficients, which come from the cubic couplings in the supergravity effective action, are given by \cite{lmrs98,cfm99,bz99}
\begin{equation}
C_{k_1 k_2 k_3} = \begin{cases}
\frac{\pi}{N^{\frac{3}{4}}}\frac{2^{-\beta-\frac{1}{4}}}{\Gamma[\frac{\beta}{2}+1]}\prod_{i=1}^3\frac{\sqrt{\Gamma[k_i+2]}}{\Gamma[\frac{\alpha_i+1}{2}]}, & d = 3 \\
\frac{\sqrt{k_1 k_2 k_3}}{N}, & d = 4 \\
\frac{2^{2\beta-2}}{(\pi N)^{\frac{3}{2}}}\Gamma[\beta]\prod_{i=1}^3 \frac{\Gamma[\alpha_i+\frac{1}{2}]}{\sqrt{\Gamma[2k_i-1]}}, & d = 6
\end{cases} \label{sugra-3pt}
\end{equation}
with 
\begin{equation}
\alpha_1 = \frac{1}{2}(k_2 + k_3 - k_1), \;\;\; \alpha_2 = \frac{1}{2}(k_3 + k_1 - k_2), \;\;\; \alpha_3 = \frac{1}{2}(k_1 + k_2 - k_3)\;, \label{alpha-def}
\end{equation}
 and $\beta=\alpha_1+\alpha_2+\alpha_3$.

\subsection{Parisi-Sourlas-like dimensional reduction}\label{sec:3.2}
The above answer gives explicit expressions for holographic correlators of four arbitrary half-BPS operators in all three maximally supersymmetry theories. The fact that they can be cast into the same form as functions of $\epsilon$ already shows an unexpected universality of amplitudes in diverse dimensions. In this subsection, we further analyze these amplitudes to uncover more interesting properties. We will show that the maximally supersymmetric Mellin amplitudes exhibit a surprising hidden structure, where a dimensionally reduced spacetime naturally arises. The emergence of a lower dimensional spacetime in these amplitudes is highly reminiscent of the Parisi-Sourlas dimensional reduction. 

Let us expose this structure by separating the amplitudes into different parts. We start from the AdS part of the amplitudes, which concerns only the dependence on the Mandelstam variables. First we observe that the Mandelstam variables appear in $K^{i,j}_{p}(t,u)$ as polynomials. We can bring this factor outside of the Mellin representation (\ref{inverse-mellin}) as differential operators via the following identification
\begin{equation}
U\partial_U\leftrightarrow (\tfrac{s}{2}-a_s)\times\;,\quad V\partial_V\leftrightarrow (\tfrac{t}{2}-a_t)\times\;.
\end{equation}
Therefore, we only need to focus on the sum over poles in $s$ which are labelled by $m$. Let us isolate the $m$-dependence from $H^{i,j}_{p,m}$, which reads
\begin{equation}\label{Aij}
\mathcal{A}_{p,m}=\frac{1}{m!\Gamma[2-\epsilon+m+\epsilon p]\Gamma[\frac{\epsilon(k_1+k_2-p)}{2} - m]\Gamma[\frac{\epsilon(k_3+k_4-p)}{2} - m]}\;,
\end{equation} 
and denote the sum as
\begin{equation}
F_p(s)=\sum_{m=0}^\infty \frac{\mathcal{A}_{p,m}}{s-\epsilon p-2m}\;.
\end{equation}
{\it A priori}, we do not expect the function $F_p(s)$ to have any special property. Remarkably, however, we find that the sum can always be written as a linear combination of exactly {\it three} scalar exchange Witten diagrams! More precisely, we have 
\begin{equation}\label{Fas3scalars}
F_p(s)=\mathcal{N}\big(\mathcal{M}^{(d)}_{\epsilon p}(s)+c_1 \mathcal{M}^{(d)}_{\epsilon p+2}(s)+c_2 \mathcal{M}^{(d)}_{\epsilon p+4}(s)\big)
\end{equation} 
where 
\begin{eqnarray}
\nonumber c_1&=&-\frac{\big(\frac{\epsilon(p+k_{12})}{2}\big)\big(\frac{\epsilon(p-k_{12})}{2}\big)\big(\frac{\epsilon(p+k_{34})}{2}\big)\big(\frac{\epsilon(p-k_{34})}{2}\big)}{2 (\epsilon p)_2\big(\frac{\epsilon(p-1)}{2}\big)_2}\;,\\
 c_2&=&\frac{\big(\frac{\epsilon(p+k_{12})}{2}\big)_2\big(\frac{\epsilon(p-k_{12})}{2}\big)_2\big(\frac{\epsilon(p+k_{34})}{2}\big)_2\big(\frac{\epsilon(p-k_{34})}{2}\big)_2}{(1+\epsilon(p-1))_3(\epsilon p)_4(2+\epsilon(p-1))}\;,
\end{eqnarray}
with $k_{ij}=k_i-k_j$. The overall coefficient $\mathcal{N}$ is not important here. $\mathcal{M}^{(d)}_\Delta(s)$ is the Mellin amplitude of a scalar exchange Witten diagram in $AdS_{d+1}$ of dimension $\Delta$, and with external dimensions $\Delta_i$. These scalar exchange Witten diagram amplitudes are given by
\begin{eqnarray}
\nonumber &&\mathcal{M}^{(d)}_\Delta(s)=\sum_{m=0}^\infty\frac{(-2)\Gamma[\Delta]\Gamma[\Delta-\epsilon]\Gamma[m+\Delta-\epsilon]^{-1}}{m!\Gamma[\frac{\Delta+\Delta_{12}}{2}]\Gamma[\frac{\Delta-\Delta_{12}}{2}]\Gamma[\frac{\Delta+\Delta_{34}}{2}]\Gamma[\frac{\Delta-\Delta_{34}}{2}]}\\
&&\quad\times \frac{1}{\Gamma[\frac{\Delta_1+\Delta_2-\Delta-2m}{2}]\Gamma[\frac{\Delta_3+\Delta_4-\Delta-2m}{2}](s-\Delta-2m)}
\end{eqnarray}
where $\Delta_{ij}=\Delta_i-\Delta_j$.

Meanwhile, we can make another striking observation: $F_p(s)$ is also secretly the scalar exchange Witten diagram $\mathcal{M}_{\epsilon p}^{(d-4)}(s)$ in {\it four dimensions lower}, with internal dimension $\epsilon p$ and the same external dimensions! The appearance of  $AdS_{d-3}$ in an $AdS_{d+1}$ supergravity amplitude is quite unexpected. However, this equivalence across dimensions can be naturally explained in terms of an emergent Parisi-Sourlas supersymmetry, as we will see below.

We first recall that Parisi-Sourlas supersymmetry is a geometrization of a hidden symmetry of stochastic equations. This supersymmetry is key to the celebrated conjecture of Parisi and Sourlas \cite{Parisi:1979ka}, and links together the IR fixed points of three seemingly unrelated models: a random field model in $d$ dimensions, a supersymmetric field theory without disorder in $d$ dimensions, and a model without disorder in $d-2$ dimensions. In particular, it claims that certain correlation functions in the supersymmetric theory should coincide with the $d-2$ dimensional theory. The Parisi-Sourlas supersymmetry is non-unitary, as the supercharges violate spin-statistics. For this reason, theories with Parisi-Sourlas supersymmetry have very unusual features, from a high energy physicist's point of view. Recently, this supersymmetry was scrutinized in \cite{Kaviraj:2019tbg} using conformal bootstrap techniques, where a number of kinematical results were derived. In particular, \cite{Kaviraj:2019tbg} showed that the Parisi-Sourlas supersymmetry leads to a beautiful identity for conformal blocks across different spacetime dimensions
\begin{equation}\label{gdimred}
g^{(d-2)}_{\Delta,\ell}=g^{(d)}_{\Delta,\ell}+C_{2,0}g^{(d)}_{\Delta+2,\ell}+C_{1,-1}g^{(d)}_{\Delta+1,\ell-1}+C_{0,-2}g^{(d)}_{\Delta,\ell-2}+C_{2,-2}g^{(d)}_{\Delta+2,\ell-2}
\end{equation}
where the $C_{i,j}$ are pure numbers found in \cite{Kaviraj:2019tbg}. The combination on the RHS can be interpreted as a single superconformal block under Parisi-Sourlas supersymmetry. Later it was pointed out in \cite{Zhou:2020ptb} that the Parisi-Sourlas supersymmetry can also be realized holographically. This gives rise to similar dimensional reduction identities for exchange Witten diagrams, obtained by replacing conformal blocks with the corresponding exchange Witten diagrams in AdS.\footnote{Similar generalizations are also found for two-point functions in boundary CFTs \cite{Zhou:2020ptb}, and CFTs on real projective space \cite{Giombi:2020xah}.} Here we will only focus on the relevant case with $\ell=0$. The relation (\ref{gdimred}) simplifies into
\begin{equation}
g^{(d-2)}_{\Delta,0}=g^{(d)}_{\Delta,0}+C_{2,0}g^{(d)}_{\Delta+2,0}\;,
\end{equation}
with 
\begin{equation}
C_{2,0}=-\frac{(\Delta-\Delta_{12})(\Delta+\Delta_{12})(\Delta-\Delta_{34})(\Delta+\Delta_{34})}{4(d-2\Delta-4)(d-2\Delta-2)\Delta(\Delta+1)}\;.
\end{equation}
The corresponding Witten diagram relation is \cite{Zhou:2020ptb}
\begin{equation}
\mathcal{M}^{(d-2)}_{\Delta}=\mathcal{M}^{(d)}_{\Delta}+C_{2,0}\mathcal{M}^{(d)}_{\Delta+2}\;.
\end{equation} 
By using the relation twice on the $AdS_{d-3}$ scalar exchange amplitude, we reach $AdS_{d+1}$ and reproduce precisely the combination (\ref{Fas3scalars}).

A similar dimensional reduction structure is also present in the R-symmetry part. Since $i$, $j$ appear in $K^{i,j}_{p}(t,u)$  polynomially, we can treat them as differential operators acting on the monomial $\sigma^i\tau^j$ in (\ref{s-channel})
\begin{equation}
\sigma\partial_\sigma \leftrightarrow i\times\;,\quad \tau\partial_\tau\leftrightarrow j\times\;,
\end{equation}
on the same footing as the Mandelstam variables. The remaining $i$-, $j$-dependence is contained in $H^{i,j}_{p,m}$, which we denote as 
\begin{equation}\label{Bij}
\mathcal{B}_{p}^{i,j}=\frac{(-1)^{i+j+\frac{2p-\kappa_t-\kappa_u}{4}} \Gamma[\frac{1}{4}(\frac{4}{\epsilon}-4+2p+\Sigma-\kappa_s-4l)]}{i!j!\Gamma[\frac{\kappa_u+2+2i}{2}]\Gamma[\frac{2(p+2)-\Sigma+\kappa_s+4l}{4}]\Gamma[\frac{\kappa_t+2+2j}{2}]}\;.
\end{equation}
We now resum $i$ and $j$ to get $\sum_{i,j}\mathcal{B}_{p}^{i,j}\sigma^i\tau^j$.
It turns out that the sum is nothing but the R-symmetry polynomial associated with the exchange of the rank-$p$ symmetric traceless representation 
\begin{equation}
\mathtt{Y}_{p}=\frac{(-1)^{\frac{p}{2}-\frac{\kappa_t+\kappa_u}{4}}}{\Gamma[\frac{8+2p-\kappa_t-\kappa_u}{4}]\Gamma[\frac{8+2p-\kappa_t+\kappa_u}{4}]\Gamma[\frac{2+\kappa_t}{2}]}\mathcal{Y}_{p,0}\bigg|_{\mathtt{d}=\frac{2}{\epsilon}}\;,
\end{equation}
but for a {\it different} R-symmetry group $SO(\tfrac{2}{\epsilon})$. Here the polynomial $\mathcal{Y}_{p,0}$ was given in \cite{az20a,az20} for any $SO(\mathtt{d})$ R-symmetry group
\begin{eqnarray}
&&\mathcal{Y}_{p,0}=\sum_{i,j}(-1)^{\frac{p}{2}-\frac{\kappa_t+\kappa_u}{4}}\sigma^i\tau^j\bigg[\frac{\Gamma[\frac{4+2p+\kappa_u-\kappa_t}{4}]}{i!\,j!\,(\frac{\kappa_u}{2})!\Gamma[\frac{\mathtt{d}}{2}+p-1]}\\
\nonumber &&\times\frac{(\frac{\kappa_t+\kappa_u-2p}{4})_{i+j}(\frac{\kappa_t+\kappa_u+2p+2\mathtt{d}-4}{4})_{i+j}\Gamma[\frac{-4+2\mathtt{d}+2p+\kappa_t+\kappa_u}{4}]}{(1+\frac{\kappa_u}{2})_i(1+\frac{\kappa_t}{2})_j}\bigg]\;,
\end{eqnarray}
and we have changed its normalization to make $\mathtt{Y}_{p}$ more symmetric. For the physical spacetime dimensions $\epsilon=\tfrac{1}{2},1,2$, the new R-symmetry dimension $\mathtt{d}=\frac{2}{\epsilon}$ is the same as shifting the original one by 4
\begin{equation}
\mathtt{d}\to\mathtt{d}-4\;.
\end{equation}
This implies the dimension of the internal sphere is reduced by four, which perfectly parallels the dimensional reduction structure for the AdS part.  

The above dimensional reduction may lead to negative dimensions. However, this does not give rise to problems. We formally view $d$ and $\mathtt{d}$ as continuous parameters, and analytically continue them to negative values. Clearly, Mellin amplitudes and R-symmetry polynomials can be defined for any values of $d$ and $\mathtt{d}$. Moreover, we regard $U$, $V$ and $\sigma$, $\tau$ as independent variables, as such is the case for $d>1$ and $\mathtt{d}>3$. 

As a side comment, the above Parisi-Sourlas-like dimensional reduction structure also explains why the prescription of \cite{az20a,az20} for recovering the full exchange amplitudes from the MRV limit is unique. Had a different prescription been used, the multiplet exchange amplitudes would differ by additional contact terms, which would ruin the above dimensional reduction structure.

\subsection{Full amplitudes from bosonic seed amplitudes}\label{sec:3.3}
It is time to take stock and put all the ingredients together. Using  observations from the last subsection, we will now give a compact new look to the results of \cite{az20a,az20}. We first restore the kinematic factor in (\ref{unbarred-correlator}) to manifest the Bose symmetry, and define 
\begin{eqnarray}
\mathbf{Y}^{(s)}_p(t_i)&&=\prod_{i<j} t_{ij}^{\gamma_{ij}^0}(t_{12}t_{34})^{\mathcal{E}}\mathtt{Y}_p(\sigma,\tau)\;,\\
\mathbf{M}^{(s)}_p(x_i)&&=\prod_{i<j} x_{ij}^{-2\epsilon\gamma_{ij}^0}(x^{2\epsilon}_{12}x^{2\epsilon}_{34})^{-\mathcal{E}}\times \int_{-i\infty}^{i\infty} \frac{dsdt}{(4\pi i)^2} U^{\frac{s}{2}-a_s}V^{\frac{t}{2}-a_t}\mathcal{M}^{(d-4)}_{\epsilon p}(s)\,\Gamma_{\{k_i\}}\;,
\end{eqnarray}
The exponents of $x^{-2\epsilon}_{ij}$ and $t_{ij}$ in the restored kinematic factor are collectively denoted here as $\gamma_{ij}^0$.
Then the four-point correlators for the $AdS_{d+1}\times S^{\mathtt{d}-1}$ background can be compactly written as 
\begin{equation}\label{GSsStSu}
G_{\{k_i\}}(x_i,t_i)=\sum_{p_s} \mathbf{S}^{(s)}_{p_s}+\sum_{p_t} \mathbf{S}^{(t)}_{p_t}+\sum_{p_u} \mathbf{S}^{(u)}_{p_u}
\end{equation}
where the contribution of each multiplet $\mathbf{S}^{(s)}_p$ is obtained from the action of a differential operator $\mathbf{K}^{(s)}_p$ on a scalar exchange diagram in $AdS_{d-3}\times S^{\mathtt{d}-5}$
\begin{equation}\label{SeqKYM}
\mathbf{S}^{(s)}_p=C_{k_1k_2p}C_{k_3k_4p} f_{\{k_i\},p}\,\mathbf{K}^{(s)}_p\circ\big(\mathbf{Y}^{(s)}_p \mathbf{M}^{(s)}_p\big)\;.
\end{equation}
Here $f_{\{k_i\},p}$ is a simple universal factor
\begin{eqnarray}
f_{\{ k_i \}, p} &=& - 2^{-\tfrac{1}{3} \left ( 2\epsilon - \tfrac{2}{\epsilon} + 12 \right )} \frac{\Gamma[\frac{p + k_{12}}{2}] \Gamma[\frac{p - k_{12}}{2}] \Gamma[\frac{p + k_{34}}{2}] \Gamma[\frac{p - k_{34}}{2}]}{\epsilon \Gamma[p] \Gamma[p + \frac{1}{\epsilon}]}
\end{eqnarray}
describing what is left after \eqref{Aij}, \eqref{Bij} and the normalization from \eqref{Fas3scalars} are taken out of $H^{i, j}_{p, m}$.\footnote{If we wanted to go back and write the original residue in the form \eqref{SeqKYM}, we would have
\begin{equation}
\mathcal{R}^{i,j}_{p,m}(t,u)= C_{k_1k_2p} C_{k_3k_4p}\, \mathcal{N}^{-1}\, f_{\{ k_i \}, p}\, K^{i,j}_{p}(t,u)\, \mathcal{A}_{p,m}\, \mathcal{B}^{i,j}_{p}
\end{equation}
with
\begin{equation}
\mathcal{N}^{-1} = -\frac{2 \Gamma[\epsilon p] \Gamma[\epsilon p - \epsilon + 2]}{\Gamma[\frac{\epsilon}{2}(p + k_{12})] \Gamma[\frac{\epsilon}{2}(p - k_{12})] \Gamma[\frac{\epsilon}{2}(p + k_{34})] \Gamma[\frac{\epsilon}{2}(p - k_{34})]}.
\end{equation}
}
The operator $\mathbf{K}^{(s)}_p$ follows from $K^{i,j}_{p}(t,u)$ in (\ref{residueR}), and reads
\begin{eqnarray}
\nonumber \mathbf{K}^{(s)}_p&=&16(D_{13}D_{24}\theta_{14}\theta_{23}+D_{14}D_{23}\theta_{13}\theta_{24}-2D_{13}D_{23}\theta_{13}\theta_{23})\\
\nonumber &-&8 D_{23}\theta_{23}(D_{24}-\epsilon \theta_{13})(\tfrac{2}{\epsilon}-2+\kappa_u)-8 D_{13}\theta_{13}(D_{14}-\epsilon \theta_{23})(\tfrac{2}{\epsilon}-2+\kappa_t)\\
 &+&(D_{24}-\epsilon \theta_{13})(D_{14}-\epsilon \theta_{23})(2p-\kappa_t-\kappa_u)(\tfrac{4}{\epsilon}-4+2p+\kappa_t+\kappa_u)\;,
\end{eqnarray}
where we have defined the shorthand notations
\begin{equation}
\theta_{ij}=t_{ij}\frac{\partial}{\partial t_{ij}}\;, \quad D_{ij}=x_{ij}^2\frac{\partial}{\partial x_{ij}^2}\;.
\end{equation}
We note that the structure of (\ref{SeqKYM}) is very reminiscent of the flat space results using supergraphs. 

The final answer (\ref{GSsStSu}) and (\ref{SeqKYM}) also suggests an effective lower dimensional ``seed'' theory which encodes all the essential data. The spectrum of the seed theory consists of only scalars, and coincides with the super-primary spectrum of the higher dimensional theory. Moreover, the scalars interact with the same cubic couplings as in the original theory. In the seed theory, correlators can be easily computed by
\begin{equation}
G^{\rm seed}_{\{k_i\}}(x_i,t_i)=\sum_p C_{k_1k_2p}C_{k_3k_4p}\,\mathbf{Y}^{(s)}_p \mathbf{M}^{(s)}_p+(t)+(u)\;.
\end{equation}
We can then simply perform the replacement
\begin{equation}
\mathbf{Y}^{(s)}_p \mathbf{M}^{(s)}_p \to  f_{\{k_i\},p}\,\mathbf{K}^{(s)}_p\circ\big(\mathbf{Y}^{(s)}_p \mathbf{M}^{(s)}_p\big)\;,
\end{equation}
to obtain the full correlator in the original theory
\begin{equation}
 G^{\rm seed}_{\{k_i\}}(x_i,t_i) \to G_{\{k_i\}}(x_i,t_i)\;.
\end{equation}

To conclude, let us mention that the Parisi-Sourlas-like dimensional reduction structure is not unique to maximally supersymmetric theories. The structure persists also in theories with half the amount of supersymmetry, {\it i.e.}, eight Poincar\'e supercharges. Some holographic four-point functions in such backgrounds were computed in \cite{Zhou:2018ofp} using bootstrap techniques. With reduced supersymmetry, the background $AdS_{d+1}$ is now reduced instead to $AdS_{d-1}$. A detailed analysis will be reported in a separate publication \cite{inprogress}.

\section{Chiral symmetry from six dimensions}
\label{sec:6d}

In this section we focus on the case of 11d supergravity on $AdS_7 \times S^4$, which is conjecturally dual to a 6d $\mathcal{N}=(2,0)$ SCFT of type $\mathfrak{g}=A_{N-1}$ at large $N$. As discussed in \cite{brv14}, the $\mathfrak{osp}(8^*|4)$ superconformal algebra contains an $\mathfrak{su}(1,1|2)$ subalgebra, and therefore according to the construction of \cite{bllprv13} such a theory admits a protected subsector of operators whose (twisted) correlation functions restricted to a plane form a chiral algebra.\footnote{The R-symmetry decomposes as $\mathfrak{su}(2) \times \mathfrak{u}(1) \subset \mathfrak{usp}(4)$ and all operators that can contribute to the chiral algebra are neutral under $\mathfrak{u}(1)$. The ``twisting-translating'' procedure
\begin{equation}
\mathcal{O}(z) = \mathcal{O}^{b_1 \dots b_{2j}}(z, \bar{z}, 0) \, u_{b_1}(\bar{z}) \dots u_{b_{2j}}(\bar{z}), \hspace{1cm} u(\bar{z}) \equiv (1, \bar{z}) \label{tw-tr}
\end{equation}
defines their representatives away from the origin.}
In \cite{brv14} it was furthermore conjectured that the chiral algebra associated to the type $\mathfrak{g}=A_{N-1}$ theory is $\mathcal{W}_{\mathfrak{g}}$: precisely the algebra appearing on the 2d side of the AGT correspondence \cite{agt09}. In this case, the full higher-spin algebra is known: it is the quantization of the classical $\mathcal{W}_N$ algebra, which in turn is the outcome of the Drinfel'd-Sokolov reduction of $\mathfrak{sl}(N)$. This also appears as the asymptotic symmetry algebra of $AdS_3$ higher-spin gravity, as discussed in \cite{cfpt10,cfp11}.

In \cite{brv14} two tests of this conjecture were proposed: the first based on the superconformal index, the second on the computation of the three-point functions between half-BPS operators at large $N$, comparing the latter with certain three-point functions in the chiral algebra $\mathcal{W}_{\mathfrak{g}}$. Here we focus on the second kind of test, providing further evidence for the validity of the conjecture. In the following, we shall focus on the sector of half-BPS operators and compute the four-point functions between the dual operators in the chiral algebra. Then, we shall use the four-point functions computed in \cite{az20,az20a} at large $N$, perform the twist described in section \ref{sec:setup}, and derive the same four-point functions directly from Mellin space, finding perfect agreement.

\subsection{$\mathcal{W}$-algebra correlators at large N}
\label{sec:4.1}

As discussed in \cite{brv14}, the ring of half-BPS operators in the 6d $\mathcal{N}=(2,0)$ theory is isomorphic to the ring that is freely generated by the Casimir invariants of $\mathfrak{g}$. This led to the conjecture that the chiral algebra associated with the 6d theory is $\mathcal{W}_{\mathfrak{g}}$, generated precisely by the meromorphic currents associated to the half-BPS operators. It was also shown that the central charge of $\mathcal{W}_{\mathfrak{g}}$ is given by
\begin{align}
c_{2d}=4\,N^3-3\,N-1\,.
\end{align} 
The chiral algebra generators, which we shall refer to as ${W}^{(k)}$, with $k\ge 2$, are bosonic primaries of dimension $k$. The three-point functions of such operators read
\begin{align}\label{6d_3pt}
\langle 
{W}^{(k_1)}(z_1)\,{W}^{(k_1)}(z_2)\,W^{(k_3)}(z_3)\rangle
=\frac{C_{k_1k_2k_3}}{z_{12}^{k_1+k_2-k_3}\,z_{13}^{k_1+k_3-k_2}\,z_{23}^{k_2+k_3-k_1}}\,,
\end{align}
where the coefficients $C_{k_1k_2k_3}$ are given in the supergravity approximation by \eqref{sugra-3pt} (in the case $d=6$). Note that these three-point functions are non-vanishing only when 
\begin{align}
k_3=|k_{12}|+2,\,|k_{12}|+4,\,...\,,k_1+k_2-2\,. \label{6d-triangle}
\end{align}
This means that, to this order, the OPE between generators is
\begin{align}
W^{(k_1)}(z_1) W^{(k_2)}(z_2) = \frac{\delta_{k_1,k_2}}{z_{12}^{2k_1}} + \sum_{\substack{p = |k_{12}| + 2 \\ \mathrm{step} \; 2}}^{k_1 + k_2 - 2} C_{k_1k_2p} \sum_{m = 0}^{k_1 + k_2 - p -1} \frac{(k_{12} + p)_m}{m!(2p)_m} \frac{\partial^m W^{(p)}(z_2)}{z_{12}^{k_1 + k_2 - p - m}} + \dots \label{6d_ope}
\end{align}
where we have specialized \eqref{z-ope} to this case and explicitly singled out the contribution of the identity. The terms hidden by $\dots$ in \eqref{6d_ope} come in two types -- those suppressed by higher powers of $N$ and/or those that are regular in $z_{12}$. A careful accounting of regular and singular terms will be important for the calculation of four-point functions which we now discuss. In particular, for operators that lead to a singular contribution to the four-point function (found by taking the OPE twice), we must keep track of whether they belong to singular terms of both OPEs or just one of them.

We are concerned with solving for the dynamical part of a four-point function $\mathcal{F}_{1234}(\chi)$ which should be expanded in the $\mathfrak{sl}(2)$ blocks given by \eqref{sl2-block}. Once this is known, the kinematic prefactor can be restored by \eqref{z-4pt}. The chiral algebra four-point functions for equal weights ($k_i=k$) have been computed in \cite{rz17b} for $k=2,3,4$, using the holomorphic bootstrap method of \cite{hmpz15}. Being a bootstrap approach, this exploits the fact that the dynamical function must satisfy
\begin{subequations}
\begin{align}
\mathcal{F}_{1234}(\chi) &= (-1)^{\Sigma}\,
\chi^{k_1+k_2}\,(1-\chi)^{-k_2-k_3}\,\mathcal{F}_{3214}(1-\chi) \label{6d-crossing-13} \\
\mathcal{F}_{1234}(\chi) &= (-1)^{\Sigma}\,\chi^{k_1+k_4}\,\mathcal{F}_{4231} \left ( \frac{1}{\chi} \right ) \label{6d-crossing-14}
\end{align}
\end{subequations}
which generate the $S_3$ group of crossing transformations.\footnote{The choice to focus on the crossed configuration \eqref{6d-crossing-13} which switches $(1 \leftrightarrow 3)$, instead of the one that switches $(2 \leftrightarrow 4)$, is such to simplify the argument that we shall present below, which allows us to neglect normal ordered products of operators both in the $s$- and in the $t$-channel OPE.} Here we compute such correlators using an alternative (but equivalent) method that we shall summarize below. This also allows us to give closed form expressions for the four-point functions with general weights.
To illustrate the idea, let us first recall the basic fact that chiral algebra correlators are meromorphic functions of the coordinates. Here, after stripping off the kinematic prefactor, we obtain meromorphic functions of the cross-ratio $\chi$, whose only possible singularities in the complex $\chi$-plane are poles at $\chi=0,1,\infty$, corresponding to singular terms in the OPE. This is, at least in principle, enough to fix the correlation functions completely. To exploit this information in a convenient way, we make an ansatz for the four-point function as
\begin{align}\label{6dansatz}
\mathcal{F}_{1234}(\chi)=\mathcal{F}^{(s)}_{1234}(\chi)+\mathcal{F}^{(t)}_{1234}(\chi)\,,
\end{align}
where $\mathcal{F}^{(s)}$ contains the singular terms coming from the $s$-channel OPE, while $\mathcal{F}^{(t)}$ is its crossing-symmetric completion. By this, we mean that $\mathcal{F}^{(t)}$ is the image of $\mathcal{F}^{(s)}$ under \eqref{6d-crossing-13}.
This ensures that \eqref{6dansatz} is crossing symmetric. To fix the correlators, all that is left to do is then to constrain the form of $\mathcal{F}^{(s)}$. As anticipated, this is done by requiring that, in the small-$\chi$ expansion, it contains the correct powers of $\chi$ with coefficients that correspond to singular terms in the $s$-channel OPE. 

A comment is now in order. We would like to find these correlators purely from the knowledge of the three-point functions \eqref{6d_3pt}, but this is possible only if, at large $N$, these are the only contributions to singular terms in the OPE. The OPE between two generators $W^{(k_1)}$ and $W^{(k_2)}$ contains, in general, two kinds of $\mathfrak{sl}(2)$ multiplets: those built on top of another chiral generator, $W^{(p)}$, but also those coming from normal ordered products of generators. At large $N$, the former have OPE coefficients that scale with $N^{-\frac{3}{2}}$, while the latter generically scale with more negative powers of $N$ starting at $N^{-3}$. There is, however, an exception: the operator $:W^{(k_1)}\,W^{(k_2)}:$, which appears with an order 1 coefficient. This will contribute to $\mathcal{F}^{(s)}$ if it is a singular term in the four-point function as a whole.

Let us then look at the four-point function $\langle W^{(k_1)}\,W^{(k_2)}\,W^{(k_3)}\,W^{(k_4)}\rangle$. There are two ways one can have a contribution of order $N^{-3}$ from potentially singular terms in the $s$-channel OPE: either the exchange of a generator $W^{(p)}$, or that of $:W^{(k_3)}\,W^{(k_4)}:$.\footnote{The prefactor chosen for the four-point function is such that
\begin{align}
\langle {W}^{(k_1)}(0)&{W}^{(k_2)}(\chi)\,{W}^{(k_3)}(1)\,{W}^{(k_4)}(\infty)\rangle=\chi^{-k_1-k_2}\,\mathcal{F}_{1234}(\chi)\,,
\end{align}
so that $:W^{(k_1)}\,W^{(k_2)}:$ is always a regular term, while  $:W^{(k_3)}\,W^{(k_4)}:$ is regular only when $k_1+k_2\le k_3+k_4$. The crossing relation chosen in \eqref{6d-crossing-13}, instead, is such that $:W^{(k_2)}\,W^{(k_3)}:$ is always a regular term, while  $:W^{(k_1)}\,W^{(k_4)}:$ is regular only when $k_2+k_3\le k_1+k_4$.} The latter appears with an OPE coefficient of order $N^{-3}$ in $W^{(k_1)}\times W^{(k_2)}$, and order $1$ in $W^{(k_3)}\times W^{(k_4)}$, hence we are allowed to ignore it only if it is a regular term, which happens if $k_1+k_2 \le k_3+k_4$. A similar story applies to the $t$-channel OPE, which leads to the requirement that $k_2+k_3 \le k_1+k_4$. Hence, we should carry out the computation with these two assumptions, so that we will be able to neglect all the normal ordered products of operators in the OPE. All other orderings can be obtained from this using crossing symmetry.

Under the assumptions above, the chiral algebra generators $W^{(p)}$ and their $\mathfrak{sl}(2)$ descendants are the only singular terms relevant to tree-level in the $s$- and the $t$-channel OPEs. Hence, we make an ansatz for $\mathcal{F}^{(s)}$ as
\begin{align}\label{6d_s_ansatz}
\mathcal{F}^{(s)}_{1234}(\chi)=\sum_{n=\max\{|k_{12}|,|k_{34}|\}+2}^{\min\{k_1+k_2,k_3+k_4\}-1}a_n\,\chi^n\,,
\end{align}
for some coefficients $a_n$ to be determined, and we demand that this satisfies the right OPE, namely that
\begin{align}\label{6d_sOPE_ansatz}
\mathcal{F}^{(s)}_{1234}(\chi)=\left[\sum_{\substack{p=\max\{|k_{12}|,|k_{34}|\}+2 \\ \text{step}\,2}}^{k_1 + k_2 - 2}C_{k_1k_2p}\,C_{k_3k_4p}\,g_p^{k_{12},k_{34}}(\chi)\right]_{\text{trunc}}\,,
\end{align}
where the subscript ``trunc'' refers to the fact that we should expand the expression enclosed in brackets for small $\chi$, and truncate it at an order corresponding to the highest power of $\chi$ that appears in \eqref{6d_s_ansatz}, namely $k_1 + k_2 - 1$.
This allows us to extract the coefficients $a_n$ in a direct way: we simply project the right hand side of \eqref{6d_sOPE_ansatz} onto each power of $\chi$ appearing in the ansatz \eqref{6d_s_ansatz}, for instance using the orthogonality relation
\begin{align}
\oint_{C_0}\frac{d\chi}{2\pi i\,\chi}\chi^{n-m}=\delta_{n,m}\,,
\end{align}
where $C_0$ is a contour the encircles $\chi=0$ in the complex plane. A direct computation then gives the result
\begin{align}
a_n=\sum_{\substack{p=\max\{|k_{12}|,|k_{34}|\}+2 \\ \text{step}\,2}}^{k_1 + k_2 - 2} C_{k_1k_2p}C_{k_3k_4p} \frac{(p-k_{12})_{n-p}\,(p+k_{34})_{n-p}}{(2h)_{n-p}\,(n-p)!}\,.
\end{align}
As discussed above, this expression is valid only under the assumption that $k_1+k_2\le k_3+k_4$ and $k_2+k_3 \le k_1+k_4$. Using crossing symmetry one can extend these results to any configuration. The four-point function that follows from \eqref{6d_s_ansatz}, with these coefficients, can now be recast into the form
\begin{align}
\left < W^{(k_1)}(z_1)W^{(k_2)}(z_2)W^{(k_3)}(z_3)W^{(k_4)}(z_4) \right > &= \left ( \frac{z_{24}}{z_{14}} \right )^{k_{12}} \left ( \frac{z_{14}}{z_{13}} \right )^{k_{34}} \frac{\mathcal{F}_{1234}(\chi)}{z_{12}^{k_1 + k_2} z_{34}^{k_3 + k_4}} \nonumber \\
k_1 + k_2 \leq k_3 + k_4 \quad &, \quad k_2 + k_3 \leq k_1 + k_4 \label{6d-answer-form}
\end{align}
with
\begin{align}
\mathcal{F}_{1234}(\chi) &= \chi^{k_1 + k_2} \left [ \sum_{\substack{p = |k_{12}| + 2 \\ \mathrm{step} \; 2}}^{k_1 + k_2 - 2} C_{k_1k_2p} C_{k_3k_4p} \sum_{m = 0}^{k_1 + k_2 - p - 1} \frac{(p - k_{12})_m (p + k_{34})_m}{m! (2p)_m \chi^{k_1 + k_2 - p - m}}  \right. \nonumber \\
& \left. + \sum_{\substack{p = |k_{23}| + 2 \\ \mathrm{step} \; 2}}^{k_2 + k_3 - 2} C_{k_2k_3p} C_{k_1k_4p} \sum_{m = 0}^{k_2 + k_3 - p - 1} \frac{(p + k_{23})_m (p + k_{14})_m}{m! (2p)_m (1 - \chi)^{k_2 + k_3 - p - m}} \right ]. \label{6d-answer}
\end{align}
This will be convenient for the next subsection.

We give a final reminder that, due to the $\delta_{k_1, k_2}$ in \eqref{6d_ope}, there is one more term that should be added when the weights are pairwise equal. This is $a_0 = 1$, in contrast to the other ones that are proportional to $C_{k_1k_2p}C_{k_3k_4p} = O(N^{-3})$. It is easy to see that this gives the same disconnected correlator that we get after applying the superconformal twist to generalized free theory in six dimensions.

\subsection{Matching from Mellin space}
\label{sec:4.2}

We now turn our attention to the Mellin amplitudes presented in \eqref{residueR} for $AdS_7\times S^4$. We wish to derive the same chiral algebra correlators \eqref{6d-answer} by evaluating \eqref{inverse-mellin} in a twisted configuration.\footnote{To explain the abbreviated notation, notice that the sums in \eqref{6d-answer} actually do not start until $p = \mathrm{max}\{ |k_{12}|, |k_{34}| \} + 2$ and $p = \mathrm{max}\{ |k_{23}|, |k_{14}| \} + 2$. This is because the $C_{k_1k_2k_3}$ are only given by \eqref{sugra-3pt} when the triangle inequality is satisfied. As stated in \eqref{6d-triangle}, they are zero otherwise.} Due to the inequalities in \eqref{6d-answer-form}, we will focus on the ordering $k_2 \leq k_1 \leq k_3 \leq k_4$.\footnote{We could just as well consider $k_2 \leq k_3 \leq k_1 \leq k_4$ since \eqref{6d-answer} is manifestly symmetric under $(1 \leftrightarrow 3)$.} Based on this, the Mellin amplitudes quoted in \eqref{s-channel} are associated with the correlator that comes from switching $(1 \leftrightarrow 2)$ in \eqref{unbarred-correlator}.
\begin{align}
& \left < \mathcal{O}_1(x_1, t_1)\mathcal{O}_2(x_2, t_2)\mathcal{O}_3(x_3, t_3)\mathcal{O}_4(x_4, t_4) \right > = \left ( \frac{t_{34}}{x_{34}^4} \right )^{\frac{1}{2}(k_3 + k_4 - k_1 - k_2)} \left ( \frac{t_{14}}{x_{14}^4} \right )^{\frac{1}{2}(k_1 + k_4 - k_2 - k_3)} \nonumber \\
& \hspace{2cm} \left ( \frac{t_{13}}{x_{13}^4} \right )^{\frac{1}{2}(k_1 + k_2 + k_3 - k_4) - \mathcal{E}} \left ( \frac{t_{24}}{x_{24}^4} \right )^{k_2 - \mathcal{E}} \left ( \frac{t_{12} t_{34}}{x_{12}^4 x_{34}^4} \right )^{\mathcal{E}} \mathcal{G}_{2134} \left ( \frac{U}{V}, \frac{1}{V}; \tau, \sigma \right ) \label{change-prefactor1}
\end{align}
On the other hand, we should change to a different prefactor to facilitate direct comparisons to \eqref{6d-answer}.
\begin{align}
& \left < \mathcal{O}_1(x_1, t_1)\mathcal{O}_2(x_2, t_2)\mathcal{O}_3(x_3, t_3)\mathcal{O}_4(x_4, t_4) \right > = \left ( \frac{t_{24}}{x_{24}^4} \right )^{-\frac{1}{2} k_{12}} \left ( \frac{t_{14}}{x_{14}^4} \right )^{\frac{1}{2} k_{12} - \frac{1}{2} k_{34}} \left ( \frac{t_{13}}{x_{13}^4} \right )^{\frac{1}{2} k_{34}} \nonumber \\
& \hspace{2.5cm} \left ( \frac{t_{12}}{x_{12}^4} \right )^{\frac{1}{2}(k_1 + k_2)} \left ( \frac{t_{34}}{x_{34}^4} \right )^{\frac{1}{2}(k_3 + k_4)} \mathcal{F}_{1234}(U, V; \sigma, \tau) \label{change-prefactor2}
\end{align}
Relating one function to the other and going to the Mellin representation gives
\begin{align}
\mathcal{F}_{1234}(U, V; \sigma, \tau) &= (\sigma U^2)^{\frac{1}{2}(k_1 + k_2) - \mathcal{E}}\, \mathcal{G}_{2134} \left ( \frac{U}{V}, \frac{1}{V}; \tau, \sigma \right ) \label{pre-twist-6d} \\
&= \int_{-i\infty}^{i\infty} \frac{\textup{d}s \textup{d}t}{(4\pi i)^2} U^{\frac{s}{2}} V^{-\frac{s}{2} - \frac{t}{2} + k_1 + k_4} \sigma^{\frac{1}{2}(k_1 + k_2) - \mathcal{E}} \mathcal{M}_{2134}(s, t; \tau, \sigma) \Gamma_{\{ k_i \}}. \nonumber
\end{align}
Again, the contour keeps poles such that the exponents on $U$ and $V$ form increasing sequences. In this case, that means the $s$-channel and the $u$-channel.

After we implement the superconformal twist, the integrand of \eqref{pre-twist-6d} will depend on $\chi^\prime$. The key observation we make is that, by meromorphy of the final answer, $\chi^\prime$ may be set to any value which makes the evaluation of \eqref{pre-twist-6d} convenient. This does not constitute any additional assumption on the four-point function apart from the statement that the superconformal Ward identity \eqref{scwi-6d} holds. We will now take $\chi^\prime \rightarrow 0$ which corresponds precisely to the $s$-channel MRV limit. We should therefore expect the resulting function to decompose into the two pieces of \eqref{6dansatz}.

After we do this, the largest pole in $s$ that we need is $s = 2k_1 + 2k_2$. Due to the ordering being considered, this is also the smallest pole in $s$. It just barely contributes if we set $\tau^{\mathcal{E} - j} \sigma^j \mapsto \chi^{\prime -2\mathcal{E}}$ in the polynomial and drop all lower degree terms. We now get
\begin{align}
\mathcal{F}_{1234}(\chi) = \frac{1}{2} \Gamma (\kappa_s) & \int_{-i\infty}^{i\infty} \frac{\textup{d}t}{2\pi i} \chi^{k_1 + k_2} (1 - \chi)^{k_{42} - \frac{t}{2}} \mathcal{M}_{2134}(2k_1 + 2k_2, t; \chi^{\prime -2}, \chi^{\prime -2}) \nonumber \\
&\times \Gamma \left [ k_1 + k_3 - \frac{t}{2} \right ] \Gamma \left [ k_2 + k_4 - \frac{t}{2} \right ] \Gamma \left [ k_{13} + \frac{t}{2} \right ] \Gamma \left [ k_{24} + \frac{t}{2} \right ]. \label{mid-twist-6d}
\end{align}
Since the function is guaranteed to have the form $\mathcal{F}(\chi) = \mathcal{F}^{(s)}(\chi) + \mathcal{F}^{(t)}(\chi)$, it is enough to compute $\mathcal{F}^{(t)}(\chi)$ by extracting the terms that have a negative power of $1 - \chi$. These come from the single-particle poles in $u$ -- namely $t = 2k_3 + 2k_4 - 2m - 2p$. Clearly, $\mathcal{F}^{(s)}(\chi)$ can then be computed using crossing symmetry. Looking at the expression for $\mathcal{R}^{p, m; 0, j}_{4132}(2k_3 + 2k_4 - 2m - 2p, 2k_1 + 2k_2)$,
\begin{align}
& \mathcal{F}^{(t)}(\chi) = \chi^{k_1 + k_2} \frac{\Gamma(\kappa_s)}{16} \sum_{\substack{p = \mathrm{max} \{ |k_{14}|, |k_{23}| \} + 2 \\ \mathrm{step} \; 2}}^{k_2 + k_3 - 2} C_{k_1k_4p} C_{k_2k_3p} \sum_{m = 0}^{k_2 + k_3 - p - 1} \frac{\pi^{-\frac{1}{2}} 2^{2p - 2}}{m! (2p)_m (1 - \chi)^{k_2 + k_3 - p - m}} \nonumber \\
& \hspace{0.25cm} \frac{\Gamma \left [ \frac{p + k_{14}}{2} \right ] \Gamma \left [ \frac{p - k_{14}}{2} \right ] \Gamma \left [ \frac{p + k_{23}}{2} \right ] \Gamma \left [ \frac{p - k_{23}}{2} \right ]}{\Gamma(p - k_{14}) \Gamma(p - k_{23}) (p + k_{14})_m^{-1} (p + k_{23})_m^{-1}} \sum_{i = 0}^\mathcal{E} \mathcal{B}^{0, j}_p K^{0, j}_p \binom{2k_3 + 2k_4 - 2m - 2p,}{2k_1 + 2k_2} \label{post-twist-6d}
\end{align}
where
\begin{align}
& K_p^{0, j} (2k_3 + 2k_4 - 2m - 2p, 2k_1 + 2k_2) = K^\prime + j K^{\prime\prime} \label{k-expansion} \\
& K^\prime \equiv -\frac{1}{2} \kappa_s (\kappa_s - \kappa_u - 2m - 2p) (2p - \kappa_s - \kappa_u) (2p - 2 + \kappa_s + \kappa_u) \nonumber \\
& K^{\prime\prime} \equiv 4\kappa_s(\kappa_s - 1)(\kappa_s + \kappa_u - 2m - 2p) + 2\kappa_s (2p - \kappa_s - \kappa_u) (2p - 2 + \kappa_s + \kappa_u). \nonumber
\end{align}
Note that, compared to \eqref{k-expression}, the permutation of the four points causes $\kappa_s$ to appear in place of $\kappa_u$ and $\kappa_u$ to appear in place of $\kappa_t$. There are now $O(1)$ and $O(j)$ contributions to the inner sum of \eqref{post-twist-6d}. For $O(1)$,
\begin{align}
\sum_{j = 0}^{\mathcal{E}} \mathcal{B}^{0,j}_p K^\prime &= \frac{\Gamma \left ( \frac{3 - \kappa_s}{2} \right ) \Gamma \left ( \frac{p - 1 - \kappa_u - 2\mathcal{E} + k_2 + k_3}{2} \right ) \Gamma \left ( \frac{p - 1 - 2\mathcal{E} + k_2 + k_3}{2} \right )}{\Gamma \left ( \frac{\kappa_s + 2}{2} \right ) \Gamma \left ( \frac{p + 2 + \kappa_u + 2\mathcal{E} - k_2 - k_3}{2} \right ) \Gamma \left ( \frac{p + 2 + 2\mathcal{E} - k_2 - k_3}{2} \right )} \frac{K^\prime}{\pi} \nonumber \\
&= 2^{\kappa_s - 1} \frac{\kappa_s - 1}{\Gamma(\kappa_s + 1)} \frac{\Gamma \left ( \frac{p - 1}{2} + \frac{\kappa_s \pm \kappa_u}{4} \right )}{\Gamma \left ( \frac{p + 2}{2} - \frac{\kappa_s \pm \kappa_u}{4} \right )} \frac{K^\prime}{\sqrt{\pi}} \nonumber \\
&= 2^{\kappa_s - 1} \frac{\kappa_s - 1}{\Gamma(\kappa_s + 1)} \frac{\Gamma \left ( \frac{p - 1 - k_{14}}{2} \right ) \Gamma \left ( \frac{p - 1 - k_{23}}{2} \right )}{\Gamma \left ( \frac{p + 2 + k_{14}}{2} \right ) \Gamma \left ( \frac{p + 2 + k_{23}}{2} \right )} \frac{K^\prime}{\sqrt{\pi}}. \label{inner-sum2}
\end{align}
In the second line, we have used a gamma function identity. We have also used the fact that, even though $\mathcal{E}$ has a piecewise dependence on the weights, it can be traded for $\kappa_u$ which also does. In the third line, we have used the fact that $\{ \kappa_s + \kappa_u, \kappa_s - \kappa_u \} = \{ -2k_{14}, -2k_{23} \}$ even if we do not know which is which. The analysis for $O(j)$ proceeds similarly with
\begin{align}
\sum_{j = 0}^{\mathcal{E}} j \mathcal{B}^{0,j}_p K^{\prime\prime} &= \frac{2^{\kappa_s - 4}}{\Gamma(\kappa_s + 1)} (2p - 2 + \kappa_s + \kappa_u)(2p - \kappa_s - \kappa_u) \nonumber \\
&\times \frac{\Gamma \left ( \frac{p - 1 - k_{14}}{2} \right ) \Gamma \left ( \frac{p - 1 - k_{23}}{2} \right )}{\Gamma \left ( \frac{p + 2 + k_{14}}{2} \right ) \Gamma \left ( \frac{p + 2 + k_{23}}{2} \right )} \frac{K^{\prime\prime}}{\sqrt{\pi}}. \label{inner-sum3}
\end{align}
Putting \eqref{inner-sum2} and \eqref{inner-sum3} back into the original formula,
\begin{align}
\mathcal{F}^{(t)}(\chi) = \chi^{k_1 + k_2} & \sum_{\substack{p = \mathrm{max} \{ |k_{14}|, |k_{23}| \} + 2 \\ \mathrm{step} \; 2}}^{k_2 + k_3 - 2} C_{k_1k_4p} C_{k_2k_3p} \sum_{m = 0}^{k_2 + k_3 - p - 1} \frac{(p + k_{14})_m (p + k_{23})_m}{m!(2p)_m (1 - \chi)^{k_2 + k_3 - p - m}} \nonumber \\
&\times \frac{\frac{1}{2}(\kappa_s - 1)K^\prime + \frac{1}{16}(2p - 2 + \kappa_s + \kappa_u)(2p - \kappa_s - \kappa_u)K^{\prime\prime}}{2\kappa_s (p + k_{14})(p + k_{23})(p - k_{14} - 1)(p - k_{23} - 1)}. \label{confirmation-6d}
\end{align}
Since this is almost the desired expression, we just have to go back to \eqref{k-expansion} to verify that the fraction at the end is indeed $1$.

As a check of our result, the inverse Mellin transform can also be arranged so that we compute the dynamical function all at once instead of $\mathcal{F}^{(s)}$ and $\mathcal{F}^{(t)}$ individually. This is explained in Appendix \ref{sec:appa}.

\section{Chiral symmetry from four dimensions}
\label{sec:4d}

Let us now move to the case of Type IIB supergravity on $AdS_5\times S^5$, which is conjecturally dual to 4d $\mathcal{N}=4$ super Yang-Mills (SYM) with gauge algebra $\mathfrak{g}=A_{N-1}$, at large $N$. The relevant superconformal algebra in this case is $\mathfrak{psu}(2,2|4)$ which, as discussed in \cite{bllprv13}, admits a chiral algebra subsector whose currents come from \textit{Schur multiplets} of the 4d theory.\footnote{Again, the twisted translation \eqref{tw-tr} must be applied to the Schur operators. This time, the maximal subalgebra is $\mathfrak{su}(2) \times \mathfrak{su}(2) \times \mathfrak{u}(1) \subset \mathfrak{su}(4)$ which means that an $\mathfrak{su}(2)$ R-symmetry survives in the chiral algebra. In general there can be Schur operators with non-zero $\mathfrak{u}(1)$ charge but not in the OPE between two scalars \cite{bllprv13}.} Furthermore, when $\mathcal{N} > 2$ in $d=4$, the chiral algebra contains supercurrents of its own. This leads to a global subalgebra of $\mathfrak{psu}(1,1|2)$ when $\mathcal{N} = 4$. The minimal extension of this to a chiral algebra is the ``small'' $\mathcal{N} = 4$ superconformal algebra, which turns out to be the full answer for $N = 2$. In more general cases, the proposed chiral algebra for 4d $\mathcal{N}=4$ SYM is an $\mathcal{N}=4$ super $\mathcal{W}$-algebra with $\text{rank}(\mathfrak{g})$ generators that are in one-to-one correspondence with the Casimir invariants of $\mathfrak{g}$. As an initial check of this conjecture, the authors of \cite{bllprv13} showed that it predicts the correct superconformal index. They also showed that the chiral algebra for a superconformal gauge theory can in principle be constructed from BRST quantization.

More recent results include free-field realizations for $N = 2, 3, 4$ in \cite{bmr18} and a duality between the chiral algebra of 4d $\mathcal{N}=4$ SYM at large $N$ and an $AdS_3$ higher-spin Chern-Simons theory in \cite{br16}. The latter echoes the discussion in section \ref{sec:6d} wherein the 6d $\mathcal{N}=(2,0)$ theory is associated with the higher-spin algebra $\mathfrak{hs}[\mu]$. Here we perform a test based on computing the four-point functions between chiral algebra generators at large $N$, and comparing them with the holographic correlators between half-BPS operators \cite{rz16,rz17} using Mellin space techniques. Again, the symmetric form of the Mellin amplitudes found in \cite{az20a,az20} will be most convenient for our purposes.

\subsection{Super $\mathcal{W}$-algebra correlators at large N}
\label{sec:5.1}

As in the case of the 6d $\mathcal{N}=(2,0)$ theory, the ring of half-BPS operators is freely generated by the Casimir invariants of the gauge algebra $\mathfrak{g}$. In the case $\mathfrak{g}=A_{N-1}$, the central charge is 
\begin{align}
c_{2d}=-12\,c_{4d}=-3\,(N^2-1)\,.
\end{align} 
The global $\mathfrak{psu}(1,1|2)$ of the full higher-spin algebra has an $\mathfrak{su}(2)$ R-symmetry, and the chiral algebra generators form non-trivial irreducible representations of this algebra. Furthermore, they are superconformal primaries of half-BPS multiplets with respect to the global $\mathfrak{psu}(1,1|2)$. We will denote these currents as $J^{(k)}_{a_1 \dots a_{2j}}$ and contract them with polarization vectors as in \eqref{polarization2}. Their $\mathfrak{sl}(2)$ and $\mathfrak{su}(2)$ spins (which must be equal) are given by $h = j = \frac{k}{2}$. Note that for odd $k$ they are fermionic, having half-integer dimension. In this notation, their three-point functions can be written as
\begin{equation}
\langle {J}^{(k_1)}(z_1,y_1)\,{J}^{(k_2)}(z_2,y_2)\,{J}^{(k_3)}(z_3,y_3)\rangle = \frac{C_{k_1k_2k_3}}{\left ( \frac{z_{12}}{y_{12}} \right )^{\frac{k_1 + k_2 - k_3}{2}} \left ( \frac{z_{13}}{y_{13}} \right )^{\frac{k_1 + k_3 - k_2}{2}} \left ( \frac{z_{23}}{y_{23}} \right )^{\frac{k_2 + k_3 - k_1}{2}}},
\end{equation}
where $C_{k_1k_2k_3}$ to the order we need should be read off from \eqref{sugra-3pt} (in the $d=4$ case). Once again, the third operator weight satisfies the selection rule \eqref{6d-triangle} in terms of the other two, which we restate as
\begin{align}
k_3=|k_{12}|+2,\,|k_{12}|+4,\,...\,,k_1+k_2-2\,. \label{4d-triangle}
\end{align}

As anticipated at the end of section \ref{sec:setup},  four-point functions in this chiral algebra must satisfy an additional constraint, namely
\begin{align} \label{4d_scwi_chiral}
\left [ \alpha^2 \frac{\partial}{\partial \alpha} - \frac{\partial}{\partial \chi} \right ] \mathcal{F}_{1234}(\chi; \alpha) \biggl |_{\alpha = \frac{1}{\chi}} = 0
\end{align}
in terms of the cross-ratios \eqref{cross-ratio3}. This is the superconformal Ward identity (SCWI) for the global superalgebra $\mathfrak{psu}(1,1|2)$, much like \eqref{scwi} for the respective superalgebras. Additionally, it can be seen as arising from a chiral algebra twist as described in appendix A of \cite{rrz19}.

Based on the structure of these half-BPS supermultiplets, we now have a way to compute their superconformal blocks -- making a suitable ansatz and then fixing any free coefficients by demanding that the SCWI \eqref{4d_scwi_chiral} is satisfied. Since the $J^{(k)}$ are annihilated by half of the supercharges of a ``small'' $\mathcal{N} = 4$ superconformal algebra, there are $2$ directions by which we can descend and therefore $2^2$ conformal primaries in their superconformal multiplet. However, when $J^{(k)}$ is Grassman even (odd), it will only be the state obtained by acting with \textit{both} supercharges which is again Grassman even (odd) and therefore admissible in the same OPE. This super descendant, which we will denote by $T^{(k)}_{a_1 \dots a_{2j}}$, has the quantum numbers $h = \frac{k + 2}{2}$ and $j = \frac{k - 2}{2}$. Notice that it is precisely the stress tensor for the case of $k = 2$. In the primary and descendant terms of the superconformal block, the cross-ratio $\chi$ must appear through the $\mathfrak{sl}(2)$ block \eqref{sl2-block}. Due to the ``isomorphism'' with $\mathfrak{su}(2)$, the cross-ratio $\alpha$ will appear in the same function, continued to negative values of the conformal dimensions. We arrive at
\begin{align}\label{blocksN=4}
\begin{split}
\mathcal{G}_h^{k_{12}, k_{34}}(\chi; \alpha) &=\Big[g_{h}^{(k_{12}/2,k_{34}/2)}(\chi)\,g_{-h}^{(-k_{12}/2,-k_{34}/2)}(\alpha^{-1})\\
&+\frac{\left(4h^2-k_{12}^2\right)  \left(4h^2-k_{34}^2\right)}{64 h^2 \left(1-4h^2\right)}\,g_{h+1}^{(k_{12}/2,k_{34}/2)}(\chi)\,g_{1-h}^{(-k_{12}/2,-k_{34}/2)}(\alpha^{-1})\Big]\,
\end{split}
\end{align}
for a $J^{(2h)}$ multiplet exchanged in $\langle J^{(k_1)}J^{(k_2)}J^{(k_3)}J^{(k_4)} \rangle$. Knowing the relative coefficient in \eqref{blocksN=4} enables us to write the singular OPE of two generators. Combining \eqref{z-ope} for space-time and \eqref{y-ope} for R-symmetry, we find
\begin{align}
& J^{(k_1)}(z_1, y_1) J^{(k_2)}(z_2, y_2) = \frac{\delta_{k_1,k_2}}{\left ( z_{12} / y_{12} \right )^{k_1}} + \sum_{\substack{p = |k_{12}| + 2 \\ \mathrm{step} \; 2}}^{k_1 + k_2 - 2} C_{k_1k_2p} \sum_{m = 0}^{\frac{k_1 + k_2 - p - 2}{2}} \frac{\left ( \frac{k_{12} + p}{2} \right )_m}{m! (p)_m} \frac{y_{12}^{\frac{k_1 + k_2 - p}{2}}}{z_{12}^{\frac{k_1 + k_2 - p}{2} - m}} \nonumber \\
& \hspace{2cm} \frac{\partial^m}{k!} \sum_{\sigma \in S_p} J^{(p)}_{a_{\sigma(1)} \dots a_{\sigma(p)}}(z_2) \, y_1^{a_1} \dots y_1^{a_{(p + k_{12}) / 2}} y_2^{a_{(p + k_{12} + 2) / 2}} \dots y_2^{a_p} \nonumber \\
& \hspace{2cm} - \sum_{\substack{p = |k_{12}| + 2 \\ \mathrm{step} \; 2}}^{k_1 + k_2 - 2} C_{k_1k_2p} \frac{(p + k_{12})(p - k_{12})}{4p \sqrt{p^2 - 1}} \sum_{m = 0}^{\frac{k_1 + k_2 - p - 4}{2}} \frac{\left ( \frac{k_{12} + p + 2}{2} \right )_m}{m! (p + 2)_m} \frac{y_{12}^{\frac{k_1 + k_2 - p + 2}{2}}}{z_{12}^{\frac{k_1 + k_2 - p - 2}{2} - m}} \nonumber \\
& \hspace{2cm} \frac{\partial^m}{(p - 2)!} \sum_{\sigma \in S_{p - 2}} T^{(p)}_{a_{\sigma(1)} \dots a_{\sigma(p - 2)}}(z_2) \, y_1^{a_1} \dots y_1^{a_{(p + k_{12} - 2) / 2}} y_2^{a_{(p + k_{12}) / 2}} \dots y_2^{a_{p - 2}} \label{4d_ope}
\end{align}
plus terms involving normal ordered products that are suppressed by at least $1 / N$.

We are now ready to compute the four-point functions in the chiral algebra. Again, it is possible to use (an adaptation of) the holomorphic bootstrap of \cite{hmpz15}, but instead it turns out to be better to use a similar strategy to the one of section \ref{sec:6d}. That is, we start by making an ansatz
\begin{align}\label{4dansatz}
\mathcal{F}_{1234}(\chi ; \alpha)=\mathcal{F}^{(s)}_{1234}(\chi ; \alpha)+\mathcal{F}^{(t)}_{1234}(\chi ; \alpha)\,,
\end{align}
where $\mathcal{F}^{(t)}$ is identified as the completion of $\mathcal{F}^{(s)}$ under crossing symmetry, which is generated by
\begin{subequations}
\begin{align}
\mathcal{F}_{1234}(\chi ; \alpha) &= (\alpha\,\chi)^{\frac{k_1+k_2}{2}}\,\left(\frac{\alpha(1-\chi)}{\alpha-1}\right)^{-\frac{k_2+k_3}{2}}\, \mathcal{F}_{3214}\left(1-\chi;\frac{\alpha}{\alpha-1}\right) \label{4d-crossing-13} \\
\mathcal{F}_{1234}(\chi ; \alpha) &= (\alpha\,\chi)^{\frac{k_1 + k_4}{2}}\,\mathcal{F}_{4231} \left ( \frac{1}{\chi} ; \frac{1}{\alpha} \right ). \label{4d-crossing-14}
\end{align}
\end{subequations}
As in section \ref{sec:6d}, we shall work under the assumption that $k_1+k_2\le k_3+k_4$ and $k_2+k_3\le k_1+k_4$, so that we can neglect normal ordered products of operators in the OPE.
By using \eqref{4d-crossing-13} and \eqref{4d-crossing-14}, all other choices for the weights are within reach. If we fix the terms in $\mathcal{F}^{(s)}$ such that each power of $\chi$ appears multiple times, once for each power of $z_{12}$ in \eqref{4d_ope}, this leads us to a four-point function with the same structure as \eqref{6d-answer}:
\begin{align}
&\left < J^{(k_1)}(z_1, y_1) J^{(k_2)}(z_2, y_2) J^{(k_3)}(z_3, y_3) J^{(k_4)}(z_4, y_4) \right > = \left ( \frac{z_{24} y_{14}}{z_{14} y_{24}} \right )^{\frac{k_{12}}{2}} \left ( \frac{z_{14} y_{13}}{z_{13} y_{14}} \right )^{\frac{k_{34}}{2}} \nonumber \\
& \hspace{3.5cm} \left ( \frac{y_{12}}{z_{12}} \right )^{\frac{k_1 + k_2}{2}} \left ( \frac{y_{34}}{z_{34}} \right )^{\frac{k_3 + k_4}{2}} \mathcal{F}_{1234}(\chi; \alpha) \nonumber \\
& \hspace{3.5cm} k_1 + k_2 \leq k_3 + k_4 \quad, \quad k_2 + k_3 \leq k_1 + k_4 \label{4d-answer-form}
\end{align}
where
\begin{align}
& \frac{\mathcal{F}_{1234}(\chi; \alpha)}{\chi^{\frac{k_1 + k_2}{2}}} = \sum_{\substack{p = |k_{12}| + 2 \\ \mathrm{step} \; 2}}^{k_1 + k_2 - 2} C_{k_1k_2p}C_{k_3k_4p} \left [ g^{\frac{k_{21}}{2}, \frac{k_{43}}{2}}_{-\frac{p}{2}} (\alpha^{-1}) \sum_{m = 0}^{\frac{k_1 + k_2 - p - 2}{2}} \frac{\left ( \frac{k_{21} + p}{2} \right )_m \left ( \frac{k_{34} + p}{2} \right )_m}{m! (p)_m \chi^{\frac{k_1 + k_2 - p}{2} - m}} \right. \label{4d-answer} \\
& \hspace{0.6cm} + \left. \frac{(p^2 - k_{12}^2)(p^2 - k_{34}^2)}{16p^2(1 - p^2)} g^{\frac{k_{21}}{2}, \frac{k_{43}}{2}}_{1-\frac{p}{2}} (\alpha^{-1}) \sum_{m = 0}^{\frac{k_1 + k_2 - p - 4}{2}} \frac{\left ( \frac{k_{21} + p + 2}{2} \right )_m \left ( \frac{k_{34} + p + 2}{2} \right )_m}{m! (p + 2)_m \chi^{\frac{k_1 + k_2 - p - 2}{2} - m}} \right ] \nonumber \\
& \hspace{0.6cm} + \frac{(\alpha - 1)^{\frac{k_2 + k_3}{2}}}{\alpha^{\frac{k_{31}}{2}}} \sum_{\substack{p = |k_{23}| + 2 \\ \mathrm{step} \; 2}}^{k_2 + k_3 - 2} C_{k_1k_4p}C_{k_2k_3p} \left [ g^{\frac{k_{23}}{2}, \frac{k_{41}}{2}}_{-\frac{p}{2}} (1 - \alpha^{-1}) \sum_{m = 0}^{\frac{k_2 + k_3 - p - 2}{2}} \frac{\left ( \frac{k_{23} + p}{2} \right )_m \left ( \frac{k_{14} + p}{2} \right )_m}{m! (p)_m (1-\chi)^{\frac{k_2 + k_3 - p}{2} - m}} \right. \nonumber \\
& \hspace{0.6cm} \left. + \frac{(p^2 - k_{14}^2)(p^2 - k_{23}^2)}{16p^2(1 - p^2)} g^{\frac{k_{23}}{2}, \frac{k_{41}}{2}}_{1-\frac{p}{2}} (1 - \alpha^{-1}) \sum_{m = 0}^{\frac{k_2 + k_3 - p - 4}{2}} \frac{\left ( \frac{k_{23} + p + 2}{2} \right )_m \left ( \frac{k_{14} + p + 2}{2} \right )_m}{m! (p + 2)_m (1-\chi)^{\frac{k_2 + k_3 - p - 2}{2} - m}} \right ]. \nonumber
\end{align}
However, computing the coefficient for a given power of $\chi$ all at once will lead us to a significant simplification. An ansatz for $\mathcal{F}^{(s)}$ which facilitates this latter approach and takes into account the singular OPE is then
\begin{align}\label{4d_sch_ansatz}
\mathcal{F}^{(s)}_{1234}(\chi ; \alpha)=\sum_{n=1+\tfrac{1}{2}\max\{|k_{12}|,|k_{34}|\}}^{\tfrac{1}{2}(\min\{k_1+k_2,k_3+k_4\}-2)}P_n(\alpha)\,\chi^n\,,
\end{align}
where $P_n(\alpha)$ are functions of $\alpha$ that we can fix by requiring consistency with the OPE, including only the chiral algebra generators. Namely, we demand that
\begin{align}\label{4d_sch_blocks}
\mathcal{F}^{(s)}_{1234}(\chi ; \alpha)=\left[\sum_{\substack{p = \max\{|k_{12}|,|k_{34}|\} + 2 \\ \mathrm{step} \; 2}}^{k_1 + k_2 - 2}C_{k_1k_2p}\,C_{k_3k_4p}\,\mathcal{G}^{k_{12}, k_{34}}_{p / 2}(\chi ; \alpha)\right]_{\text{trunc}}\,,
\end{align}
where the subscript ``trunc'' refers to the fact that we should expand the expression enclosed in brackets for small $\chi$, and truncate it at an order corresponding to the highest power of $\chi$ that appears in \eqref{4d_sch_ansatz}, namely $\tfrac{1}{2}\left(k_1+k_2-2\right)$.
Projecting \eqref{4d_sch_blocks} onto each power of $\chi$, we can extract the coefficients $P_n(\alpha)$, which turn out to be given by
\begin{align}\label{4d_def_Pn}
P_n(\alpha)=\sum_{\substack{p = \max\{|k_{12}|,|k_{34}|\} + 2 \\ \mathrm{step} \; 2}}^{k_1 + k_2 - 2} C_{k_1k_2p}\,C_{k_3k_4p}\, \mathcal{I}_{p/2,n}(\alpha)\,,
\end{align}
where we have introduced
\begin{align} \label{def_I_h,n}
\begin{split}
\mathcal{I}_{h,n}(\alpha)=&\oint_{C_0}\frac{d\chi}{2\pi i\,\chi}\chi^{-n}\,\mathcal{G}^{k_{12}, k_{34}}_h(\chi ; \alpha)\\
=&\frac{\left(h-k_{12}/2\right)_{n-h} \left(h+k_{34}/2\right)_{n-h}}{(n-h)! (2 h)_{n-h}}\,\Big[g_{-h}^{(-k_{12}/2,-k_{34}/2)}(\alpha^{-1})\\
&+\frac{(h-n) (2 h+k_{12}) (2 h-k_{34})}{8 h (2 h-1) (h+n)}\,g_{-h-1}^{(-k_{12}/2,-k_{34}/2)}(\alpha^{-1})\Big]\,.
\end{split}
\end{align}
While the sum over $\mathcal{I}_{h,n}(\alpha)$ looks complicated \textit{a priori}, with $P_n(\alpha)$ being in principle arbitrary polynomials of degree $n$,\footnote{This can be argued from the relation between the $\mathfrak{sl}(2)$ blocks with negative arguments appearing in \eqref{def_I_h,n} and the Jacobi polynomials (see \eqref{y-4pt}).} for the specific values of $n$ that enter the sum \eqref{4d_def_Pn} they turn out to have a particularly simple form, in which the dependence on $\alpha$ is almost completely factorized:
\begin{align}\label{nice-pn}
P_n(\alpha)=\frac{\sqrt{k_1\,k_2\,k_3\,k_4}}{N^2}\,\alpha^{n-1}\,\left(\frac{2+k_1-k_2-k_3+k_4}{2}-2\,n\,(1-\alpha)\right)\,.
\end{align}
As in section \ref{sec:6d}, using crossing, we can continue the results above to all configurations of the $\{k_i\}$, and therefore this fixes all the four-point functions between chiral generators at tree-level. 

\subsection{Matching from Mellin space}
\label{sec:5.2}

The next step is to show that the same answer in encoded in the Mellin amplitudes \eqref{residueR} for $AdS_5 \times S^5$. Our discussion here will run parallel to that of subsection \ref{sec:4.2}. As explained there, we will consider weights that satisfy $k_2 \leq k_1 \leq k_3 \leq k_4$. The correlators we derive will initially take the form \eqref{4d-answer} but the simplification apparent in \eqref{nice-pn} will turn out to have an interpretation in Mellin space as well.

Since the convention that will allow the most direct comparison is
\begin{align}
& \left < \mathcal{O}_1(x_1, t_1)\mathcal{O}_2(x_2, t_2)\mathcal{O}_3(x_3, t_3)\mathcal{O}_4(x_4, t_4) \right > = \left ( \frac{t_{24}}{x_{24}^2} \right )^{-\frac{1}{2} k_{12}} \left ( \frac{t_{14}}{x_{14}^2} \right )^{\frac{1}{2} k_{12} - \frac{1}{2} k_{34}} \left ( \frac{t_{13}}{x_{13}^2} \right )^{\frac{1}{2} k_{34}} \nonumber \\
& \hspace{3cm} \left ( \frac{t_{12}}{x_{12}^2} \right )^{\frac{1}{2}(k_1 + k_2)} \left ( \frac{t_{34}}{x_{34}^2} \right )^{\frac{1}{2}(k_3 + k_4)} \mathcal{F}_{1234}(U, V; \sigma, \tau), \label{change-prefactor3}
\end{align}
we can solve for the dynamical part as
\begin{align}
\mathcal{F}_{1234}(U, V; \sigma, \tau) &= (\sigma U)^{\frac{1}{2}(k_1 + k_2) - \mathcal{E}} \mathcal{G}_{2134} \left ( \frac{U}{V}, \frac{1}{V}; \tau, \sigma \right ) \label{pre-twist-4d} \\
&= \int_{-i\infty}^{i\infty} \frac{\textup{d}s \textup{d}t}{(4\pi i)^2} U^{\frac{s}{2}} V^{-\frac{s}{2} - \frac{t}{2} + \frac{1}{2}(k_1 + k_4)} \sigma^{\frac{1}{2}(k_1 + k_2) - \mathcal{E}} \mathcal{M}_{2134}(s, t; \tau, \sigma) \Gamma_{\{ k_i \}}. \nonumber
\end{align}
Evaluating \eqref{pre-twist-4d} in the twisted configuration will again be based on taking $\chi^\prime \rightarrow 0$. Clearly, the freedom to set $\chi^\prime$ to a convenient value is a consequence of the superconformal Ward identity. We have been able to prove analytically that it holds for all weights by showing that the Mellin amplitudes in \eqref{residueR} can be written in the form
\begin{equation}
\mathcal{M}(s, t; \sigma, \tau) = \hat{R} \circ \widetilde{\mathcal{M}}(s, t; \sigma, \tau) \label{aux-m}
\end{equation}
where the operator $\hat{R}$ is the Mellin space version of $(1 - \chi\alpha)(1 - \chi\alpha^\prime)(1 - \chi^\prime \alpha)(1 - \chi^\prime \alpha^\prime)$. It should come as no surprise that the auxiliary Mellin amplitude on the right hand side which accomplishes this is the one conjectured in \cite{rz16,rz17} with the normalization from \cite{Aprile:2018efk}.

Based on our chosen ordering, the superconformal twist with $\chi^\prime \rightarrow 0$ ensures that $s = k_1 + k_2$ is the only pole that contributes (effectively giving $\tau^{\mathcal{E} - j} \sigma^j \mapsto \chi^{\prime -\mathcal{E}} \alpha^j (\alpha - 1)^{\mathcal{E} - j}$). After this, the $u$-channel poles of $t = k_3 + k_4 - 2m - p$ that lead to singularities can be extracted to give the second term in $\mathcal{F}(\chi; \alpha) = \mathcal{F}^{(s)}(\chi; \alpha) + \mathcal{F}^{(t)}(\chi; \alpha)$. Continuing, the expression for $\mathcal{R}^{p,m;0,j}_{4132}(k_3 + k_4 - 2m - p, k_1 + k_2)$ allows us to write
\begin{align}
& \frac{\mathcal{F}^{(t)}(\chi; \alpha)}{(\alpha \chi)^{\frac{k_1 + k_2}{2}}} = \frac{\Gamma(\frac{\kappa_s}{2})}{16} \sum_{\substack{p = \mathrm{max} \{ |k_{14}|, |k_{23}| \} + 2 \\ \mathrm{step} \; 2}}^{k_2 + k_3 - 2} C_{k_1k_4p} C_{k_2k_3p} \sum_{m = 0}^{\frac{k_2 + k_3 - p - 2}{2}} \frac{\Gamma[p + m + 1]^{-1}}{m!(1 - \chi)^{\frac{k_2 + k_3 - p}{2} - m}} \nonumber \\
& \Gamma \left [ \frac{k_{14} + p + 2m}{2} \right ] \Gamma \left [ \frac{k_{23} + p + 2m}{2} \right ] \sum_{j = 0}^\mathcal{E} (1-\alpha^{-1})^{\mathcal{E} - j} \mathcal{B}_p^{0,j} K_p^{0,j}\binom{k_3 + k_4 - 2m - p,}{k_1 + k_2}
\end{align}
in terms of the factors from \eqref{residueR}. By scrutinizing the gamma functions in \eqref{Bij}, we can see that the inner sum only goes up to $\mathcal{E} - \frac{k_2 + k_3 - p}{2}$ instead of $\mathcal{E}$. This allows us to perform the reflection $j \mapsto \mathcal{E} - \frac{k_2 + k_3 - p}{2} - j$ which yields
\begin{align}
& \sum_{j = 0}^{\mathcal{E}} (1 - \alpha^{-1})^{\mathcal{E} - j} \mathcal{B}_p^{0,j} K_p^{0,j}\binom{k_3 + k_4 - 2m - p,}{k_1 + k_2} = \frac{\Gamma(p) \Gamma \left ( \frac{\kappa_s + 2}{2} \right )^{-1}}{\Gamma \left ( \frac{p + k_{14} + 2}{2} \right ) \Gamma \left ( \frac{p + k_{23} + 2}{2} \right )} \label{inner-sum4} \\
& \left ( \frac{\alpha}{\alpha - 1} \right )^{\frac{p - k_2 - k_3}{2}} \sum_{j = 0}^{\mathcal{E} - \frac{k_2 + k_3 - p}{2}} (1 - \alpha^{-1})^j \frac{\left ( \frac{k_{41} - p}{2} \right )_j \left ( \frac{k_{32} - p}{2} \right )_j}{j!(1 - p)_j} K_p^{0,\mathcal{E} - \frac{k_2 + k_3 - p}{2} - j}\binom{k_3 + k_4 - 2m - p,}{k_1 + k_2}. \nonumber
\end{align}
To evaluate \eqref{inner-sum4}, we need to plug in
\begin{align}
K_p^{0,\mathcal{E} - \frac{k_2 + k_3 - p}{2} - j} & (k_3 + k_4 - 2m - p, k_1 + k_2) \label{k-expansion2} \\
=& \frac{1}{2} \kappa_s (m + p) [(\kappa_s - 2p)^2 - \kappa_u^2] + \frac{1}{2} \kappa_s j [\kappa_u^2 - \kappa_s^2 + 4\kappa_s(2m + p) - 4p^2] \nonumber \\
=& \kappa_s \left [ \frac{2(p - j)(p + m)}{p}(p + k_{14})(p + k_{23}) + \frac{2jm}{p}(p - k_{14})(p - k_{23}) \right ]. \nonumber
\end{align}
The fact that it is an even function of $\kappa_u$ is what allows us to write it in terms of the weights analytically. It is now straightforward to see that our expression for the $t$-channel singularity becomes
\begin{align}
& \frac{\mathcal{F}^{(t)}(\chi; \alpha)}{(\alpha \chi)^{\frac{k_1 + k_2}{2}}} = \frac{1}{8} \sum_{\substack{p = \mathrm{max} \{ |k_{14}|, |k_{23}| \} + 2 \\ \mathrm{step} \; 2}}^{k_2 + k_3 - 2} C_{k_1k_4p} C_{k_2k_3p} \sum_{m = 0}^{\frac{k_2 + k_3 - p - 2}{2}} (1 - \chi)^{\frac{p + 2m - k_2 - k_3}{2}} \nonumber \\
&\times \frac{\Gamma(p) \Gamma \left ( \frac{k_{14} + 2m + p}{2} \right ) \Gamma \left ( \frac{k_{23} + 2m + p}{2} \right ) \left ( \frac{\alpha}{\alpha - 1} \right )^{\frac{p - k_2 - k_3}{2}}}{m!\Gamma(p + m + 1) \Gamma \left ( \frac{k_{14} + 2 + p}{2} \right ) \Gamma \left ( \frac{k_{23} + 2 + p}{2} \right )} \sum_{j = 0}^{\infty} \frac{\left ( \frac{k_{41} - p}{2} \right )_j \left ( \frac{k_{32} - p}{2} \right )_j}{j! (1 - p)_j} \left ( \frac{\alpha - 1}{\alpha} \right )^j \nonumber \\
&\times \left [ \frac{2(p - j)(p + m)}{p}(p + k_{14})(p + k_{23}) + \frac{2jm}{p}(p - k_{14})(p - k_{23}) \right ]. \label{k-expansion3}
\end{align}
In the splitting that we have chosen for \eqref{k-expansion2}, the first term has the property that all factors may be readily absorbed into the gamma functions. This shifts their arguments to exactly the values that we expect for a superconformal primary. From this point of view, it is reassuring that the second term is proportional to $m$. This allows us to reindex the sum over $m$ so that the $(1 - z)^{\frac{p - k_2 - k_3}{2}}$ singularity, which cannot come from a super descendant, is manifestly absent. When this is done,
\begin{align}
& \frac{\mathcal{F}^{(t)}(\chi; \alpha)}{(\alpha \chi)^{\frac{k_1 + k_2}{2}}} = \sum_{\substack{p = |k_{23}| + 2 \\ \mathrm{step} \; 2}}^{k_2 + k_3 - 2} C_{k_1k_4p}C_{k_2k_3p} \left [ \frac{g^{\frac{k_{23}}{2}, \frac{k_{41}}{2}}_{-\frac{p}{2}} (1 - \alpha^{-1})}{(1 - \alpha^{-1})^{\frac{- k_2 - k_3}{2}}} \sum_{m = 0}^{\frac{k_2 + k_3 - p - 2}{2}} \frac{\left ( \frac{k_{23} + p}{2} \right )_m \left ( \frac{k_{14} + p}{2} \right )_m}{m! (p)_m (1-\chi)^{\frac{k_2 + k_3 - p}{2} - m}} + \right. \nonumber \\
& \hspace{0.15cm} \left. \frac{(p^2 - k_{14}^2)(p^2 - k_{23}^2)}{16p^2(1 - p^2)} \frac{g^{\frac{k_{23}}{2}, \frac{k_{41}}{2}}_{1-\frac{p}{2}} (1 - \alpha^{-1})}{(1 - \alpha^{-1})^{\frac{- k_2 - k_3}{2}}} \sum_{m = 0}^{\frac{k_2 + k_3 - p - 4}{2}} \frac{\left ( \frac{k_{23} + p + 2}{2} \right )_m \left ( \frac{k_{14} + p + 2}{2} \right )_m}{m! (p + 2)_m (1-\chi)^{\frac{k_2 + k_3 - p - 2}{2} - m}} \right ]
\end{align}
which is nothing but the second part of \eqref{4d-answer}.

We have already mentioned in \eqref{aux-m} that these Mellin amplitudes can be obtained from the action of a difference operator $\hat{R}$ on an auxiliary Mellin amplitude $\widetilde{\mathcal{M}}(s, t; \sigma, \tau)$. Returning to this point can help us understand the absence of $\alpha^{n - 2}, \alpha^{n - 3}, \dots$ terms in \eqref{nice-pn}. Consider the poles of the $s$-channel amplitude delineated by \eqref{s-channel}. Although every value in $(\mathrm{max} \{ |k_{12}|, |k_{34}| \} + 2, \dots, \mathrm{min} \{ k_1 + k_2, k_3 + k_4 \} - 2)$ appears as a pole of $\mathcal{M}^{(s)}(s, t; \sigma, \tau)$, closer inspection reveals that each one has at most two R-symmetry monomials $\sigma^i \tau^j$ in its residue. This non-trivial fact follows from a key property of the auxiliary amplitude bootstrapped in \cite{rz16,rz17}. Namely that, for a fixed monomial $\sigma^i \tau^j$, it has exactly one pole in each Mandelstam variable. The action of the difference operator is then only able to produce one additional pole in the full result.

We can see how this structure appears by changing the order of the sums in \eqref{s-channel}. Specifically, we should focus on a specific pole $s = s_0$ (taken to be even for simplicity) and then set $m = \frac{s_0 - p}{2}$. Performing the sum over $p$ and keeping $u$ as an independent variable leads us to
\begin{align}
\underset{s \rightarrow s_0}{\rm Res} \mathcal{M}^{i, j}(s, t) &= \frac{(-1)^{i + j - \frac{\kappa_t + \kappa_u}{4}} \sqrt{k_1k_2k_3k_4}\, \Gamma \left ( \frac{s_0 + 2}{2} \right )^{-1}}{2 i! j! \Gamma \Big ( \frac{\kappa_u + 2 + 2i}{2} \Big ) \Gamma \Big ( \frac{\kappa_t + 2 + 2j}{2} \Big ) \Gamma \left ( \frac{k_1 + k_2 - s_0}{2} \right ) \Gamma \left ( \frac{k_3 + k_4 - s_0}{2} \right )} \nonumber \\
&\times \left [ \frac{K^{i, j}_0(t, u) \Gamma \left ( 1 + i + j - \mathcal{E} + \frac{\Sigma - \kappa_s}{4} \right ) \Gamma \left ( \frac{s_0}{2} \right )^{-1}}{(4i + 4j - 4\mathcal{E} - \kappa_s + \Sigma - 2s_0) \Gamma \left ( 1 - i - j + \mathcal{E} - \frac{\Sigma - \kappa_s}{4} \right )} \right. \nonumber \\
&- \frac{32s_0(s_0 + 2) - 16(4i + 4j - 4\mathcal{E} + \Sigma - \kappa_s)(s_0 + 4)}{(4i + 4j - 4\mathcal{E} - \kappa_s + \Sigma - 2s_0)^2 (4 + 4i + 4j - 4\mathcal{E} - \kappa_s + \Sigma - 2s_0)^2} \nonumber \\
&\left. \frac{(t^- - 2j)(u^- - 2i) \Gamma \left ( \frac{s_0 + 2}{2} + i + j - \mathcal{E} + \frac{\Sigma - \kappa_s}{4} \right ) \Gamma \left ( -\frac{s_0}{2} \right )^{-1}}{\Gamma(s_0) \Gamma \left ( \frac{s_0 - 2}{2} - i - j + \mathcal{E} - \frac{\Sigma - \kappa_s}{4} \right )} \right ].
\end{align}
We now observe that the first term forces us to have $2s_0 = 4i + 4j - 4\mathcal{E} - \kappa_s + \Sigma$ because $1 - i - j + \mathcal{E} - \frac{\Sigma - \kappa_s}{4}$, which appears in the gamma function, is a non-positive number.\footnote{The only exception occurs when $i = j = 0$ and all the weights are identical. But then, $K^{0,0}_0(t, u) = 0$ anyway.} By a similar argument, the second term allows us to have either this value or $2s_0 = 4 + 4i + 4j - 4\mathcal{E} - \kappa_s + \Sigma$.

Now that we know $\mathcal{M}^{i,j}(s,t)$ can be written as a sum of only two residues, it is this form of the Mellin amplitude, when subjected to the procedure of this subsection, that produces the polynomial \eqref{nice-pn}. Indeed, we can look at all of the single-particle poles in $s$ which can contribute singular terms to $\mathcal{F}^{(s)}$. These have the form $s_0 = p + 2m$ which, as shown above, can only appear with $\tau^j \sigma^{(p + 2m + 2\mathcal{E} - k_1 - k_2)/2 - j}$ and $\tau^j \sigma^{(p + 2m + 2\mathcal{E} - k_1 - k_2 - 2)/2 - j}$. If we then perform the superconformal twist with $\chi^\prime \rightarrow 1$, we pick out only those monomials with $j = 0$. This causes the integrand of \eqref{pre-twist-4d} to become a linear combination of $\alpha^{\frac{p}{2} + m}$ and $\alpha^{\frac{p}{2} + m - 1}$ as required.

\section{Intermezzo: What makes topological correlators different?}
\label{sec:topological}

For the SCFTs studied in sections \ref{sec:6d} and \ref{sec:4d}, we managed to find order $1 / c_T$ four-point functions of arbitrary strong generators in the associated chiral algebra. Much like in higher dimensions, four-point functions in a 2d CFT (whether chiral or not) receive a contribution from infinitely many quasiprimary operators. This is in agreement with the fact that there are infinitely many 4d and 6d superconformal multiplets in the cohomology of the supercharge used in the chiral algebra construction.

We can easily list those 6d $\mathcal{N} = (2, 0)$ multiplets determined in \cite{brv14}. In the notation of \cite{cdi16}, they are $B[j_1, j_2, 0]_\Delta^{[r_1, r_2]}$ for $r_2 = 0$, $C[j_1, 0, 0]_\Delta^{[r_1, r_2]}$ for $r_2 \leq 1$ and $D[0, 0, 0]_\Delta^{[r_1, r_2]}$ for $r_2 \leq 2$. Similarly, the Schur multiplets of 4d $\mathcal{N} = 4$ super Yang-Mills are $B\bar{B}[0, 0]_\Delta^R$, $A\bar{A}[j, \bar{\jmath}]_\Delta^R$, $A\bar{B}[j, 0]_\Delta^R$ and $B\bar{A}[0, \bar{\jmath}]_\Delta^R$ where there are no restrictions on the R-symmetry representation beyond those required for unitarity.\footnote{In CFTs with less supersymmetry, this way of writing the Schur multiplets is still correct as long as we additionally specify that the primaries have zero $\mathfrak{u}(1)$ r-charge. The detailed calculation was done in \cite{bllprv13} for $\mathcal{N} = 2$, \cite{llmm16} for $\mathcal{N} = 3$ and \cite{bmr18} for $\mathcal{N} = 4$.} In particular, certain Lorentz quantum numbers are allowed to be arbitrarily large. Of course, the derivations of \eqref{6d-answer} and \eqref{4d-answer} did not require us to deal with infinite sums over the spin explicitly. We instead exploited the fact that chiral algebra correlators, being meromorphic functions, are uniquely fixed by a finite number of singularities.

This situation is quite different for the topological correlators predicted by the 3d superconformal Ward identity \eqref{scwi-3d}. On the one hand, one sees from the multiplet structure of $\mathfrak{osp}(\mathcal{N} | 4)$ that there is no longer any room for spinning operators to contribute to an absolutely protected subsector. As a result, adding up every term in the OPE becomes a viable way to compute twisted four-point functions. On the other hand, due to the lack of any nice meromorphic property, it might be the \textit{only} viable way. We can see a simple example of spinning operators dropping out in the four-point functions at $c_T = \infty$:
\begin{align}
\left < \mathcal{O}_{\mathcal{E}}(x_1,t_1) \mathcal{O}_{\mathcal{E}}(x_2,t_2) \mathcal{O}_k(x_3,t_3) \mathcal{O}_k(x_4,t_4) \right > &= \left ( \frac{t_{34}}{|x_{34}|} \right )^{k - \mathcal{E}} \left ( \frac{t_{12}t_{34}}{|x_{12}||x_{34}|} \right )^{\mathcal{E}} \nonumber \\
& \left [ 1 + \delta_{k\mathcal{E}} \, U^{\frac{\mathcal{E}}{2}} \left ( \sigma^\mathcal{E} + \tau^\mathcal{E} V^{-\frac{\mathcal{E}}{2}} \right ) \right ]. \label{gff-example}
\end{align}
Here, without loss of generality, we have labelled two operators by $\mathcal{E}$ since the extremality is the smaller of the two weights. The superconformal twist instructs us to take
\begin{align}
\begin{split}
\frac{t_{12}t_{34}}{|x_{12}||x_{34}|} U^{\frac{1}{2}} \sigma \quad &\rightarrow \quad \mathrm{sgn}(z_{13}) \mathrm{sgn}(z_{24}) \, y_{13} y_{24} \\
\frac{t_{12}t_{34}}{|x_{12}||x_{34}|} \frac{U^{\frac{1}{2}}}{V^{\frac{1}{2}}} \tau \quad &\rightarrow \quad \mathrm{sgn}(z_{14}) \mathrm{sgn}(z_{23}) \, y_{14} y_{23}
\end{split}
\end{align}
which no longer require infinitely many conformal blocks in the $s$-channel to reproduce. Instead, we just need one exchanged multiplet for each of the $\mathfrak{su}(2)$ spins from $0$ to $\mathcal{E}$.

\vspace{0.5cm}
\noindent{\bf The way forward}
\vspace{0.3cm}

To compute twisted four-point functions in a way which does not rely on crossing symmetry, we have already mentioned the method in Appendix \ref{sec:appa}. While this has been presented mainly as a check of the results in sections \ref{sec:6d} and \ref{sec:4d}, one could imagine using it as a starting point for 3d calculations as well. Unfortunately, we can already see that this approach will not allow us to compute a four-point function all at once anymore. Consider setting $\chi = \chi^\prime$ and taking them both to zero. In Appendix \ref{sec:appa}, we needed this step to make one of the integrals localize onto a single term. For the 3d case, however, such a limit would prevent us from seeing the $\mathrm{sgn}(z_{12})$ and $\mathrm{sgn}(z_{34})$ terms which provide a non-zero contribution above.

\begin{table}[h]
\centering
\begin{tabular}{|l|l|l|l|}
\hline
Name & Primary & Unitarity Bound & Null State \\
\hline
\hline
$L$ & $[j]_\Delta^{[r_1,r_2,r_3,r_4]}$ & $\Delta > \frac{1}{2}j + r_1 + r_2 + \frac{1}{2}(r_3 + r_4) + 1$ & \\
\hline
\hline
$A_1$ & $[j]_\Delta^{[r_1,r_2,r_3,r_4]}$, $j \geq 1$ & $\Delta = \frac{1}{2}j + r_1 + r_2 + \frac{1}{2}(r_3 + r_4) + 1$ & $[j - 1]_{\Delta + \frac{1}{2}}^{[r_1 + 1, r_2, r_3, r_4]}$ \\
\hline
$A_2$ & $[0]_\Delta^{[r_1,r_2,r_3,r_4]}$ & $\Delta = r_1 + r_2 + \frac{1}{2}(r_3 + r_4) + 1$ & $[0]_{\Delta + 1}^{[r_1 + 2, r_2, r_3, r_4]}$ \\
\hline
\hline
$B_1$ & $[0]_\Delta^{[r_1,r_2,r_3,r_4]}$ & $\Delta = r_1 + r_2 + \frac{1}{2}(r_3 + r_4)$ & $[1]_{\Delta + \frac{1}{2}}^{[r_1 + 1, r_2, r_3, r_4]}$ \\
\hline
\end{tabular}
\caption{Unitary multiplets of the 3d $\mathcal{N} = 8$ superconformal algebra which is a real form of $\mathfrak{osp}(8 | 4)$. We follow the terminology of \cite{cdi16} -- short multiplets of A-type are ``at threshold'' while those of B-type are ``below threshold''. One could further refine the classification by counting how many supercharges annihilate each one.}
\label{3d-multiplets}
\end{table}

Since four-point functions in the twisted configuration will have to be assembled from multiple pieces either way, we have found it more intuitive to do this using the OPE. For this reason, we will devote the rest of this paper to understanding the OPE coefficients that couple two half-BPS operators to a third (exchanged) operator in three dimensions. The 3d $\mathcal{N} = 8$ multiplets, which will feature heavily in this analysis, are listed in Table \ref{3d-multiplets}. They obey the R-symmetry selection rule
\begin{equation}
[0,0,k_1,0] \otimes [0,0,k_2,0] = \bigoplus_{j_{\rm I} = \frac{|k_{12}|}{2}}^{\frac{k_1 + k_2}{2}} \bigoplus_{j_{\rm II} = \frac{|k_{12}|}{2}}^{j_{\rm I}} [0, j_{\rm I} - j_{\rm II}, 2 j_{\rm II}, 0] \label{tensor-product}
\end{equation}
which covers all external operators dual to KK-modes as they are primaries of $B[0]^{[0,0,k,0]}_{\frac{1}{2}k}$.\footnote{Apart from the Roman numerals, there can still be external labels that run from $1$ to $4$ in a four-point function. The half-BPS operator at position $i$ has $j_i^{\rm I} = j_i^{\rm II} = \frac{1}{2} k_i = \Delta_i$.}

To summarize, we wish to solve for topological four-point functions by identifying the finitely many operators that contribute in a given channel and then computing their OPE coefficients in a $1 / c_T$ expansion. Section \ref{sec:matrix-model} will approach this question from the TQFT point of view while in section \ref{sec:ope-coeffs} we will return to Mellin space. In both cases, the $\mathcal{E} = 2$ correlators $\left < \mathcal{O}_2\mathcal{O}_2\mathcal{O}_k\mathcal{O}_k \right >$, and permutations thereof, will be our primary interest. Before moving onto the next section, we will get the process started with generalized free theory.

This exercise consists of setting $\mathcal{E} = 2$ in \eqref{gff-example} and projecting onto the $SO(8)$ harmonic polynomials of \cite{no04} in the $t$-channel. A straightforward calculation leads to
\begin{gather}
\lambda^2_{22B[0]_0^{[0000]}} = 1, \; \lambda^2_{22B[0]_2^{[0200]}} = \frac{32}{3}, \; \lambda_{22B[0]_2^{[0040]}} = \frac{16}{3} \label{order0-res2} \\
\lambda^2_{2kB[0]_{\frac{k+2}{2}}^{[0,2,k-2,0]}} = 2^{k+2} \frac{k-1}{k+1}, \; \lambda^2_{2kB[0]_{\frac{k+2}{2}}^{[0,1,k,0]}} = \frac{2^{k + 3}}{k + 2}, \; \lambda^2_{2kB[0]_{\frac{k+2}{2}}^{[0,0,k+2,0]}} = \frac{2^{k + 3}}{(k+1)(k+2)} \nonumber
\end{gather}
as the non-zero protected OPE coefficients.\footnote{Note that we are taking the ``numerics-inspired'' convention $g_{\Delta, \ell}(z, \bar{z}) \sim \left ( \frac{z\bar{z}}{16} \right )^{\frac{\Delta}{2}} \sim (\rho \bar{\rho})^{\frac{\Delta}{2}}$ instead of $g_{\Delta, \ell}(z, \bar{z}) \sim \left ( z\bar{z} \right )^{\frac{\Delta}{2}} \sim ( 16 \rho \bar{\rho})^{\frac{\Delta}{2}}$ which is also common.} We should now investigate how the 3d/1d correspondence allows us to recover \eqref{order0-res2} along with some OPE data at the next order.

\section{The topological subsector with maximal supersymmetry}
\label{sec:matrix-model}

As shown in \cite{clpy14,bpr16}, any 3d $\mathcal{N} = 4$ SCFT admits correlators that are topological on the line. In order to construct them, one passes to the cohomology of a certain linear combination of Poincar\'{e} and conformal supercharges. Consider the following subalgebra of the $\mathcal{N} = 8$ R-symmetry.
\begin{equation}
\mathfrak{su}(2)_{\rm I} \times \mathfrak{su}(2)_{\rm II} \times \mathfrak{su}(2)_{\rm III} \times \mathfrak{su}(2)_{\rm IV} \subset \mathfrak{so}(8) \label{subalgebra}
\end{equation}
We will regard these factors as
R, flavour, flavour, R from the $\mathcal{N} = 4$ point of view. Standard branching rules show that the $\mathfrak{su}(2)$ spins are
\begin{equation}
j_{\rm I} = \frac{r_1 + 2r_2 + r_3 + r_4}{2}, \; j_{\rm II} = \frac{r_3}{2}, \; j_{\rm III} = \frac{r_4}{2}, \; j_{\rm IV} = \frac{r_1}{2} \label{su2-spins}
\end{equation}
in terms of the $[r_1,r_2,r_3,r_4]$ Dynkin labels of $\mathfrak{so}(8)$. After focusing on irreps that can be exchanged by two half-BPS operators, we find $j_{\rm III} = j_{\rm IV} = 0$ while $j_{\rm I}$ and $j_{\rm II}$ are precisely the quantities indicated in \eqref{tensor-product}. A result of \cite{clpy14,bpr16} is that superconformal primaries of the $B[0]_{j_{\rm I}}^{[0, j_{\rm I} - j_{\rm II}, 2 j_{\rm II}, 0]}$ multiplets survive the cohomological prescription. Since they are invariant under $\mathfrak{su}(2)_{\rm IV}$, one may refer to them as Higgs branch operators. Their position dependence is given by \eqref{polarization2} after a ``twisting-translating'' procedure with $u(x) = \left ( 1, \frac{x}{2r} \right )$.
\begin{align}
\mathcal{O}(x, y) &= \mathcal{O}_{a_1 \dots a_{2j_{\rm II}}}(x) \; y^{a_1} \dots y^{a_{2j_{\rm II}}} \nonumber \\
&= \mathcal{O}^{b_1 \dots b_{2j_{\rm I}}}_{a_1 \dots a_{2j_{\rm II}}}(x, 0, 0) \; u_{b_1}(x) \dots u_{b_{2j_{\rm I}}}(x) \; y^{a_1} \dots y^{a_{2j_{\rm II}}} \label{twisted-translated-3d}
\end{align}
In what follows, we will compactify the $x$-direction on a circle of radius $r$.

\subsection{Finite sum rules}
If the scaling dimension of the original operator is a half-integer, it turns out that the corresponding twisted-translated operator on the circle is effectively fermionic. In particular, their two-point and three-point functions may be expressed as
\begin{align}
\left < \mathcal{O}_A(\varphi_1, y_1) \, \mathcal{O}_B(\varphi_2, y_2) \right > &= \delta_{AB} \, B_A\, y_{12}^{2j_A}\, (\mathrm{sgn} \varphi_{21})^{2\Delta_A} \label{top-2pt-3pt} \\
\left < \mathcal{O}_A(\varphi_1, y_1)\, \mathcal{O}_B(\varphi_2, y_2)\, \mathcal{O}_C(\varphi_3, y_3) \right > &= C_{ABC} \, y_{12}^{j_A + j_B - j_C}  \,y_{23}^{j_B + j_C - j_A} \, y_{31}^{j_C + j_A - j_B} \nonumber \\
& \hspace{-3.2cm} (\mathrm{sgn}\varphi_{21})^{\Delta_B + \Delta_A - \Delta_C}\,  (\mathrm{sgn}\varphi_{13})^{\Delta_A + \Delta_C - \Delta_B}\, (\mathrm{sgn}\varphi_{32})^{\Delta_C + \Delta_B - \Delta_A}. \nonumber
\end{align}
This structure leads to the OPE
\begin{align}
\mathcal{O}_A(\varphi_1, y_1)	\, \mathcal{O}_B(\varphi_2, y_2) =& \sum_{\mathcal{O}} \frac{C_{AB\mathcal{O}}}{B_{\mathcal{O}}} (-1)^{j + j_{AB} + \Delta + \Delta_{AB}} \left < y_1, y_2 \right >^{j_A + j_B - j} (\mathrm{sgn}\varphi_{21})^{\Delta_A + \Delta_B - \Delta} \nonumber \\
& \frac{1}{(2j)!} \sum_{\sigma \in S_{2j}} \mathcal{O}_{a_{\sigma(1)} \dots a_{\sigma(2j)}}(\varphi) \; y_1^{a_1} \dots y_1^{a_{j + j_{AB}}} y_2^{a_{j + j_{AB} + 1}} \dots y_2^{2j} \label{top-ope}
\end{align}
which is a straightforward modification of \eqref{y-ope}. We can solve for four-point functions by using it twice. In terms of the cross-ratio $\alpha$, the $s$-channel result is
\begin{align}
& \left < \mathcal{O}_A(\varphi_1, y_1)\,\mathcal{O}_B(\varphi_2, y_2) \,\mathcal{O}_C(\varphi_3, y_3)\,\mathcal{O}_D(\varphi_4, y_4) \right > = \sum_{\mathcal{O}} \frac{C_{AB\mathcal{O}} \,C_{CD\mathcal{O}}}{B_{\mathcal{O}} (\alpha - 1)^{j_{DC}}} (-1)^j g_{-j}^{j_{AB}, j_{DC}} \left ( \frac{1}{1 - \alpha} \right ) \nonumber \\
& \hspace{3cm} \left ( \frac{y_{14}}{y_{24}} \right )^{j_{AB}} \left ( \frac{y_{13}}{y_{14}} \right )^{j_{CD}} \frac{\alpha^{j_{DC}} \,y_{12}^{j_A + j_B} \,y_{34}^{j_C + j_D}}{(\mathrm{sgn} \varphi_{21})^{\Delta_A + \Delta_B - \Delta}\, (\mathrm{sgn} \varphi_{43})^{\Delta_C + \Delta_D - \Delta}}. \label{top-4pt}
\end{align}
Recalling \eqref{y-4pt}, the first line of \eqref{top-4pt} can be written as
\begin{equation}
\frac{C_{AB\mathcal{O}} \,C_{CD\mathcal{O}}}{B_{\mathcal{O}}} \frac{(j + j_{CD})!}{(j - j_{CD} + 1)_{j + j_{CD}}} P^{j_{AB} - j_{CD}, j_{BA} - j_{CD}}_{j + j_{CD}}(2\alpha - 1). \label{top-4pt-rewriting}
\end{equation}
It is shown in \cite{clpy14} that the coefficient of the Jacobi polynomial in \eqref{top-4pt-rewriting} is nothing but the superconformal block coefficient $4^{-\Delta} \lambda_{AB\mathcal{O}} \lambda_{CD\mathcal{O}}$ in three dimensions. This allows us to write down crossing equations that only involve the operators in \eqref{top-4pt}.

Let us pick an $\mathcal{E} = 2$ (next-to-next-to-extremal) correlator which has 6 multiplets in the topological four-point function. More precisely, we will pick a family of them given by $\Delta_A = \Delta_C = j_A = j_C = 1$, $\Delta_B = \Delta_D = j_B = j_D = \frac{k}{2}$. To derive a crossing equation, let us choose the $\varphi_1 < \varphi_2 < \varphi_3 < \varphi_4$ ordering so that all of the sign functions above come out positive. Equating \eqref{top-4pt} to the one other channel that preserves this cyclic ordering tells us that
\begin{align}
\alpha^{-2} & \left [ \left ( 16\lambda^2_{2kB[0]_{\frac{k-2}{2}}^{[0,0,k-2,0]}} + 4\lambda^2_{2kB[0]_{\frac{k}{2}}^{[0,1,k-2,0]}} + \lambda^2_{2kB[0]_{\frac{k + 2}{2}}^{[0,2,k-2,0]}} \right ) P^{0,k-2}_0(2\alpha - 1) \right. \label{raw-crossing} \\
& \left. + \left ( 4\lambda^2_{2kB[0]_{\frac{k}{2}}^{[0,0,k,0]}} + \lambda^2_{2kB[0]_{\frac{k + 2}{2}}^{[0,1,k,0]}} \right ) P^{0,k-2}_1(2\alpha - 1) + \lambda^2_{2kB[0]_{\frac{k + 2}{2}}^{[0,0,k+2,0]}} P^{0,k-2}_2(2\alpha - 1) \right ] \nonumber
\end{align}
must be invariant under $\alpha \leftrightarrow \frac{\alpha}{\alpha - 1}$. This constraint is solved by
\begin{align}
& 32\lambda^2_{2kB[0]_{\frac{k-2}{2}}^{[0,0,k-2,0]}} + 8\lambda^2_{2kB[0]_{\frac{k}{2}}^{[0,1,k-2,0]}} + 8\lambda^2_{2kB[0]_{\frac{k}{2}}^{[0,0,k,0]}} \nonumber \\
& + 2\lambda^2_{2kB[0]_{\frac{k + 2}{2}}^{[0,2,k-2,0]}} + 2\lambda^2_{2kB[0]_{\frac{k + 2}{2}}^{[0,1,k,0]}} - k(k + 3)\lambda^2_{2kB[0]_{\frac{k + 2}{2}}^{[0,0,k+2,0]}} = 0. \label{1d-crossing}
\end{align}
This does not need to be modified for $k = 2$ which has the additional consideration of Bose symmetry. In this case, we will just get $\lambda^2_{22B[0]_1^{[0100]}} = \lambda^2_{22B[0]_2^{[0120]}} = 0$ automatically. One can permute the correlator with $\Delta_A = \Delta_B = j_A = j_B = 1$ and $\Delta_C = \Delta_D = j_C = j_D = \frac{k}{2}$. Studying this next,
\begin{align}
\frac{\alpha^{-2}}{2^{k - 2}} & \left [ \left ( 16\lambda^2_{2kB[0]_{\frac{k-2}{2}}^{[0,0,k-2,0]}} - 4\lambda^2_{2kB[0]_{\frac{k}{2}}^{[0,1,k-2,0]}} + \lambda^2_{2kB[0]_{\frac{k + 2}{2}}^{[0,2,k-2,0]}} \right ) P^{0,k-2}_0(2\alpha - 1) \right. \label{raw-crossing2} \\
&+ \left. \left ( 4\lambda^2_{2kB[0]_{\frac{k}{2}}^{[0,0,k,0]}} - \lambda^2_{2kB[0]_{\frac{k + 2}{2}}^{[0,1,k,0]}} \right ) P^{0,k-2}_1(2\alpha - 1) + \lambda^2_{2kB[0]_{\frac{k + 2}{2}}^{[0,0,k+2,0]}} P^{0,k-2}_2(2\alpha - 1) \right ] \nonumber \\
=& \left ( \frac{\alpha - 1}{\alpha} \right )^2 \left [ \left ( 16 \lambda_{22B[0]_0^{[0000]}}\lambda_{kkB[0]_0^{[0000]}} + \lambda_{22B[0]_2^{[0200]}}\lambda_{kkB[0]_2^{[0200]}} \right ) P_0^{0,0} \left ( \frac{\alpha + 1}{\alpha - 1} \right ) \right. \nonumber \\
&\left. \hspace{1.5cm} + 4 \lambda_{22B[0]_1^{[0020]}} \lambda_{kkB[0]_1^{[0020]}} P_1^{0,0} \left ( \frac{\alpha + 1}{\alpha - 1} \right ) + \lambda_{22B[0]_2^{[0040]}} \lambda_{kkB[0]_2^{[0040]}} P_2^{0,0} \left ( \frac{\alpha + 1}{\alpha - 1} \right ) \right ]. \nonumber
\end{align}
The left hand side of \eqref{raw-crossing2} has some sign differences compared to \eqref{raw-crossing}. This is because $C_{22\mathcal{O}}C_{kk\mathcal{O}}$ becomes $C_{k2\mathcal{O}}C_{2k\mathcal{O}}$ under crossing which needs to be put back in order. Solving \eqref{raw-crossing2} and taking the identity to be unit-normalized, we have
\begin{subequations}
\label{1d-crossing2}
\begin{align}
& \text{\footnotesize $2^k \lambda_{22B[0]_2^{[0040]}} \lambda_{kkB[0]_2^{[0040]}} = \frac{32}{3} \lambda^2_{2kB[0]_{\frac{k-2}{2}}^{[0,0,k-2,0]}} - \frac{8}{3} \lambda^2_{2kB[0]_{\frac{k}{2}}^{[0,1,k-2,0]}} + \frac{2}{3} \lambda^2_{2kB[0]_{\frac{k+2}{2}}^{[0,2,k-2,0]}}$} \label{1d-crossing2a} \\
& \hspace{3cm} \text{\footnotesize $+ \frac{8(k-1)}{3} \lambda^2_{2kB[0]_{\frac{k}{2}}^{[0,0,k,0]}} - \frac{2(k-1)}{3} \lambda^2_{2kB[0]_{\frac{k+2}{2}}^{[0,1,k,0]}} + \frac{k(k-1)}{3} \lambda^2_{2kB[0]_{\frac{k+2}{2}}^{[0,0,k+2,0]}}$} \nonumber \\
& \text{\footnotesize $2^{k+2} \lambda_{22B[0]_1^{[0020]}} \lambda_{kkB[0]_1^{[0020]}} = -32\lambda^2_{2kB[0]_{\frac{k-2}{2}}^{[0,0,k-2,0]}} + 8\lambda^2_{2kB[0]_{\frac{k}{2}}^{[0,1,k-2,0]}} - 2 \lambda^2_{2kB[0]_{\frac{k+2}{2}}^{[0,2,k-2,0]}}$} \label{1d-crossing2b} \\
& \hspace{3cm} \text{\footnotesize $+ 8\lambda^2_{2kB[0]_{\frac{k}{2}}^{[0,0,k,0]}} - 2\lambda^2_{2kB[0]_{\frac{k+2}{2}}^{[0,1,k,0]}} + k(k+3) \lambda^2_{2kB[0]_{\frac{k+2}{2}}^{[0,0,k+2,0]}}$} \nonumber \\
& \text{\footnotesize $2^{k-1} \lambda_{22B[0]_2^{[0200]}} \lambda_{kkB[0]_2^{[0200]}} = \frac{32}{3} \lambda^2_{2kB[0]_{\frac{k-2}{2}}^{[0,0,k-2,0]}} - \frac{8}{3} \lambda^2_{2kB[0]_{\frac{k}{2}}^{[0,1,k-2,0]}} + \frac{2}{3} \lambda^2_{2kB[0]_{\frac{k+2}{2}}^{[0,2,k-2,0]}} - 2^{k + 3}$} \label{1d-crossing2c} \\
& \hspace{3cm} \text{\footnotesize $- \frac{4(k+2)}{3} \lambda^2_{2kB[0]_{\frac{k}{2}}^{[0,0,k,0]}} + \frac{k+2}{3} \lambda^2_{2kB[0]_{\frac{k+2}{2}}^{[0,1,k,0]}} + \frac{(k+2)(k+3)}{3} \lambda^2_{2kB[0]_{\frac{k+2}{2}}^{[0,0,k+2,0]}}$}. \nonumber
\end{align}
\end{subequations}
The specialization of \eqref{1d-crossing2} to $k = 3$ agrees with the set of equations given in \cite{acp19} even though we have chosen a different basis. It is convenient that we get enough crossing equations to completely fix the new OPE coefficients in terms of the ones that already appear in \eqref{1d-crossing}.

The number of topological crossing equations we obtain, whether of linear or quadratic type, becomes arbitrarily large as we increase the external weights. Consider $\Delta_A = \Delta_C = j_A = j_C = \frac{\mathcal{E}}{2}$, $\Delta_B = \Delta_D = j_B = j_D = \frac{k}{2}$ which is not next-to-next-to-extremal anymore. This is a more general setup which still has only squared OPE coefficients (linear type). If we define the weighted sums
\begin{equation}
\bar{\lambda}_m \equiv \sum_{p = m + \frac{k - \mathcal{E}}{2}}^{\frac{k + \mathcal{E}}{2}} 4^{-p} \lambda^2_{\mathcal{E}kB[0]_p^{\left [ 0, p - m - \frac{k - \mathcal{E}}{2}, 2m + k - \mathcal{E}, 0 \right ]}},
\end{equation}
crossing symmetry for this correlator may be compactly stated in terms of a generalized hypergeometric function.\footnote{Hypergeometric functions with the argument suppressed are understood to be evaluated at $1$.}
\begin{align}
\bar{\lambda}_n &= \sum_{m = 0}^\mathcal{E} K_{nm} \bar{\lambda}_m \label{more-general} \\
K_{nm} &\equiv \frac{(-1)^\mathcal{E} \mathcal{E}! (-\mathcal{E})_n (2n + k - \mathcal{E} + 1)}{n! (n + k - \mathcal{E} + 1)_{\mathcal{E} + 1}} {}_4F_3 \left [ \begin{tabular}{c} \begin{tabular}{cccc} $-m$, & $n - \mathcal{E}$, & $-1 - n - k$, & $1 + m + k - \mathcal{E}$ \end{tabular} \\ \begin{tabular}{ccc} $1$, & $-\mathcal{E}$, & $-\mathcal{E}$ \end{tabular} \end{tabular} \right ] \nonumber
\end{align}
This comes from explicitly projecting one Jacobi polynomial onto another term-by-term.\footnote{Since the conformal blocks for this problem are the same as those for a spatial cross-ratio, analytically continued to negative integer weights, it should be possible to obtain \eqref{more-general} as a limit of the continuous case in \cite{hv17}.} An empirical observation is that this crossing matrix gives half of its geometric multiplicity to the $+1$ eigenvalue rounded up which leads to $\left \lceil \frac{\mathcal{E}}{2} \right \rceil$ crossing equations.

\subsection{A matrix integral for ABJM theory}
Basic consistency conditions like \eqref{1d-crossing} and \eqref{1d-crossing2} need to hold (order-by-order when $c_T$ can be tuned) in any 3d $\mathcal{N} = 8$ SCFT. A more precise check of AdS/CFT can be done if we compute topologically protected data that are unique to ABJM theory at level 1. An important tool for this is an explicit Lagrangian, derived in \cite{dpy16} via localization, which describes the topological subsector for the fixed-point of any theory consisting of $\mathcal{N} = 4$ hyper and vector multiplets. Even though it is a Chern-Simons matter theory, ABJM can be realized in this way due to its duality with a $U(N)$ gauge theory consisting of two hypermultiplets \cite{bk10,bk11}.

Since each representation of the gauge group must be accompanied by its conjugate, the superconformal primary of a hypermultiplet contributes two fields to a gauge theory Lagrangian. Instead of writing down these fields in three dimensions, we will jump right to their twisted-translated counterparts in the topological theory. To wit, a fundamental hypermultiplet gives rise to 1d fields $Q$ and $\tilde{Q}$ -- one for each $N$ dimensional representation of $U(N)$ -- and an adjoint hypermultiplet analogously gives rise to $X$ and $\tilde{X}$. These are the ingredients needed for ABJM theory. Their actions are given by
\begin{align}
S_Q &= -4\pi r \int_{-\pi}^{\pi} \textup{d}\varphi \; \tilde{Q}_\alpha \left [ \dot{Q} + \sigma Q \right ]^\alpha \nonumber \\
S_X &= -4\pi r \int_{-\pi}^{\pi} \textup{d}\varphi \; \tilde{X}^\alpha_{\;\;\beta} \left [ \dot{X} + [\sigma, X] \right ]^\beta_{\;\;\alpha} \label{fixed-sigma-s}
\end{align}
where
\begin{equation}
\sigma = \mathrm{diag}(\sigma_1, \dots, \sigma_N) \label{fixed-sigma-def}
\end{equation}
describes the Cartan of the $U(N)$ gauge group. Computing functional determinants with the proper reality condition, the one for $S_Q$ depends on individual matrix elements of \eqref{fixed-sigma-def} while the one for $S_X$ depends on their differences \cite{dpy16}.
\begin{align}
Z_\sigma &= \left [ \prod_{\alpha = 1}^N \prod_{n \in \mathbb{Z} + \frac{1}{2}} (in + \sigma_\alpha) \right ]^{-1} \left [ \prod_{\alpha, \beta = 1}^N \prod_{n \in \mathbb{Z} + \frac{1}{2}} (in + \sigma_{\alpha\beta}) \right ]^{-1} \nonumber \\
&= \left [ \prod_{\alpha = 1}^N 2\cosh(\pi \sigma_\alpha) \prod_{\alpha, \beta = 1}^N 2\cosh(\pi \sigma_{\alpha\beta}) \right ]^{-1} \label{fixed-sigma-z}
\end{align}
Integrating out $Q$ and $\tilde{Q}$, local operators on the circle will be built from gauge invariant products of $X$ and $\tilde{X}$. Their correlators at fixed $\sigma$, which should be fed into
\begin{equation}
\left < \mathcal{O}_1(\varphi_1) \dots \mathcal{O}_n(\varphi_n) \right > = \frac{1}{Z_N N!} \int \textup{d}^N \sigma \prod_{\alpha < \beta} 4\sinh^2(\pi \sigma_{\alpha\beta}) Z_\sigma \left < \mathcal{O}_1(\varphi_1) \dots \mathcal{O}_n(\varphi_n) \right >_\sigma, \label{fixed-sigma-to-full}
\end{equation}
follow from Wick's theorem. Since $\sigma$ plays the role of a mass, the relevant propagator should be read off from the inverse Fourier transform of $(in + \sigma)^{-1}$. The result of this is
\begin{align}
\left < X^\alpha_{\;\;\beta}(\varphi_1) \tilde{X}^\delta_{\;\;\gamma}(\varphi_2) \right >_\sigma &= -\delta^\alpha_\gamma \delta^\delta_\beta \frac{e^{-\sigma_{\alpha\beta} \varphi_{12}}}{8\pi r} [\mathrm{sgn} \varphi_{12} + \tanh(\pi \sigma_{\alpha\beta})] \label{fixed-sigma-propagator} \\
\left < \mathcal{X}^\alpha_{\;\;\beta}(\varphi_1, y_1) \mathcal{X}^\delta_{\;\;\gamma}(\varphi_2, y_2) \right >_\sigma &= \delta^\alpha_\gamma \delta^\delta_\beta \frac{e^{-\sigma_{\alpha\beta} \varphi_{12}}}{8\pi r} [\mathrm{sgn} \varphi_{12} + \tanh(\pi \sigma_{\alpha\beta})] y_{12}. \nonumber
\end{align}
In the second line, we have recognized that $(X, \tilde{X})$ forms a doublet under $\mathfrak{su}(2)_{\rm II}$ which means we can define $\mathcal{X}$ by saturating this doublet with a polarization vector. Although most of our calculations will be done at large $N$, it is important to note that operators from the interacting theory should only involve the traceless part
\begin{equation}
\hat{\mathcal{X}} \equiv \mathcal{X} - \frac{1}{N} \mathrm{Tr} \mathcal{X}. \label{traceless-x}
\end{equation}
The trace is necessarily a decoupled free multiplet as can be seen from the action or the representation theory of $\mathfrak{osp}(8 | 4)$.

In general, there are many ways to write gauge-invariant monomials that have degree $2j_{\rm I}$ in $\hat{\mathcal{X}}$ and degree $2j_{\rm II}$ in $y$. For us, this degeneracy is resolved by the fact that each trace suppresses an operator further at large $N$.\footnote{Notice that there would be trace relations if we worked at finite $N$. Even if one is careful not to overcount operators, such relations make a grading of the spectrum by the number of traces ambiguous.} We have already mentioned that the most sensible external operators in a Witten diagram expansion are primaries of a half-BPS multiplet. To reflect their single-particle character, we will write
\begin{equation}
\mathcal{O}_k(\varphi, y) = \mathrm{Tr} \left [ \hat{\mathcal{X}}(\varphi, y)^k \right ] \label{single-trace-def}
\end{equation}
in this section. However, it is important to remember that multi-trace operators also contribute to single-particle states. Their contribution just receives additional powers of $1 / N$. The internal operators come in various types which we will now discuss for $\left < \mathcal{O}_2\mathcal{O}_k\mathcal{O}_2\mathcal{O}_k \right >$. In fact, it is not difficult to predict the powers of $N$ at which they first appear.
\begin{gather}
\lambda^2_{2kB[0]_{\frac{k-2}{2}}^{[0,0,k-2,0]}}, \; \lambda^2_{2kB[0]_{\frac{k}{2}}^{[0,1,k-2,0]}} = O \left ( N^{-3} \right ), \; \lambda^2_{2kB[0]_{\frac{k}{2}}^{[0,0,k,0]}} = O \left ( N^{-\frac{3}{2}} \right ) \label{power-laws} \\
\lambda^2_{2kB[0]_{\frac{k+2}{2}}^{[0,2,k-2,0]}}, \; \lambda^2_{2kB[0]_{\frac{k+2}{2}}^{[0,1,k,0]}}, \; \lambda^2_{2kB[0]_{\frac{k+2}{2}}^{[0,0,k+2,0]}} = O(1) \nonumber
\end{gather}
First, the $O(1)$ operators are the ones that survive in generalized free theory so this means they are double-trace. However, $\mathcal{O}_{k + 2}$ has the same quantum numbers as $\left [ \mathcal{O}_2 \mathcal{O}_k \right ]^{[0,0,k+2,0]}$ so, already at tree-level, the squared coefficient for $B[0]_{\frac{k+2}{2}}^{[0,0,k+2,0]}$ should really be considered a sum of two squared OPE coefficients from the field theory point of view. It once again becomes a single squared OPE coefficient (in the right basis), if we regard the external operator not as $\mathcal{O}_k$ but as the single-particle state having maximal overlap with $\mathcal{O}_k$. This is because extremal correlators of operators dual to KK-modes vanish at all orders \cite{Aprile:2019rep,az19}.

Next, $B[0]_{\frac{k}{2}}^{[0,0,k,0]}$ first contributes at tree-level because its dimension is too low to include a double-trace between $\mathcal{O}_2$ and $\mathcal{O}_k$ as one of its components. The two remaining coefficients vanish even at tree-level. For $B[0]_{\frac{k}{2}}^{[0,1,k-2,0]}$, this is because double-trace components are ruled out by scaling dimension while single-trace components are ruled out by the fact that $j_{\rm I} \neq j_{\rm II}$.\footnote{Indeed, once the indices of \eqref{single-trace-def} are restored, there is no way to contract them because $\mathfrak{su}(2)$ does not have any invariant tensors that are totally symmetric.} For $B[0]_{\frac{k-2}{2}}^{[0,0,k-2,0]}$, this is because extremal correlators vanish in the supergravity basis. As we will see, we indeed get a contribution at tree-level if we consider the external operator to be purely $\mathcal{O}_k$ rather than a linear combination of $\mathcal{O}_k$ and $\left [ \mathcal{O}_2 \mathcal{O}_{k-2} \right ]^{[0,0,k,0]}$. It is instructive to check that this linear combination does not affect the $O \left ( N^{-\frac{3}{2}} \right )$ coefficient.

\subsection{Single-trace OPE coefficients}
The next step is to explore what the integral \eqref{fixed-sigma-to-full} can tell us in detail. After solving for single-trace OPE coefficients at the first non-vanishing order following \cite{mpw17}, we will turn to the analogous calculation for double-trace operators. This will reproduce \eqref{order0-res2} but also extend this result to the most general $O(1)$ four-point function.

To understand the partition function
\begin{equation}
Z_N = \frac{1}{N!} \int \textup{d}^N \sigma \prod_{\alpha < \beta} 4\sinh^2(\pi \sigma_{\alpha\beta}) Z_\sigma, \label{0pt-form1}
\end{equation}
we will change to the variable
\begin{equation}
x_\alpha = \frac{\sigma_\alpha}{\sqrt{N}} \label{trick1}
\end{equation}
which should be treated as continuously indexed. Sums over the eigenvalues then become integrals over $x$ where the measure is $N \rho(x) \textup{d}x$. Correspondingly, \eqref{0pt-form1} becomes a functional integral over all of the possible choices for this measure. In the saddle-point approximation, this partition function is
\begin{equation}
Z_N \sim e^{-F[\rho_*]} \label{0pt-form2}
\end{equation}
where $-F[\rho]$ is the log of the integrand in \eqref{0pt-form1} and $\rho_*$ is the density that extremizes it. To proceed, we will need the approximations
\begin{align}
\log 2\cosh \left ( \pi \sqrt{N} x \right ) &\sim \pi \sqrt{N} |x| \nonumber \\
\log \tanh^2 \left ( \pi \sqrt{N} x \right ) &\sim \frac{\pi}{2\sqrt{N}} \delta(x) \label{trick2}
\end{align}
where the latter coefficient follows from Taylor expanding $\log \left ( 1 \pm e^{\pi \sqrt{N} x} \right )$ and integrating term-by-term. This allows us to write the free energy as
\begin{align}
F &= \sum_{\alpha = 1}^N \log 2\cosh \left ( \pi \sqrt{N} x_\alpha \right ) - \frac{1}{2} \sum_{\alpha, \beta = 1}^N \log \tanh^2 \left ( \pi \sqrt{N} x_{\alpha\beta} \right ) \nonumber \\
&\sim \pi N^{\frac{3}{2}} \int \textup{d}x \; \left [ |x| \rho(x) + \frac{1}{4} \rho(x)^2 \right ]. \label{free-energy}
\end{align}
Importantly, the power of $N$ in \eqref{trick1} has been chosen self-consistently so that \eqref{free-energy} can indeed be extremized. Following \cite{hkpt11,mp13}, the solution is
\begin{equation}
\rho_*(x) = \mathrm{max} \left ( \sqrt{2} - 2|x|, 0 \right ) \label{free-dos}
\end{equation}
which leads to the well known result
\begin{equation}
Z_N \sim e^{\frac{\pi \sqrt{2}}{3} N^{\frac{3}{2}}}. \label{0pt-large-n}
\end{equation}
Although the partition function itself will cancel out from correlation functions at leading order, the remaining calculations will depend crucially on the density \eqref{free-dos}.

\begin{figure}[h]
\centering
\includegraphics[scale=0.5]{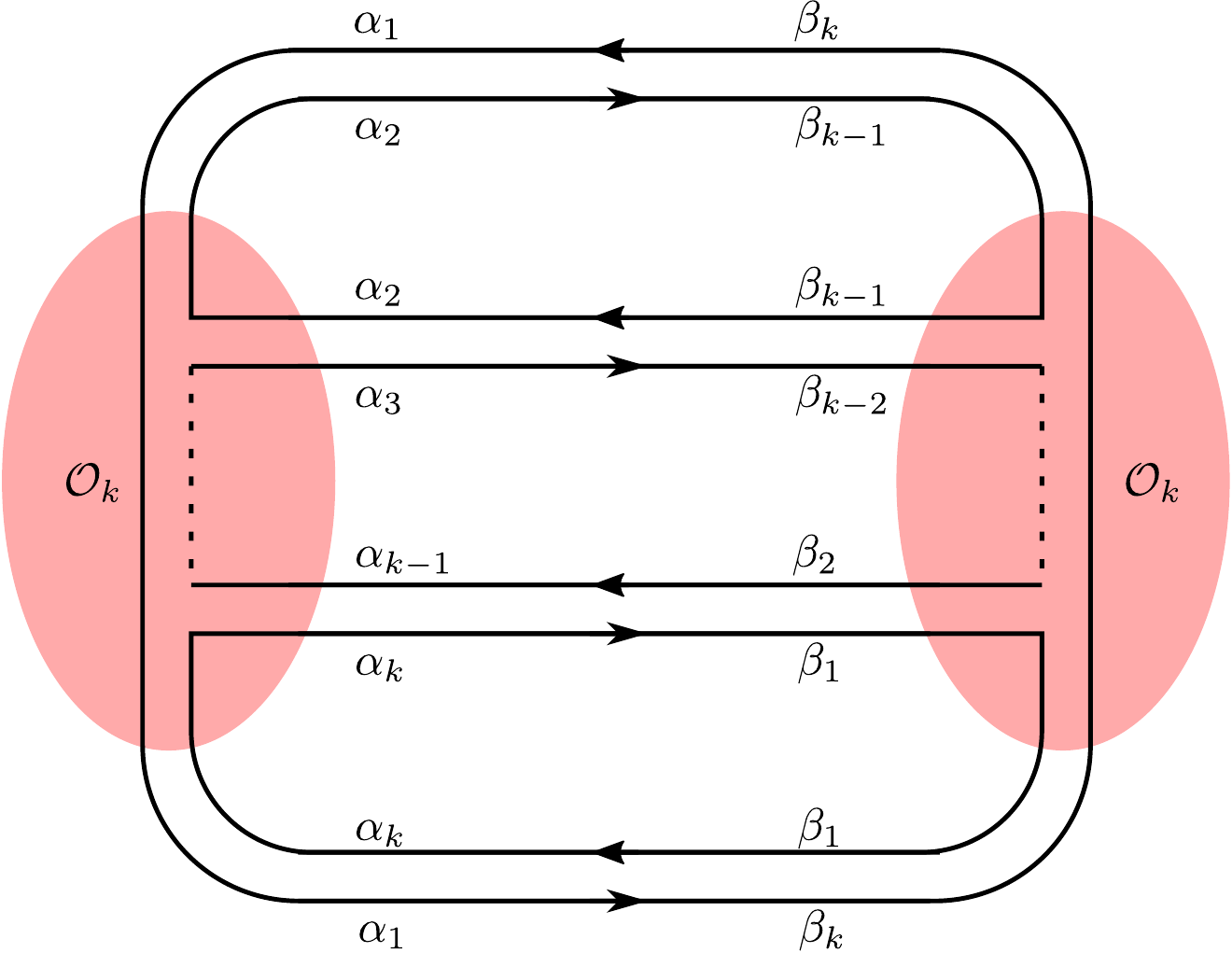}
\caption{A planar diagram representing the $B_k$ two-point function with $k$ contractions of double-lines. The shaded areas, from which the lines emanate, represent the single-trace operators $\mathcal{O}_k$. To translate between the diagram and \eqref{planar-non-diagram}, we explicitly show the $\alpha_j$ indices and the $\beta_{\tau(j) + 1}$ indices to which they are being set equal.}
\label{fig:2pt-st}
\end{figure}
For the correlators at fixed $\sigma$, we will set $\varphi_1 > \varphi_2$ and express the leading term using
\begin{equation}
G_\sigma(\varphi) = e^{-\sigma \varphi} [1 + \tanh(\pi \sigma)] \label{propagator-function}
\end{equation}
which appears in \eqref{fixed-sigma-propagator}. When tracing over a product of $k$ of these functions (and taking $x_{k + 1} \equiv x_1$ for brevity), it is a simple exercise to show that
\begin{align}
\sum_{\alpha_1, \dots, \alpha_k = 1}^N G_{-\sigma_{\alpha_1 \alpha_2}}(\varphi) \dots G_{-\sigma_{\alpha_k \alpha_1}}(\varphi) &= \prod_{j = 1}^k \frac{1}{\cosh \left [ \pi \sqrt{N} (x_j - x_{j + 1}) \right ]} \nonumber \\
&\sim \frac{N^{\frac{1 - k}{2}} \Gamma \left ( \frac{k}{2} \right )}{\sqrt{\pi} \Gamma \left ( \frac{k + 1}{2} \right )} \prod_{j = 1}^{k - 1} \delta(x_j - x_{j + 1}). \label{trick3}
\end{align}
The dominant contribution to $\left < \mathcal{O}_k \mathcal{O}_k \right >$ will indeed have this form. This can be seen by picking an arbitrary $\tau \in S_k$ and considering
\begin{align}
\prod_{j = 1}^k \left < \hat{\mathcal{X}}^{\alpha_j}_{\;\;\;\alpha_{j + 1}}(\varphi_1, y_1) \hat{\mathcal{X}}^{\beta_{\tau(j)}}_{\;\;\;\;\beta_{\tau(j) + 1}}(\varphi_2, y_2) \right > = \left ( \frac{y_{12}}{8\pi r} \right )^k & \sum_{\alpha_1, \dots, \alpha_k = 1}^N G_{-\sigma_{\alpha_1 \alpha_2}}(\varphi_{12}) \dots G_{-\sigma_{\alpha_k \alpha_1}}(\varphi_{12}) \nonumber \\
& \sum_{\beta_1, \dots, \beta_k = 1}^N \prod_{j = 1}^k \delta^{\alpha_j}_{\beta_{\tau(j) + 1}} \delta^{\beta_{\tau(j)}}_{\alpha_{j + 1}}. \label{planar-non-diagram}
\end{align}
The most important permutation is the one that allows every term to survive in the inner sum -- namely
\begin{equation}
\tau(j) = \tau(j + 1) + 1 \implies \tau(j) + j \in \{ 1, \dots, k \} \label{planar-permutation}
\end{equation}
which follows from examining the Kronecker deltas.\footnote{Unsurprisingly, this corresponds to the planar diagram when we use double-line notation as a bookkeeping device. The standard genus expansion would then suggest the scaling $N^k$ since a planar diagram with $2$ vertices and $k$ edges also has $k$ faces. Instead of this, the matrix integral causes $N$ to appear with a non-trivial power.} Clearly, all $k$ of these permutations contribute equally. Figure \ref{fig:2pt-st} shows the result of \eqref{planar-permutation} diagrammatically. As expected in a large $N$ approximation, this pattern of contracting indices leads to the largest possible number of closed loops.

Now that we know the most important term in a two-point function, we can integrate it to get
\begin{align}
B_k = \frac{k \Gamma \left ( \frac{k}{2} \right )}{\sqrt{\pi} \Gamma \left ( \frac{k + 1}{2} \right )} \frac{N^{\frac{k + 1}{2}}}{(8\pi r)^k} \int \textup{d}x \; \rho(x)^k = \frac{\Gamma \left ( \frac{k + 2}{2} \right )}{\sqrt{\pi} \Gamma \left ( \frac{k + 3}{2} \right )} \frac{(2N)^{\frac{k + 1}{2}}}{(8\pi r)^k} \label{2pt-st}
\end{align}
at leading order. The single-trace three-point function will only survive if $\alpha_1$, $\alpha_2$ and $\alpha_3$ from \eqref{alpha-def} are all non-negative integers. When this holds, the leading contraction can be deduced in a similar manner yielding
\begin{align}
C_{k_1k_2k_3} &= \frac{k_1k_2k_3}{N} \left ( \frac{N}{8\pi r} \right )^{\frac{k_1 + k_2 + k_3}{2}} \int \textup{d}x_1 \textup{d}x_2 \; \rho(x_1) \rho(x_2) \cosh \left ( \pi \sqrt{N} x_{12} \right ) \label{3pt-st} \\
& \hspace{4cm} \frac{\prod_{j = 1}^{\alpha_1 - 1} \int \textup{d}s_j \; \rho(s_j)}{\cosh \left [ \pi \sqrt{N} \left ( x_1 - s_1 \right ) \right ] \dots \cosh \left [ \pi \sqrt{N} \left ( s_{\alpha_1 - 1} - x_2 \right ) \right ]} \nonumber \\
& \hspace{4cm} \frac{\prod_{j = 1}^{\alpha_2 - 1} \int \textup{d}t_j \; \rho(t_j)}{\cosh \left [ \pi \sqrt{N} \left ( x_1 - t_1 \right ) \right ] \dots \cosh \left [ \pi \sqrt{N} \left ( t_{\alpha_2 - 1} - x_2 \right ) \right ]} \nonumber \\
& \hspace{4cm} \frac{\prod_{j = 1}^{\alpha_3 - 1} \int \textup{d}u_j \; \rho(u_j)}{\cosh \left [ \pi \sqrt{N} \left ( x_1 - u_1 \right ) \right ] \dots \cosh \left [ \pi \sqrt{N} \left ( u_{\alpha_3 - 1} - x_2 \right ) \right ]} \nonumber \\
&= \frac{16}{k_1 + k_2 + k_3} \frac{\Gamma \left ( \frac{k_1 + 2}{2} \right )\Gamma \left ( \frac{k_2 + 2}{2} \right )\Gamma \left ( \frac{k_3 + 2}{2} \right )}{\sqrt{\pi} \Gamma \left ( \frac{\alpha_1 + 1}{2} \right ) \Gamma \left ( \frac{\alpha_2 + 1}{2} \right ) \Gamma \left ( \frac{\alpha_3 + 1}{2} \right ) \Gamma \left ( \frac{k_1 + k_2 + k_3}{4} \right )} \left ( \frac{\sqrt{2N}}{8\pi r} \right )^{\frac{k_1 + k_2 + k_3}{2}}. \nonumber
\end{align}
The evaluation of this integral requires another formula like \eqref{trick3} which is proven in \cite{mpw17}. Putting \eqref{2pt-st} and \eqref{3pt-st} together, we have the squared OPE coefficient
\begin{align}
\lambda^2_{k_1k_2k_3} &= \frac{2^{k_3} \alpha_2!}{(\alpha_1 + 1)_{\alpha_2}} \frac{C^2_{k_1k_2k_3}}{B_{k_1}B_{k_2}B_{k_3}} \label{ope-st} \\
&= \frac{2^{5-k_1-k_2} \alpha_1! \alpha_2!}{(\alpha_1 + \alpha_2)!} \left ( \frac{\pi}{2\beta} \right )^2 \frac{(k_1 + 1)!(k_2 + 1)!(k_3 + 1)!}{\Gamma \left ( \frac{\alpha_1 + 1}{2} \right )^2 \Gamma \left ( \frac{\alpha_2 + 1}{2} \right )^2 \Gamma \left ( \frac{\alpha_3 + 1}{2} \right )^2 \Gamma \left ( \frac{\beta}{2} \right )^2 (2N)^{\frac{3}{2}}}. \nonumber
\end{align}
This only differs from \eqref{sugra-3pt} by a normalization.

\subsection{Double-trace OPE coefficients}
The possibility of more R-symmetry representations opens up when we move onto double-trace operators. Let us keep track of them in an index-free way with
\begin{equation}
\left [ \mathcal{O}_{\mathcal{E}} \mathcal{O}_k \right ]^{[0, j, k + \mathcal{E} - 2j, 0]}(\varphi, y) = \frac{(k - j)! (\mathcal{E} - j)!}{k! \mathcal{E}!} \left < \partial_y, \partial_{y^\prime} \right >^j
\mathcal{O}_{\mathcal{E}}(\varphi, y) \mathcal{O}_k(\varphi, y^\prime) \biggl |_{y^\prime = y} \label{double-trace-def}
\end{equation}
where we have defined $\left < \partial_y, \partial_{y^\prime} \right > \equiv \partial_y^a \partial_{y^\prime}^b \epsilon_{ab}$. A straightforward calculation shows that the three-point functions we are interested in are essentially squares of two-point functions. This means that, at leading order, $\mathcal{O}_{\mathcal{E}}(\varphi_1, y_1)$ is always contracted with $\mathcal{O}_{\mathcal{E}}(\varphi_3, y_3)$ and $\mathcal{O}_k(\varphi_2, y_2)$ is always contracted with $\mathcal{O}_k(\varphi_3, y_3^\prime)$.\footnote{Clearly, the permutation \eqref{planar-permutation} should be used in both cases.} OPEs with $\mathcal{E} = k$ also allow the opposite type of contraction at the cost of a $(-1)^j$ sign. As a result,
\begin{equation}
C_{\mathcal{E}k\left [ \mathcal{O}_{\mathcal{E}} \mathcal{O}_k \right ]^{[0, j, k + \mathcal{E} - 2j, 0]}} = \frac{1 + (-1)^j \delta_{k\mathcal{E}}}{\pi} \frac{\Gamma \left ( \frac{k + 2}{2} \right )\Gamma \left ( \frac{\mathcal{E} + 2}{2} \right )}{\Gamma \left ( \frac{k + 3}{2} \right )\Gamma \left ( \frac{\mathcal{E} + 3}{2} \right )} \frac{(2N)^{\frac{k + \mathcal{E} + 2}{2}}}{(8\pi r)^{k + \mathcal{E}}}. \label{3pt-dr}
\end{equation}

\begin{figure}[h]
\centering
\subfloat[][$C_{34[\mathcal{O}_3\mathcal{O}_4]}$]{\includegraphics[scale=0.5]{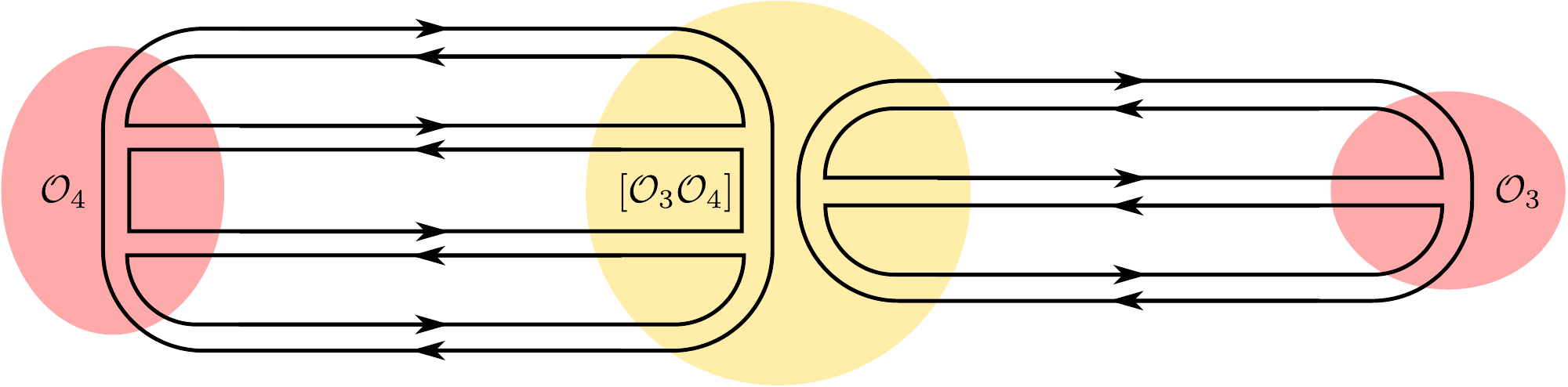}} \\
\subfloat[][$B_{[\mathcal{O}_3\mathcal{O}_4]}$]{\includegraphics[scale=0.5]{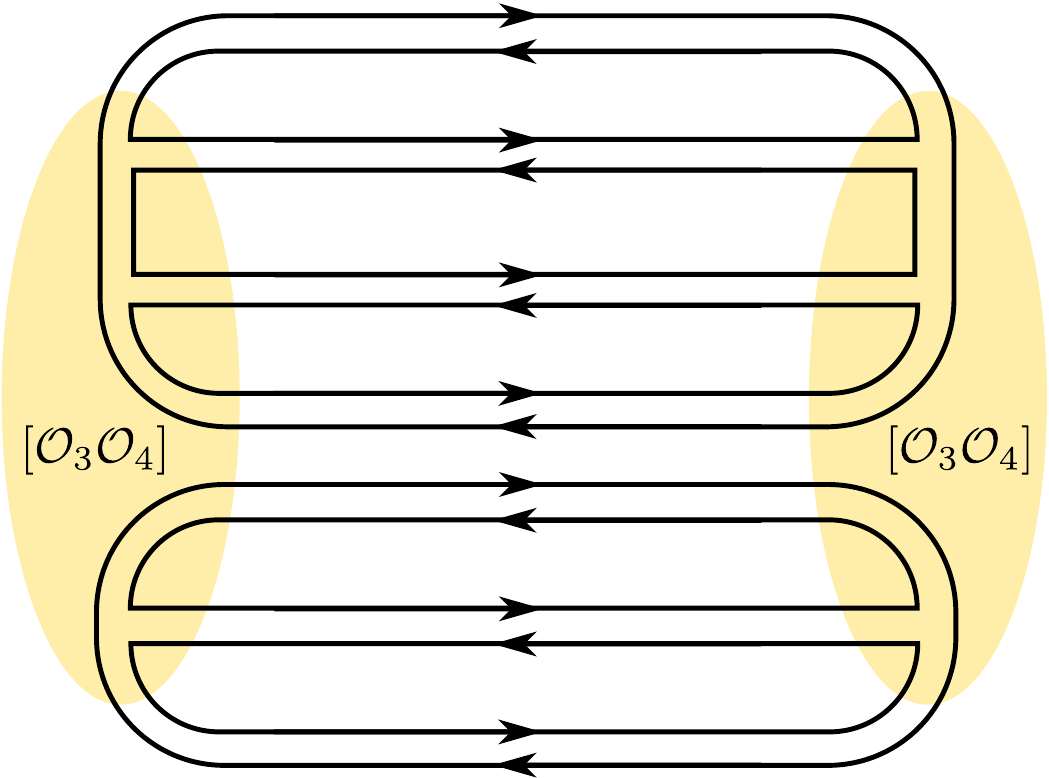}}
\caption{Diagrams showing the correlation functions that effectively reduce to products of two-point functions of the type shown in Figure \ref{fig:2pt-st}. Single-trace operators are shaded in red while double-trace operators are shaded in yellow. We have not attempted to distinguish between the different R-symmetry representations that $[\mathcal{O}_3\mathcal{O}_4]$ can have in the diagram. The main technical challenge of this subsection is that each one leads to derivatives being distributed in different ways.}
\label{2pt-3pt-dt}
\end{figure}
This time, it is the calculation of the norm that requires more work. The key piece of algebra is
\begin{align}
& \text{\footnotesize $\epsilon_{a_1b_1} \partial_{y_2}^{a_1} \dots \epsilon_{a_jb_j} \partial_{y_2}^{a_j} \left < y_1, y_2 \right >^{k - j} \left < y_2, y_2^\prime \right >^p y_1^{\prime b_1} \dots y_1^{\prime b_p} y_2^{b_{p + 1}} \dots y_2^{b_j} \biggl |_{y^\prime = y} = \epsilon_{a_{p + 1}b_{p + 1}} \partial_{y_2}^{a_{p + 1}} \dots \epsilon_{a_jb_j} \partial_{y_2}^{a_j}$} \nonumber \\
& \text{\footnotesize $\left < y_1, y_2 \right >^{k - j} \sum_{q = 0}^p \binom{p}{q}^2 \frac{q! (j - p)!}{(j - 2p + q)!} \left < y_2, y_2^\prime \right >^{p - q} \left < y_1^\prime, y_2^\prime \right >^q y_1^{\prime b_{p + 1}} \dots y_1^{\prime b_{2p - q}} y_2^{b_{2p - q + 1}} \dots y_2^{b_j} \biggl |_{y^\prime = y}$} \nonumber \\
& \text{\footnotesize $= \left < y_1, y_2 \right >^{k - j + p} \sum_{q = 0}^p \binom{p}{q} \frac{p! (j - p)! (k - p + 1)!}{(j - 2p + q)! (k - j + p - q + 1)!}$}. \label{tedious1}
\end{align}
This in turn allows us to compute
\begin{align}
& \text{\footnotesize $\left < \partial_{y_1}, \partial_{y_1^\prime} \right >^j \left < \partial_{y_2}, \partial_{y_2^\prime} \right >^j \left < y_1, y_2 \right >^k \left < y_1^\prime, y_2^\prime \right >^{\mathcal{E}} \biggl |_{y^\prime = y} = \frac{k! \mathcal{E}! \left < \partial_{y_2}, \partial_{y_2^\prime} \right >^j}{(k - j)! (\mathcal{E} - j)!} \left < y_1, y_2 \right >^{k - j} \left < y_1^\prime, y_2^\prime \right >^{\mathcal{E} - j} \left < y_2, y_2^\prime \right >^j \biggl |_{y^\prime = y}$} \nonumber \\
& \text{\footnotesize $= \frac{k! \mathcal{E}!}{(k - j)!} \epsilon_{a_1b_1} \partial_{y_2}^{a_1} \dots \epsilon_{a_jb_j} \partial_{y_2}^{a_j} \left < y_1, y_2 \right >^{k - j} \sum_{p = 0}^j \binom{p}{j} \frac{j! \left < y_1^\prime, y_2^\prime \right >^{\mathcal{E} - j - p}}{p! (\mathcal{E} - j - p)!} \left < y_2, y_2^\prime \right >^p y_1^{\prime b_1} \dots y_1^{\prime b_p} y_2^{b_{p + 1}} \dots y_2^{b_j} \biggl |_{y^\prime = y}$} \nonumber \\
& \text{\footnotesize $= \frac{k! \mathcal{E}!}{(k - j)!} \left < y_1, y_2 \right >^{k + \mathcal{E} - 2j} \sum_{p = 0}^j \binom{j}{p} \frac{j! (k + 1)!}{(\mathcal{E} - j - p)! (k + p - j + 1)!}$}. \label{tedious2}
\end{align}
With \eqref{tedious1} and \eqref{tedious2} in hand, the desired two-point function is
\begin{align}
B_{\left [ \mathcal{O}_{\mathcal{E}} \mathcal{O}_k \right ]^{[0, j, k + \mathcal{E} - 2j, 0]}} = \frac{1 + \delta_{k\mathcal{E}}}{\pi} \frac{(k - j)!(\mathcal{E} - j)!}{k! \mathcal{E}!} \frac{j! (k + \mathcal{E} - j + 1)!}{(k + \mathcal{E} - 2j + 1)!} \frac{\Gamma \left ( \frac{k + 2}{2} \right )\Gamma \left ( \frac{\mathcal{E} + 2}{2} \right )}{\Gamma \left ( \frac{k + 3}{2} \right )\Gamma \left ( \frac{\mathcal{E} + 3}{2} \right )} \frac{(2N)^{\frac{k + \mathcal{E} + 2}{2}}}{(8\pi r)^{k + \mathcal{E}}} \label{2pt-dt}
\end{align}
which means
\begin{align}
\lambda^2_{\mathcal{E}k\left [ \mathcal{O}_{\mathcal{E}} \mathcal{O}_k \right ]^{[0, j, k + \mathcal{E} - 2j, 0]}} &= \frac{2^{k + \mathcal{E}} (\mathcal{E} - j)!}{(k - j + 1)_{\mathcal{E} - j}} \frac{C^2_{\mathcal{E}k\left [ \mathcal{O}_{\mathcal{E}} \mathcal{O}_k \right ]^{[0, j, k + \mathcal{E} - 2j, 0]}}}{B_{\mathcal{E}} B_k B_{\left [ \mathcal{O}_{\mathcal{E}} \mathcal{O}_k \right ]^{[0, j, k + \mathcal{E} - 2j, 0]}}} \nonumber \\
&= 2^{k + \mathcal{E}} \left [ 1 + (-1)^j \delta_{k\mathcal{E}} \right ] \frac{k! \mathcal{E}!}{j!} \frac{k + \mathcal{E} - 2j + 1}{(k + \mathcal{E} - j + 1)!}. \label{ope-dt}
\end{align}
These OPE coefficients, which satisfy \eqref{more-general}, look like they would have been quite hard to derive otherwise. By demanding agreement with the standard methods for generalized free theories, we predict the identity
\begin{equation}
\sigma^{\mathcal{E}} = \sum_{j = 0}^{\mathcal{E}} \frac{k! \mathcal{E}!}{j!} \frac{k + \mathcal{E} - 2j + 1}{(k + \mathcal{E} - j + 1)!} Y^{0, k - \mathcal{E}}_{\mathcal{E}, \mathcal{E} - j}(\sigma, \tau) + O \left ( \sigma^p \tau^q \right ), \;\;\; p + q < \mathcal{E} \label{poly-identity}
\end{equation}
which holds for all of the harmonic polynomials we have checked. The matrix model cannot determine the lower degree terms because they describe double-trace operators that live in A-type multiplets.

Due to a remarkable correspondence with a Fermi gas, all of the perturbative terms in the partition function have been computed and resummed into an Airy function \cite{mp11, h15}. This is the origin of the expression \eqref{ct-3d}.\footnote{These Fermi gas methods also have applications to the protected subsector in four dimensions \cite{bdf15}.} It would be very interesting to see if these higher order techniques could be generalized to cases with operator insertions. Due to the connection between stress tensor correlators and the partition function on a squashed sphere, some initial progress in this direction was made in \cite{acp18}.\footnote{More recently, a framework for expanding more general Higgs and Coulomb branch correlators was developed in \cite{ga20}. We thank Jihwan Oh for bringing this paper to our attention.}

\section{Extracting OPE coefficients in three dimensions}
\label{sec:ope-coeffs}

Among the results of the last section are tree-level single-trace OPE coefficients, namely \eqref{power-laws} and \eqref{ope-st}. Although the limited set of matrix model techniques we have reviewed is not powerful enough to determine tree-level double-trace OPE coefficients outright, we still have \eqref{1d-crossing} and \eqref{1d-crossing2} as highly non-trivial constraints. In this section, we will extract the necessary single-particle and double-particle OPE coefficients using Mellin space and find perfect agreement.

\subsection{Proof of non-degeneracy}
To solve for generic CFT data encoded in a Mellin amplitude one needs detailed knowledge of the superconformal blocks. A question for us is whether the situation is more favorable for the short multiplets of B-type. As we will explain shortly, three necessary steps in this calculation are: projecting onto the R-symmetry polynomial for $[0, j_{\rm I} - j_{\rm II}, 2 j_{\rm II}, 0]$, choosing a particular pole in $s$ to set $\Delta - \ell = j_{\rm I}$ and taking $V \rightarrow 1$ to set $\ell = 0$. Calculations at low weight, such as those in \cite{c18}, have demonstrated that this is sufficient to describe a unique conformal primary. We will now show that the same conclusion continues to hold for arbitrary external weights.

First, we must show that $B[0]_{j_{\rm I}}^{[0, j_{\rm I} - j_{\rm II}, 2 j_{\rm II}, 0]}$ cannot overlap with a super descendant of any other $B[0]_{j_{\rm I}^\prime}^{[0, j_{\rm I}^\prime - j_{\rm II}^\prime, 2 j_{\rm II}^\prime, 0]}$. To achieve degeneracy in the scaling dimension (which can only happen for $j_{\rm I}^\prime < j_{\rm I}$), we start off with the $\Delta = j_{\rm I}^\prime$ primary and act with $2(j_{\rm I} - j_{\rm I}^\prime)$ supercharges. To see that the $\mathfrak{so}(8)$ representations thus produced stay away from $[0, j_{\rm I} - j_{\rm II}, 2 j_{\rm II}, 0]$, it is enough to show that the scaling dimension will never again be equal to the first Cartan. This becomes more transparent if we use the basis
\begin{gather}
Q_\alpha^{[1,0,0,0]}, \; Q_\alpha^{[-1,1,0,0]}, \; Q_\alpha^{[0,-1,1,1]}, \; Q_\alpha^{[0,0,1,-1]} \nonumber \\
Q_\alpha^{[0,0,-1,1]}, \; Q_\alpha^{[0,1,-1,-1]}, \; Q_\alpha^{[1,-1,0,0]}, Q_\alpha^{[-1,0,0,0]} \label{supercharge-weights}
\end{gather}
where we have left the $\mathfrak{su}(2)$ weights implicit. The only supercharge in this list that has a positive first Cartan is $Q_\alpha^{[1,0,0,0]}$. According to table \ref{3d-multiplets}, this is precisely the one that annihilates the primary. We can therefore be certain that the first Cartan of a generic representation will not increase if we constantly use the algebra to move $Q_\alpha^{[1,0,0,0]}$ to the left in the states that we generate.

There are also non-generic representations to consider which have at least one vanishing Dynkin label. If one of the supercharges in \eqref{supercharge-weights} lowers this Dynkin label again, the resulting weight vector $\lambda$ must be transformed according to
\begin{equation}
\lambda \mapsto \omega(\lambda + \rho) - \rho, \;\;\;\;\;\;\; \omega \in \mathrm{Weyl}(\mathfrak{so}(8)) \label{weyl-reflection}
\end{equation}
for some $\omega$ which makes all of the Dynkin labels non-negative. The Weyl vector of $\mathfrak{so(8)}$ is $\rho = [1, 1, 1, 1]$ and the Weyl group of $\mathfrak{so}(8)$ is $S_4 \ltimes (S_2)^3$. The $S_4$ acts by permuting the Cartan eigenvalues and an $S_2$ can be taken to flip the sign of any two of them. We therefore find that under \eqref{weyl-reflection},
\begin{equation}
h_1 \mapsto \pm(h_i + 4 - i) - 3, \;\;\;\;\;\;\; i \in \{1, 2, 3, 4\}. \label{weyl-on-cartan}
\end{equation}
To verify that \eqref{weyl-on-cartan} can never be larger than the original $h_1$, we simply go through all eight possibilities and apply the inequalities that follow from
\begin{equation}
h_1 = r_1 + r_2 + \frac{r_3 + r_4}{2}, \; h_2 = r_2 + \frac{r_3 + r_4}{2}, \; h_3 = \frac{r_3 + r_4}{2}, \; h_4 = \frac{r_3 - r_4}{2}. \label{dynkin-to-cartan}
\end{equation}
In the highest-weight case, \eqref{dynkin-to-cartan} is what ensures $h_i - h_{i + 1} \geq 0$ as befits a Young tableau. Due to the fact that the $\lambda$ considered here is slightly negative, we must weaken this condition to $h_i - h_{i + 1} \geq -1$ which is still enough to prove the claim.

The logic used above is essentially the Racah-Speiser algorithm which has become a powerful tool for building superconformal multiplets \cite{do02,bhp16,cdi16,ll19,aw20}. When applying it in the second step, we must consider primaries that are not necessarily annihilated by a single supercharge. What helps us this time is a ``headstart'' equal to $\ell + 1$ for $A[2\ell]_{j_{\rm I}^\prime + \ell + 1}^{[0, j_{\rm I}^\prime - j_{\rm II}^\prime, 2 j_{\rm II}^\prime, 0]}$ and greater than $\ell + 1$ for a long multiplet. Clearly, $Q_\alpha^{[1,0,0,0]}$ is the only supercharge which raises $h_1$ faster than it raises $\Delta$. We can therefore imagine that one catches up to the other after we descend from the primary twice. However, this leads to Dynkin labels of $[2, j_{\rm I}^\prime - j_{\rm II}^\prime, 2 j_{\rm II}^\prime, 0]$ which cannot appear in \eqref{tensor-product}. To keep us in an admissible $\mathfrak{so}(8)$ representation, each appearance of $Q_\alpha^{[1,0,0,0]}$ must be accompanied by either $Q_\alpha^{[-1,1,0,0]}$ or $Q_\alpha^{[-1,0,0,0]}$. This has the net effect of ensuring that $h_1$ never increases by $1$ unless $\Delta$ increases by $1$ as well.

\subsection{A master formula}
We will now fill in the technical steps of taking a four-point function in Mellin space and extracting a particular set of quantum numbers using \eqref{inverse-mellin}.

If we wanted to solve for the position space correlator in the $0 < U < 1$ regime, we could close the contour to the right and add up the residues from all poles in the $s$ variable. In practice, we only need to look at one pole for each term in the tensor product \eqref{tensor-product}. These come in two distinct types. When $j_{\rm I} < \mathrm{min} \left ( \frac{k_1 + k_2}{2}, \frac{k_3 + k_4}{2} \right )$, all of the gamma functions in \eqref{inverse-mellin} stay finite when $s$ is set equal to the twist of $B[0]_{j_{\rm I}}^{[0, j_{\rm I} - j_{\rm II}, 2 j_{\rm II}, 0]}$. It is only the single-trace function $\mathcal{M}^{(s)}$ which can possibly diverge at this value of $s$. This leads to an especially simple $t$ integral once the $s = j_{\rm I}$ residue is isolated. Setting $V = 1$ brings it into a form that yields immediately to the first Barnes lemma
\begin{equation}
\int_{-i\infty}^{i\infty} \frac{\textup{d}t}{2\pi i} \Gamma(a_1 - t)\Gamma(a_2 - t)\Gamma(b_1 + t)\Gamma(b_2 + t) = \frac{\Gamma(a_1 + b_1)\Gamma(a_1 + b_2)\Gamma(a_2 + b_1)\Gamma(a_2 + b_2)}{\Gamma(a_1 + a_2 + b_1 + b_2)}. \label{barnes}
\end{equation}
It is worth pointing out that adding up the residues in the $t$ variable is not enough to derive \eqref{barnes} as this would miss a contribution from the arc at infinity. One way to see this is to start off with $V$ as a parameter and only take $V \rightarrow 1$ after the residues are added. The term proportional to $\log V$, which is neglected in the naive approach, generically multiplies a sum that diverges as $V \rightarrow 1$ such that the overall correction is finite.

This approach, based on the first Barnes lemma, is enough to handle every term in the outer sum of \eqref{tensor-product} except the last. Setting $j_{\rm I}$ to its maximum possible value is how we first reach a vanishing gamma function argument. As explained in footnote \ref{overlapping}, $s = j_{\rm I}$ is strictly a double-particle pole. In fact, a slightly closer look at the amplitudes reveals that only $\mathcal{M}^{(t)}$ and $\mathcal{M}^{(u)}$ contribute in this case.\footnote{\label{only-tu} This is because $j_{\rm I} = \mathrm{min} \left ( \frac{k_1 + k_2}{2}, \frac{k_3 + k_4}{2} \right )$ is associated with the harmonic polynomials of maximal degree but all of the monomials $\sigma^{\mathcal{E}}, \dots, \tau^{\mathcal{E}}$ are absent from $\mathcal{M}^{(s)}(s, t; \sigma, \tau)$.} The contour of the resulting $t$ integral encloses the single-particle poles of the former and avoids the single-particle poles of the latter. We must therefore proceed according to two slightly different methods. Both of them are based on
\begin{align}
{}_2F_1(a_1, a_2; b_1; z) = & \frac{\Gamma(b_1)}{\Gamma(a_1)\Gamma(a_2)\Gamma(b_1-a_1)\Gamma(b_1-a_2)} \nonumber \\
& \int_{-i\infty}^{i\infty} \frac{\textup{d}t}{2\pi i} \Gamma(a_1-t)\Gamma(a_2-t)\Gamma(b_1-a_1-a_2+t)\Gamma(t)(1 - z)^{-t} \label{master1}
\end{align}
which generalizes \eqref{barnes}. Notably, \eqref{master1} was part of the original justification for \cite{gkss16}. We have not set $z = 0$ above because we want to use it to accommodate simple poles in $t$ that come from the Mellin amplitudes. This is possible thanks to the integral representations
\begin{equation}
\frac{1}{t - 2m - \delta} = -\frac{1}{2} \int_0^1 (1 - z)^{m - 1 + \frac{\delta - t}{2}} \textup{d}z, \;\;\;\;\;\;\; \frac{1}{t + 2m + \delta} = \frac{1}{2} \int_0^1 (1 - z)^{m - 1 + \frac{\delta + t}{2}} \textup{d}z. \label{master2}
\end{equation}
Importantly, we have arranged the exponents so that only the first integral can possibly diverge when the $t$ contour is closed to the right. This reflects the fact that single-particle poles of $\mathcal{M}^{(t)}$ must be counted while those of $\mathcal{M}^{(u)}$ must be skipped. We can now derive two very powerful inverse Mellin transformations from \eqref{master1} and \eqref{master2}. The first is
\begin{align}
& \int_{-i\infty}^{i\infty} \frac{\textup{d}t}{2\pi i} \frac{\Gamma\left ( a_1 - \frac{t}{2} \right )\Gamma\left ( a_2 - \frac{t}{2} \right )\Gamma\left ( b_1 + \frac{t}{2} \right )\Gamma\left ( b_2 + \frac{t}{2} \right )}{t - 2m - \delta} \label{master3} \\
&= -\frac{\Gamma(a_1 + b_1)\Gamma(a_1 + b_2)\Gamma(a_2 + b_1)\Gamma(a_2 + b_2)}{\Gamma(a_1 + a_2 + b_1 + b_2) \left [ b_1 + m + \frac{\delta}{2} \right ]} {}_3F_2 \left [ \begin{tabular}{c} \begin{tabular}{ccc} $1$, & $a_1 + b_1$, & $a_2 + b_1$ \end{tabular} \\ \begin{tabular}{cc} $a_1 + a_2 + b_1 + b_2$, & $1 + b_1 + m + \frac{\delta}{2}$ \end{tabular} \end{tabular} \right ] \nonumber
\end{align}
while the second is
\begin{align}
& \int_{-i\infty}^{i\infty} \frac{\textup{d}t}{2\pi i} \frac{\Gamma\left ( a_1 - \frac{t}{2} \right )\Gamma\left ( a_2 - \frac{t}{2} \right )\Gamma\left ( b_1 + \frac{t}{2} \right )\Gamma\left ( b_2 + \frac{t}{2} \right )}{t + 2m + \delta} \label{master4} \\
&= \frac{\Gamma(a_1 + b_1)\Gamma(a_1 + b_2)\Gamma(a_2 + b_1)\Gamma(a_2 + b_2)}{\Gamma(a_1 + a_2 + b_1 + b_2) \left [ a_1 + m + \frac{\delta}{2} \right ]} {}_3F_2 \left [ \begin{tabular}{c} \begin{tabular}{ccc} $1$, & $a_1 + b_1$, & $a_1 + b_2$ \end{tabular} \\ \begin{tabular}{cc} $a_1 + a_2 + b_1 + b_2$, & $1 + a_1 + m + \frac{\delta}{2}$ \end{tabular} \end{tabular} \right ]. \nonumber
\end{align}
To see that the right hand sides of \eqref{master3} and \eqref{master4} are both separately invariant under $a_1 \leftrightarrow a_2$ and $b_1 \leftrightarrow b_2$, we need to use the Thomae relations.

There is still work to do since these hypergeometric functions depend on $m$. For this, we refer to Appendix \ref{sec:appb} where it is shown that the sum over descendants (an infinite sum for $AdS_4 \times S^7$) can always be written as a \textit{finite} sum of integrals of the form
\begin{equation}
W = \int_{-i\infty}^{i\infty} \frac{\textup{d}t}{2\pi i} \Gamma(-t)\Gamma(1 + a - b_1 - b_2 - b_3 - t) \frac{(b_1)_t(b_2)_t(b_3)_t(1 + a - c_1 - c_2)_t}{(1 + a - c_1)_t(1 + a - c_2)_t}. \label{bailey-integral}
\end{equation}
This integral has attracted a great deal of interest since it appears in the crossing kernel for collinear blocks \cite{lprs18,gs18}. One of its interesting mathematical properties is a representation in terms of a single very well-poised hypergeometric function.
\begin{align}
& \frac{\Gamma(1 + a - b_1)\Gamma(1 + a - b_2)\Gamma(1 + a - b_3) }{\Gamma(1 + a)\Gamma(1 + a - b_1 - b_2)\Gamma(1 + a - b_1 - b_3)\Gamma(1 + a - b_2 - b_3)}\, W \label{bailey-theorem} \\
&= {}_7F_6 \left [ \begin{tabular}{c} \begin{tabular}{ccccccc} $a$, & $1 + \frac{1}{2}a$, & $b_1$, & $b_2$, & $b_3$, & $c_1$, & $c_2$ \end{tabular} \\ \begin{tabular}{cccccc} $\frac{1}{2}a$, & $1 + a - b_1$, & $1 + a - b_2$, & $1 + a - b_3$, & $1 + a - c_1$, & $1 + a - c_2$ \end{tabular} \end{tabular} \right ]. \nonumber
\end{align}
Fortunately, we have been able to find closed form expressions for all instances of \eqref{bailey-integral} that appear in the rest of this section. As we will see, these evaluations make use of \eqref{bailey-theorem} and also more elementary contour manipulations. For reasons that are not yet clear to us, it appears that one method leads to simple expressions if and only if the other leads to complicated expressions.

\subsection{Tree-level results}
We are now in a good position to return to the $\mathcal{E} = 2$ correlators that involve squared OPE coefficients. The first tree-level result we can derive is very simple. Since the Mellin amplitudes \eqref{s-channel} are regular at $s = \frac{k - 2}{2}$, we must have
\begin{equation}
\delta \lambda^2_{2kB[0]_{\frac{k-2}{2}}^{[0,0,k-2,0]}} = 0. \label{order1-res1}
\end{equation}
Moving onto the harmonic polynomials of degree 1,
\begin{align}
& \mathcal{M}_{2k2k}^{[0,1,k-2,0]}(s,t) = -\frac{8}{\pi c_T} \frac{\Gamma \left ( \frac{k+2}{2} \right )}{\Gamma \left ( \frac{k-1}{2} \right )} \sum_{m = 0}^\infty \frac{(2s+k+2)(2s+2kt-k^2-k-2)}{m! \Gamma \left ( \frac{1}{2} - m \right )^2 \Gamma \left ( \frac{k + 3}{2} + m \right ) \left ( s - 2m - \frac{k}{2} \right )} + \dots \nonumber \\
& \mathcal{M}_{2k2k}^{[0,0,k,0]}(s,t) = \frac{8}{\pi c_T} \frac{\Gamma \left ( \frac{k+2}{2} \right )}{\Gamma \left ( \frac{k-1}{2} \right )} \sum_{m = 0}^\infty \frac{(2s+k+2)(2s+k-2)}{m! \Gamma \left ( \frac{1}{2} - m \right )^2 \Gamma \left ( \frac{k + 3}{2} + m \right ) \left ( s - 2m - \frac{k}{2} \right )} + \dots \label{2k2k-degree1}
\end{align}
where the omitted terms only have poles in $t$ and $u$. Clearly, we must take $m = 0$ in \eqref{2k2k-degree1} to get a pole at $s = \frac{k}{2}$. Performing the $t$ integral with \eqref{barnes} leads to
\begin{equation}
\delta \lambda^2_{2kB[0]_{\frac{k}{2}}^{[0,1,k-2,0]}} = 0, \;\;\;\;\; \delta \lambda^2_{2kB[0]_{\frac{k}{2}}^{[0,0,k,0]}} = 2^k \frac{32k}{c_T}. \label{order1-res2}
\end{equation}
Another way to derive this result would have been to use the Ward identity with Bose symmetry. We now come to the more challenging calculation which is the one where the R-symmetry polynomials have maximal degree.
\begin{subequations}\label{2k2k-degree2}
\begin{align}
& \mathcal{M}_{2k2k}^{[0,2,k-2,0]}(s,t) = \frac{8}{\pi(k+1)c_T} \left [ \frac{\Gamma \left ( \frac{k+2}{2} \right )}{\Gamma \left ( \frac{k-1}{2} \right )} \sum_{m = 0}^\infty \frac{(2s - k - 2)(4s+2t+2kt+k^2+k-10)}{m! \Gamma \left ( \frac{1}{2} - m \right )^2 \Gamma \left ( \frac{k + 3}{2} + m \right ) \left ( t - 2m - \frac{k}{2} \right )} \right. \nonumber \\
& \hspace{2.1cm} \left. + \frac{k-1}{\sqrt{\pi}} \sum_{m = 0}^\infty \frac{(2s-k-2)(2s+2t+2kt-k^2-4k-8) \Gamma \left ( \frac{5}{2} + m \right )^{-1}}{m! \Gamma \left ( \frac{1}{2} - m \right ) \Gamma \left ( \frac{k-1}{2} - m \right )(s + t + 2m - k - 1)} \right ] \label{2k2k-degree2a} \\
& \mathcal{M}_{2k2k}^{[0,1,k,0]}(s,t) = -\frac{8}{\pi(k+2)c_T} \left [ \frac{\Gamma \left ( \frac{k+2}{2} \right )}{\Gamma \left ( \frac{k-1}{2} \right )} \sum_{m = 0}^\infty \frac{(2s - k - 2)(8s+4t+2kt+k^2-20)}{m! \Gamma \left ( \frac{1}{2} - m \right )^2 \Gamma \left ( \frac{k + 3}{2} + m \right ) \left ( t - 2m - \frac{k}{2} \right )} \right. \nonumber \\
& \hspace{2.1cm} \left. - \frac{2}{\sqrt{\pi}} \sum_{m = 0}^\infty \frac{(2s-k-2)(2s-ks+2t+kt-8) \Gamma \left ( \frac{5}{2} + m \right )^{-1}}{m! \Gamma \left ( \frac{1}{2} - m \right ) \Gamma \left ( \frac{k-1}{2} - m \right ) (s + t + 2m - k - 1)} \right ] \label{2k2k-degree2b} \\
& \mathcal{M}_{2k2k}^{[0,0,k+2,0]}(s,t) = \frac{16k}{\pi(k+1)(k+2)c_T} \left [ \frac{\Gamma \left ( \frac{k+2}{2} \right )}{\Gamma \left ( \frac{k-1}{2} \right )} \sum_{m = 0}^\infty \frac{(2s - k - 2)(2s - k - 6)}{m! \Gamma \left ( \frac{1}{2} - m \right )^2 \Gamma \left ( \frac{k + 3}{2} + m \right ) \left ( t - 2m - \frac{k}{2} \right )} \right. \nonumber \\
& \hspace{2.1cm} \left. - \frac{1}{\sqrt{\pi}} \sum_{m = 0}^\infty \frac{(2s-k-2)(2s-k-6) \Gamma \left ( \frac{5}{2} + m \right )^{-1}}{m! \Gamma \left ( \frac{1}{2} - m \right ) \Gamma \left ( \frac{k-1}{2} - m \right ) (s + t + 2m - k - 1)} \right ] \label{2k2k-degree2c}
\end{align}
\end{subequations}
The common zero in all of these projected amplitudes is no accident. Despite the squared gamma function, $s = \frac{k + 2}{2}$ must be only a simple pole since the short multiplets below threshold are protected by recombination rules. After extracting this pole, we can manually compute the $t$ integral with \eqref{master3} and \eqref{master4}. Note that this requires a partial fraction decomposition which reintroduces explicit contact terms (to be treated with \eqref{barnes}). The resulting expressions are
\begin{subequations}\label{order1-res2}
\begin{align}
& \text{\footnotesize $\delta\lambda^2_{2kB[0]_{\frac{k+2}{2}}^{[0,2,k-2,0]}} = \frac{2^{k + 9}}{\pi^2 c_T} \frac{k - 1}{k}$} \nonumber \\
& \hspace{0.8cm} \text{\footnotesize $-\frac{2^{k + 8}}{\pi(k+1)c_T} \left [ \frac{\Gamma \left ( \frac{k}{2} \right )^2}{\Gamma \left ( \frac{k-1}{2} \right )} \sum_{m = 0}^\infty \frac{2m + 2mk + k^2 + 2k - 3}{m! \Gamma \left ( \frac{1}{2} - m \right )^2 \Gamma \left ( \frac{k + 3}{2} + m \right )} \frac{{}_3F_2 \left [ \begin{tabular}{c} \begin{tabular}{ccc} $1$, & $1$, & $1$ \end{tabular} \\ \begin{tabular}{cc} $\frac{k}{2} + 1$, & $\frac{3}{2} + m$ \end{tabular} \end{tabular} \right ]}{2m + 1} \right.$} \nonumber \\
& \hspace{0.8cm} \text{\footnotesize $\left. + \frac{k-1}{\sqrt{\pi}} \frac{\Gamma \left ( \frac{k}{2} \right )^2}{\Gamma \left ( \frac{k+2}{2} \right )} \sum_{m = 0}^\infty \frac{2m + 2mk + k + 3}{m! \Gamma \left ( \frac{1}{2} - m \right ) \Gamma \left ( \frac{k-1}{2} - m \right ) \Gamma \left ( \frac{5}{2} + m \right )} \frac{{}_3F_2 \left [ \begin{tabular}{c} \begin{tabular}{ccc} $1$, & $1$, & $\frac{k}{2}$ \end{tabular} \\ \begin{tabular}{cc} $\frac{k}{2} + 1$, & $\frac{3}{2} + m$ \end{tabular} \end{tabular} \right ]}{2m + 1} \right ]$} \label{order1-res2a} \\
& \text{\footnotesize $\delta\lambda^2_{2kB[0]_{\frac{k+2}{2}}^{[0,1,k,0]}} = \frac{2^{k + 8}}{\pi(k+2)c_T} \left [ \frac{\Gamma \left ( \frac{k}{2} \right )^2}{\Gamma \left ( \frac{k-1}{2} \right )} \sum_{m = 0}^\infty \frac{4m + 2mk + k^2 + 3k - 6}{m! \Gamma \left ( \frac{1}{2} - m \right )^2 \Gamma \left ( \frac{k + 3}{2} + m \right )} \frac{{}_3F_2 \left [ \begin{tabular}{c} \begin{tabular}{ccc} $1$, & $1$, & $1$ \end{tabular} \\ \begin{tabular}{cc} $\frac{k}{2} + 1$, & $\frac{3}{2} + m$ \end{tabular} \end{tabular} \right ]}{2m + 1} \right.$} \nonumber \\
& \hspace{0.8cm} \text{\footnotesize $\left. - \frac{1}{\sqrt{\pi}} \frac{\Gamma \left ( \frac{k}{2} \right )^2}{\Gamma \left ( \frac{k+2}{2} \right )} \sum_{m = 0}^\infty \frac{4m + 2mk - k + 6}{m! \Gamma \left ( \frac{1}{2} - m \right ) \Gamma \left ( \frac{k-1}{2} - m \right ) \Gamma \left ( \frac{5}{2} + m \right )} \frac{{}_3F_2 \left [ \begin{tabular}{c} \begin{tabular}{ccc} $1$, & $1$, & $\frac{k}{2}$ \end{tabular} \\ \begin{tabular}{cc} $\frac{k}{2} + 1$, & $\frac{3}{2} + m$ \end{tabular} \end{tabular} \right ]}{2m + 1} \right ]$} \label{order1-res2b} \\
& \text{\footnotesize $\delta\lambda^2_{2kB[0]_{\frac{k+2}{2}}^{[0,0,k+2,0]}} = \frac{2^{k + 10}k}{\pi c_T} \left [ \frac{\Gamma \left ( \frac{k}{2} \right )^2}{\Gamma \left ( \frac{k-1}{2} \right )} \sum_{m = 0}^\infty \frac{(k+1)^{-1}(k+2)^{-1}}{m! \Gamma \left ( \frac{1}{2} - m \right )^2 \Gamma \left ( \frac{k + 3}{2} + m \right )} \frac{{}_3F_2 \left [ \begin{tabular}{c} \begin{tabular}{ccc} $1$, & $1$, & $1$ \end{tabular} \\ \begin{tabular}{cc} $\frac{k}{2} + 1$, & $\frac{3}{2} + m$ \end{tabular} \end{tabular} \right ]}{2m + 1} \right.$} \nonumber \\
& \hspace{0.8cm} \text{\footnotesize $\left. + \frac{1}{\sqrt{\pi}} \frac{\Gamma \left ( \frac{k}{2} \right )^2}{\Gamma \left ( \frac{k+2}{2} \right )} \sum_{m = 0}^\infty \frac{(k+1)^{-1}(k+2)^{-1}}{m! \Gamma \left ( \frac{1}{2} - m \right ) \Gamma \left ( \frac{k-1}{2} - m \right ) \Gamma \left ( \frac{5}{2} + m \right )} \frac{{}_3F_2 \left [ \begin{tabular}{c} \begin{tabular}{ccc} $1$, & $1$, & $\frac{k}{2}$ \end{tabular} \\ \begin{tabular}{cc} $\frac{k}{2} + 1$, & $\frac{3}{2} + m$ \end{tabular} \end{tabular} \right ]}{2m + 1} \right ]$}. \label{order1-res2c}
\end{align}
\end{subequations}
The middle line vanishes when $k = 2$ which makes sense because of Bose symmetry.

As a warm-up to computing \eqref{order1-res2} in closed form, let us first verify that crossing symmetry holds at tree-level in the topological subsector. This is somewhat simpler due to cancellations that take place when \eqref{order1-res2a}, \eqref{order1-res2b} and \eqref{order1-res2c} are added with the right coefficients. Defining
\begin{align}
S_1 &= \sum_{m = 0}^\infty \frac{\left ( \frac{1}{2} \right )_m^2}{m! \left ( \frac{k + 3}{2} \right )_m \left ( m + \frac{1}{2} \right )} {}_3F_2 \left [ \begin{tabular}{c} \begin{tabular}{ccc} $1$, & $1$, & $1$ \end{tabular} \\ \begin{tabular}{cc} $\frac{k}{2} + 1$, & $\frac{3}{2} + m$ \end{tabular} \end{tabular} \right ] \label{shorthand-sum} \\
S_2 &= \sum_{m = 0}^\infty \frac{\left ( \frac{1}{2} \right )_m \left ( \frac{3 - k}{2} \right )_m}{m! \left ( \frac{3}{2} \right )_m \left ( m + \frac{1}{2} \right )} {}_3F_2 \left [ \begin{tabular}{c} \begin{tabular}{ccc} $1$, & $1$, & $\frac{k}{2}$ \end{tabular} \\ \begin{tabular}{cc} $\frac{k}{2} + 1$, & $\frac{3}{2} + m$ \end{tabular} \end{tabular} \right ], \nonumber
\end{align}
the crossing equation \eqref{1d-crossing} takes the form
\begin{equation}
\frac{k}{2} + \frac{2(k - 1)}{k\pi^2} - \frac{k \Gamma \left ( \frac{k}{2} \right )^2}{\pi^2 \Gamma \left ( \frac{k-1}{2} \right ) \Gamma \left ( \frac{k+3}{2} \right )} S_1 - \frac{2k \Gamma \left ( \frac{k}{2} \right )^2}{\pi^{\frac{5}{2}} \Gamma \left ( \frac{k-1}{2} \right ) \Gamma \left ( \frac{k+2}{2} \right )} S_2 = 0. \label{shorthand-crossing}
\end{equation}
Our task is now to evaluate \eqref{shorthand-sum}.

Starting with $S_1$, Appendix \ref{sec:appb} instructs us to write it as the following Mellin-Barnes integral.
\begin{align}
S_1 &= \frac{\Gamma \left ( \frac{k+3}{2} \right )}{\sqrt{\pi}\Gamma \left ( \frac{k}{2} \right )} \int_{-i\infty}^{i\infty} \frac{\textup{d}t}{2\pi i} \frac{\Gamma(-t)\Gamma \left ( -\frac{1}{2} - t \right )\Gamma(1 + t)^2\Gamma \left ( \frac{k}{2} + t \right )\Gamma \left ( \frac{k+3}{2} + t \right )}{\Gamma \left ( \frac{k+2}{2} + t \right )\Gamma \left( \frac{k+4}{2} + t \right )} \label{sum-easy1} \\
&= \frac{\Gamma \left ( \frac{k+3}{2} \right )^2 \Gamma \left ( \frac{k-1}{2} \right )\Gamma(k+2)}{\Gamma \left ( \frac{k+4}{2} \right )^2\Gamma \left ( \frac{k+1}{2} \right )\Gamma(k+1)} {}_7F_6 \left [ \begin{tabular}{c} \begin{tabular}{ccccccc} $k + 1$, & $\frac{k+3}{2}$, & $\frac{k}{2}$, & $\frac{k}{2}$, & $\frac{k+2}{2}$, & $\frac{k+3}{2}$, & $1$ \end{tabular} \\ \begin{tabular}{cccccc} $\frac{k+1}{2}$, & $\frac{k+4}{2}$, & $\frac{k+4}{2}$, & $\frac{k+2}{2}$, & $\frac{k+1}{2}$, & $k+1$ \end{tabular} \end{tabular} \right ]. \nonumber
\end{align}
We have turned it into a single hypergeometric function using \eqref{bailey-theorem}. Even though there are three ways to do this, the other two are much less favorable. The important property of \eqref{sum-easy1} is that many of the numerator parameters can be paired with denominator parameters that are smaller by non-negative integers. After reducing the order four times, \eqref{sum-easy1} becomes
\begin{align}
S_1 &= \frac{\Gamma \left ( \frac{k+3}{2} \right )^2 \Gamma \left ( \frac{k-1}{2} \right )\Gamma(k+2)}{\Gamma \left ( \frac{k+4}{2} \right )^2\Gamma \left ( \frac{k+1}{2} \right )\Gamma(k+1)} \left [ \frac{4k^2(k+2)^2}{(k+1)^2(k+4)^2(k+6)^2} {}_3F_2 \left [ \begin{tabular}{c} \begin{tabular}{ccc} $\frac{k+4}{2}$, & $\frac{k+4}{2}$, & $3$ \end{tabular} \\ \begin{tabular}{cc} $\frac{k+8}{2}$, & $\frac{k+8}{2}$ \end{tabular} \end{tabular} \right ] \right. \nonumber \\
&\left. + \frac{4k^2(k+2)}{(k+1)^2(k+4)^2} {}_3F_2 \left [ \begin{tabular}{c} \begin{tabular}{ccc} $\frac{k+2}{2}$, & $\frac{k+2}{2}$, & $2$ \end{tabular} \\ \begin{tabular}{cc} $\frac{k+6}{2}$, & $\frac{k+6}{2}$ \end{tabular} \end{tabular} \right ] + {}_3F_2 \left [ \begin{tabular}{c} \begin{tabular}{ccc} $\frac{k}{2}$, & $\frac{k}{2}$, & $1$ \end{tabular} \\ \begin{tabular}{cc} $\frac{k+4}{2}$, & $\frac{k+4}{2}$ \end{tabular} \end{tabular} \right ] \right ] \nonumber \\
&= \frac{k^2 - 1}{16} \frac{\Gamma \left ( \frac{k-1}{2} \right )^2}{\Gamma \left ( \frac{k+2}{2} \right )^2} \left [ 2(k - 1) + k^2 \psi^{(1)} \left ( \frac{k}{2} \right ) \right ] \label{sum-easy2}
\end{align}
where we have used some of the tricks in \cite{m11}.

The same technique does not appear to work for $S_2$. The approach we will take instead starts with the substitution $s = t + \frac{k - 1}{2}$. Equivalently, the iterated Mellin-Barnes integrals that appear in Appendix \ref{sec:appb} should be evaluated in the opposite order. This leads to
\begin{equation}
S_2 = \frac{\sqrt{\pi} \Gamma \left ( \frac{k+2}{2} \right )}{2\Gamma \left ( \frac{k}{2} \right )^2 \Gamma \left ( \frac{3-k}{2} \right )} \int_{-i\infty}^{i\infty} \frac{\textup{d}s}{2\pi i} \frac{\Gamma(-s)\Gamma \left ( \frac{k-1}{2} - s \right )\Gamma \left ( \frac{3-k}{2} + s \right )\Gamma \left ( \frac{1}{2} + s \right )^2\Gamma \left ( 1 + s \right )}{\Gamma \left ( \frac{3}{2} + s \right )^2} \label{sum-hard1}
\end{equation}
after we choose the right ordering for the numerator parameters in \eqref{shorthand-sum}. Crucially, the two gamma function arguments in \eqref{sum-hard1} that depend on $k$ also add up to unity. This allows us to remove the $k$ dependence from the integrand entirely after using the reflection formula. The important caveat to keep in mind is that, after we do this, the natural contour for the resulting integral will live in the critical strip $-\frac{1}{2} < \Re(s) < 0$. Since this is not the contour for \eqref{sum-hard1}, we need to go between them by subtracting the residues at $\frac{1}{2}, \frac{3}{2}, \dots, \frac{k - 3}{2}$. An arbitrary residue from this set is given by
\begin{align}
\lim_{s \rightarrow \frac{k-3}{2} - j} \frac{\Gamma(-s)\Gamma \left ( \frac{k-1}{2} - s \right )\Gamma \left ( \frac{3-k}{2} + s \right )\Gamma \left ( \frac{1}{2} + s \right )^2\Gamma \left ( 1 + s \right )}{(s + \frac{3-k}{2} + j)^{-1} \Gamma \left ( \frac{3}{2} + s \right )^2} = \frac{\Gamma \left ( \frac{3-k}{2} \right )\Gamma \left ( \frac{k-1}{2} \right )}{\left ( \frac{2-k}{2} + j \right )^2} \label{nice-residue}
\end{align}
which is a nice object to have in a finite sum. Putting the finite sum and the massaged integral together,
\begin{align}
S_2 &= \frac{\Gamma \left ( \frac{k+2}{2} \right ) \Gamma \left ( \frac{k-1}{2} \right )}{2 \sqrt{\pi} \Gamma \left ( \frac{k}{2} \right )^2} \left [ \int_{-i\infty}^{i\infty} \frac{\textup{d}s}{2\pi i} \frac{\Gamma(-s)\Gamma \left ( \frac{1}{2} - s \right )\Gamma \left ( \frac{1}{2} + s \right )^3\Gamma \left ( 1 + s \right )}{\Gamma \left ( \frac{3}{2} + s \right )^2} + \sum_{j = 0}^{\frac{k-4}{2}} \frac{\pi}{\left ( \frac{2 - k}{2} + j \right )^2} \right ] \nonumber \\
&= \frac{\Gamma \left ( \frac{k+2}{2} \right ) \Gamma \left ( \frac{k-1}{2} \right )}{2 \sqrt{\pi} \Gamma \left ( \frac{k}{2} \right )^2} \left [ \frac{\pi^3}{3} + \pi \left ( \frac{\pi^2}{6} - \psi^{(1)} \left ( \frac{k}{2} \right ) \right ) \right ]. \label{sum-hard2}
\end{align}
From \eqref{sum-easy2} and \eqref{sum-hard2}, we can now see that crossing symmetry holds exactly in the 1d topological sector for all $k$.

\begin{figure}[h]
\centering
\includegraphics[scale=0.5]{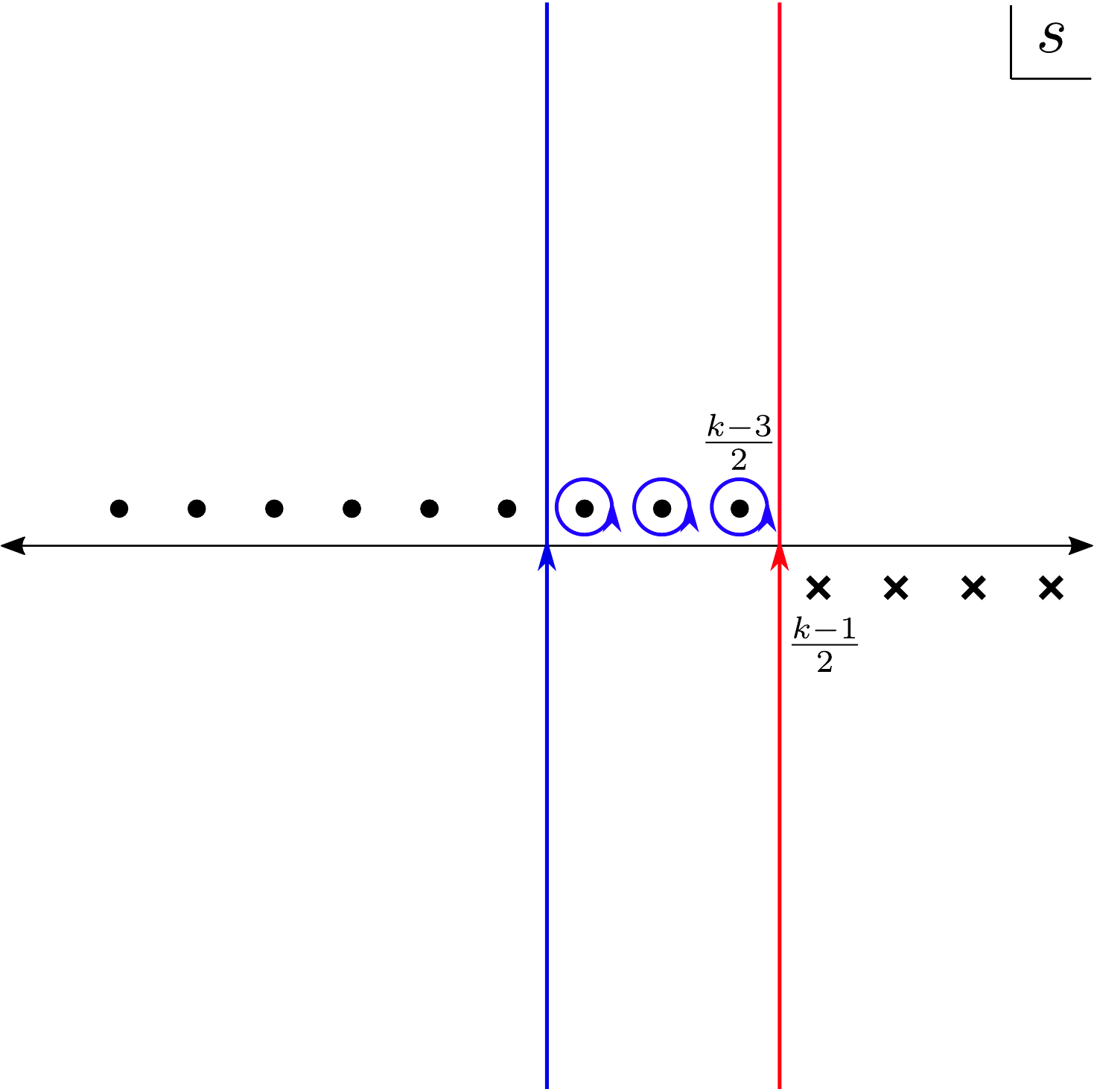}
\caption{The complex $s$-plane showing the poles of $\Gamma \left ( \frac{k - 1}{2} - s \right ) \Gamma \left ( \frac{3 - k}{2} + s \right )$ offset from the real axis for clarity. The red contour encloses the sequence that increases from $s = \frac{k - 1}{2}$ (crosses) and avoids the sequence that decreases from $s = \frac{k - 3}{2}$ (dots). It is therefore the Mellin-Barnes contour for \eqref{sum-hard1}. It must be deformed until it reaches the imaginary axis which is the Mellin-Barnes contour for the integral in \eqref{sum-hard2} that involves $\Gamma \left ( \frac{1}{2} - s \right )\Gamma \left ( \frac{1}{2} + s \right )$. The blue contour therefore includes residues at a discrete set of points. Note that we have suppressed the poles of other gamma functions that are irrelevant to this discussion.}
\label{fig:contours}
\end{figure}
Extending this analysis to all of \eqref{order1-res2} is now conceptually straightforward -- it leads to three sums similar to \eqref{sum-easy2} and three sums similar to \eqref{sum-hard2}. Performing this calculation, we arrive at
\begin{align}
\delta \lambda^2_{2kB[0]_{\frac{k+2}{2}}^{[0,2,k-2,0]}} &= \frac{2^{k + 6}(k - 1)}{k^2 (k + 1) \pi^2 c_T} \left [ 4(k^3 + k^2 + 2k + 4) - k^2 (k + 2) \pi^2 - 2k^2 (k + 2) \psi^{(1)} \left ( \frac{k}{2} \right ) \right ] \nonumber \\
\delta \lambda^2_{2kB[0]_{\frac{k+2}{2}}^{[0,1,k,0]}} &= \frac{2^{k + 8}}{k^2(k + 2) \pi^2 c_T} \left [ 2(k-1)(k^2-4) - k^2 \pi^2 + k^2 (k^2 + 2k - 2) \psi^{(1)} \left ( \frac{k}{2} \right ) \right ] \nonumber \\
\delta \lambda^2_{2kB[0]_{\frac{k+2}{2}}^{[0,0,k+2,0]}} &= \frac{2^{k + 7}}{k(k + 1)(k + 2)\pi^2 c_T} \left [ 4(k - 1)(k + 2) + k^2 \pi^2 + 2k^2 \psi^{(1)} \left ( \frac{k}{2} \right ) \right ]. \label{order1-res3}
\end{align}

\subsection{The other channel}
To learn even more about the $\mathcal{E} = 2$ correlators, we can look at the channel that does not produce squared OPE coefficients. There is one protected operator whose contribution we can extract from
\begin{align}
\mathcal{M}_{22kk}^{[0020]}(s, t) = \frac{16k}{\pi^{\frac{3}{2}}c_T} \sum_{m = 0}^\infty \frac{s(s + 2)}{m!\Gamma \left ( \frac{1}{2} - m \right )\Gamma \left ( \frac{k-1}{2} - m \right )\Gamma \left ( \frac{5}{2} + m \right )(s - 2m - 1)} + \dots \label{22kk-degree1}
\end{align}
by taking the residue at $s = 1$. Again, the omitted terms only have poles in $t$ and $u$. Setting $m = 0$ and applying the familiar steps with \eqref{barnes} leads to
\begin{equation}
\delta \lambda_{22B[0]_1^{[0020]}} \lambda_{kkB[0]_1^{[0020]}} = \frac{128 k}{c_T}. \label{order1-res4}
\end{equation}
Comparing to \eqref{1d-crossing2}, this is exactly what it should be. There are two more checks to be done and these involve the Mellin amplitudes projected onto degree 2 harmonic polynomials.
\begin{align}
\mathcal{M}_{22kk}^{[0200]}(s, t) &= \frac{8k}{3\pi c_T} \frac{\Gamma \left ( \frac{k+2}{2} \right )}{\Gamma \left ( \frac{k-1}{2} \right )} \sum_{m = 0}^\infty \frac{s - k}{m!\Gamma \left ( \frac{1}{2} - m \right )^2 \Gamma \left ( \frac{k + 3}{2} + m \right )} \nonumber \\
& \hspace{2.5cm} \left [ \frac{4s + 6t - k - 2}{t - 2m - \frac{k}{2}} + \frac{2s + 6t - 5k - 10}{s + t + 2m - 2 - \frac{k}{2}} \right ] \nonumber \\
\mathcal{M}_{22kk}^{[0040]}(s, t) &= \frac{16k}{3\pi c_T} \frac{\Gamma \left ( \frac{k+2}{2} \right )}{\Gamma \left ( \frac{k-1}{2} \right )} \sum_{m = 0}^\infty \frac{(s - k)(s - k - 2)}{m!\Gamma \left ( \frac{1}{2} - m \right )^2 \Gamma \left ( \frac{k + 3}{2} + m \right )} \nonumber \\
& \hspace{2.5cm} \left [ \frac{1}{t - 2m - \frac{k}{2}} - \frac{1}{s + t + 2m - 2 - \frac{k}{2}} \right ] \label{22kk-degree2}
\end{align}
The analogue of \eqref{order1-res2}, found by taking the residue at $s = 2$, is much simpler now because the hypergeometric functions immediately reduce to polygamma functions.
\begin{align}
\delta \lambda_{22B[0]_2^{[0200]}} \lambda_{kkB[0]_2^{[0200]}} &= \frac{1024(k - 1)}{\pi^2 c_T} - \frac{256 k^2}{3\pi c_T} \frac{\Gamma \left ( \frac{k}{2} \right )^2}{\Gamma \left ( \frac{k-1}{2} \right )} \sum_{m = 0}^\infty \frac{(6m + k + 3) \psi^{(1)} \left ( \frac{1}{2} + m \right )}{m! \Gamma \left ( \frac{1}{2} - m \right )^2\Gamma \left ( \frac{k+3}{2} + m \right )} \nonumber \\
\delta \lambda_{22B[0]_2^{[0040]}} \lambda_{kkB[0]_2^{[0040]}} &= \frac{256k^3}{3\pi c_T}  \frac{\Gamma \left ( \frac{k}{2} \right )^2}{\Gamma \left ( \frac{k-1}{2} \right )} \sum_{m = 0}^\infty \frac{\psi^{(1)} \left ( \frac{1}{2} + m \right )}{m! \Gamma \left ( \frac{1}{2} - m \right )^2\Gamma \left ( \frac{k+3}{2} + m \right )} \label{order1-res5}
\end{align}
Because of this, we do not need the full power of Appendix \ref{sec:appb} to get a closed form. The fastest way is to write
\begin{equation}
\psi^{(1)} \left ( \frac{1}{2} + m \right ) = \lim_{\eta,\zeta \rightarrow \frac{1}{2}} \left [ \frac{\partial^2}{\partial \eta^2} \frac{\Gamma(\eta + m)}{\Gamma \left ( \frac{1}{2} + m \right )} - \frac{\partial^2}{\partial \eta \partial \zeta} \frac{\Gamma(\eta + m)\Gamma(\zeta + m)}{\Gamma \left ( \frac{1}{2} + m \right )^2} \right ] \label{polygamma-def}
\end{equation}
and use Gauss' theorem on the sums that have $\eta$ and $\zeta$ as free parameters. We arrive at
\begin{align}
\delta \lambda_{22B[0]_2^{[0200]}} \lambda_{kkB[0]_2^{[0200]}} &= \frac{256}{3k \pi^2 c_T} \left [ 4(k - 1)(3k + 4) - k^2 (k + 2) \pi^2 + 2k^2 (k + 2) \psi^{(1)} \left ( \frac{k}{2} \right ) \right ] \nonumber \\
\delta \lambda_{22B[0]_2^{[0040]}} \lambda_{kkB[0]_2^{[0040]}} &= \frac{256(k - 1)}{3k \pi^2 c_T} \left [ 8 + k^2 \pi^2 - 2k^2 \psi^{(1)} \left ( \frac{k}{2} \right ) \right ] \label{order1-res6}
\end{align}
which is the same as what comes out of \eqref{1d-crossing2}.

\section{Conclusion}
\label{sec:conc}

In this paper, we performed a detailed study of the structures of tree-level four-point holographic correlators in maximally superconformal theories. We made essential use of two important ingredients, namely the MRV limit and residue symmetrization procedure introduced in \cite{az20a,az20}. The former was demonstrated in \cite{az20a,az20} to lead to remarkable simplifications, and in this paper we find further use of it in applying the superconformal twist. The latter allows the holographic correlators to be written purely in terms of exchange amplitudes, and makes it easier to recognize underlying structures. In the first of our main results, the symmetrization prescription led us to a tantalizing formula \eqref{SeqKYM} which exhibits a remarkable dimensional reduction structure. It is also interesting that the differential operator appearing there acts on a \textit{scalar} exchange Witten diagram, thereby avoiding the contact term ambiguity associated with spinning vertices. Note this structure would be spoiled if additional contact terms in the amplitudes were present.\footnote{Similar absence of intrinsic contact interactions was also observed for the stress tensor multiplet five-point function for $AdS_5\times S^5$ IIB supergravity \cite{Goncalves:2019znr}.} Our second main result was that the chiral algebra four-point functions predicted by \cite{brv14,bllprv13} can all be extracted from Mellin amplitudes of the parent theory. This calculation exploited the fact that one of the cross-ratios drops out which means we can access the full chiral algebra correlator without ever leaving the MRV limit. This insight failed in an interesting way for the topological sector of ABJM theory which had to be treated by other means. We will now recap these facets of the paper and comment on future directions.

In section \ref{sec:ps}, we showed that the basic ingredients of holographic Mellin amplitudes take the form of a differential operator acting on a simple linear combination of scalar exchange Witten diagrams. This turned out to be related to the linear combination one needs in order to repackage bosonic conformal blocks into \textit{superblocks} of the Parisi-Sourlas superconformal algebra. As shown in \cite{Kaviraj:2019tbg,Zhou:2020ptb}, correlation functions in $d$ dimensions with this supersymmetric property can be interpreted as those of a theory in $d - 2$ dimensions. For the theories studied here, we were able to use this relation twice so that the $AdS_{d + 1} \times S^{\mathtt{d} - 1}$ amplitudes exhibited features of $AdS_{d - 3} \times S^{\mathtt{d} - 5}$. To better understand this phenomenon, one must confront the fact that the dimensional reduction is rather formal. In particular, negative space-time dimensions can appear. Nevertheless, this is not the first time that a seed theory in a fictitious number of dimensions has been proposed in the holographic context. By a very different argument, \cite{ct18,rrz19,grtw20} showed that $SO(d, 2) \times SO(\mathtt{d})$ enhances to the \textit{higher} dimensional conformal group $SO(d + \mathtt{d}, 2)$ for the case of $AdS_5 \times S^5$ and $AdS_3 \times S^3$. We also note that our application of Parisi-Sourlas supersymmetry was somewhat indirect as it required a differential operator to be extracted first. This operator, which is different for each multiplet, led to expressions that only involve the simplest representations (scalar primaries of $SO(d, 2)$ and symmetric traceless tensors of $SO(\mathtt{d})$). It therefore has some similarities to weight-shifting operators \cite{kks17,ch18,bdjlp19} which have been introduced as a tool to compute spinning conformal blocks and spinning Witten diagrams from knowledge of their scalar counterparts. In the future, it would also be very important to check whether the dimensional reduction structure \eqref{SeqKYM} is a coincidence for four-point functions, or a universal feature shared by all $n$-point functions ({\it e.g.} in the $AdS_5\times S^5$ five-point function computed in \cite{Goncalves:2019znr}). If the latter scenario were true, then the  computation of tree-level higher-point holographic correlators would be greatly facilitated by uplifting the much simpler correlators of the underlying scalar seed theory.

In sections \ref{sec:6d} and \ref{sec:4d}, we studied four-point functions in the chiral algebras that have been conjectured for the 6d $\mathcal{N} = (2, 0)$ theory and 4d $\mathcal{N} = 4$ SYM respectively. Our large $N$ expressions \eqref{6d-answer} and \eqref{4d-answer}, which are new for all but a handful of choices for the external weights, were derived by two methods. First, by directly applying the singular OPE in two dimensions and second, by writing down the corresponding half-BPS holographic correlator and going to the twisted configuration. The fact that these two methods agree is a feather in the cap of AdS/CFT. Our results also reveal another property of the MRV limit which, as explained in \cite{az20a,az20}, projects out the exchanges of long multiplets with low twist. What we have seen is that the MRV limit is, at the same time, a refined enough tool that crucial short multiplets (\textit{e.g.} all Schur operators) are still there.

Of considerable interest to us is that chiral algebra four-point functions provide a rich framework for constraining holographic correlators beyond tree-level. The simplest demonstration of this is that the OPE data in $\left < \mathcal{O}_2\mathcal{O}_2\mathcal{O}_k\mathcal{O}_k \right >$ that survives the superconformal twist is $1 / c_{2d}$-exact. Let us now raise the stakes with a loop-level prediction for a correlator that has not been computed yet. Consider $J^{(3)} \times J^{(3)}$ in the super $\mathcal{W}$-algebra for $AdS_5 \times S^5$. There are two chiral operators missing from \eqref{4d_ope} which are allowed by symmetry and low enough in dimension to provide a singular term.
\begin{equation}
\Theta(z) = \frac{\left < \partial_y, \partial_{y^\prime} \right >^2}{4\sqrt{6}} : J^{(2)}(z; y) J^{(2)}(z; y^\prime) : \biggl |_{y = y^\prime}, \;\; \Omega(z; y) = \frac{1}{\sqrt{2}} : J^{(2)}(z; y) J^{(2)}(z; y) : \label{new-ops}
\end{equation}
The overall coefficients ensure that these are unit normalized. Knowing the four-point functions of strong generators, three-point functions involving \eqref{new-ops} follow by taking a coincident limit. Since $\Omega$ and $J^{(4)}$ are degenerate, we only need the one for the Sugawara stress tensor.
\begin{equation}
\left < \Theta(z_1) J^{(3)}(z_2; y_2) J^{(3)}(z_3; y_3) \right > = \frac{45}{\sqrt{6} c_{2d}} \frac{y_{23}^3}{z_{12}^2 z_{23} z_{31}^2}
\end{equation}
Adding this piece to the OPE that defines $\left < J^{(3)}J^{(3)}J^{(3)}J^{(3)} \right >$, we find that the four-point function so obtained is only crossing symmetric if the corrections to existing OPE coefficients satisfy
\begin{equation}
C_{334}^2 + C_{33\Omega}^2 = -\frac{108}{c_{2d}} - \frac{675}{c_{2d}^2}.
\end{equation}
Putting everything together, we have bootstrapped
\begin{align}
\mathcal{F}_{3333}(\chi; \alpha) &= 1 + (\alpha \chi)^3 + \chi^3 \left ( \frac{\alpha - 1}{1 - \chi} \right )^3 + \frac{27\chi (1 + \alpha \chi) (1 - 2\alpha + \alpha\chi) (1 - 2\chi + \alpha\chi)}{c_{2d} (1 - \chi)^2} \nonumber \\
&- \frac{2025 \chi^2 \alpha (\alpha - 1)}{c_{2d}^2 (1 - \chi)} + O \left ( \frac{1}{c_{2d}^3} \right ). \label{3333-prediction}
\end{align}
It will be important to undertake a systematic extension of these results as the AdS unitarity method \cite{aabp16,ac17} continues to progress.

Finally, sections \ref{sec:matrix-model} and \ref{sec:ope-coeffs} focused on 3d $\mathcal{N} = 8$ ABJM theory whose half-BPS correlators, under the superconformal twist, are topological rather than meromorphic. This constitutes a significant obstacle to the goal of making loop-level predictions along the lines of \eqref{3333-prediction}. Indeed, even at tree-level, only single-trace OPE coefficients were within reach of the matrix model techniques we used. Nevertheless, we have the double-trace OPE coefficients \eqref{order1-res3} and \eqref{order1-res6} from Mellin space, and there is a strong indication that the TQFT results will match these once they are computed. This is because the 1d spectrum is sparse enough that all-order crossing equations like \eqref{1d-crossing} are finite and therefore easy to check. In order to pursue subleading calculations in matrix quantum mechanics, the statistical methods of \cite{mp11,h15,ga20} appear promising and it is also worth noting that bootstrap approaches have recently been developed for these theories in \cite{l20,hhk20}.

The last property of these special OPE coefficients that should be highlighted is their connection to the crossing kernel. The Lorentzian inversion formula \cite{c17} establishes that the crossing kernel determines CFT data along double-twist trajectories and the nicest version of this should occur when there is enough supersymmetry to have exact double-twist operators in the spectrum. It is all the more pleasing that we see the one for $\mathfrak{sl}(2)$ as all of the operators discussed here contribute to a lower dimensional subsector. Although the integrals used to derive this are quite general, they are especially useful for Mellin amplitudes with infinitely many poles. In the maximally supersymmetric case, this means $AdS_4 \times S^7$. Existing techniques, which we have generalized, appear to have started with \cite{z17,c18} after the stress tensor four-point function first became available. With an expanded toolkit, and more correlators at our disposal, it will be worthwhile to see if some of the interesting patterns observed for $AdS_5 \times S^5$ \cite{ab17,adhp17,Aprile:2018efk} have analogues in other dimensions. This could suggest interplay between the superconformal twist and other organizing principles for these theories.

\acknowledgments
We thank Fernando Alday for discussions and collaboration in the initial stages of this work. This project has received funding from the European Research Council (ERC) under the European Union's Horizon 2020 research and innovation programme (grant agreement No 787185). The work of X.Z. is supported in part by Simons Foundation Grant No. 488653.

\appendix

\section{An alternate derivation}
\label{sec:appa}

In general, the transformation of a tree-level Mellin amplitude back to position space does not yield a closed form result. Conversely, we have seen that the integral \eqref{inverse-mellin} does lead to simple expressions for the special case of a four-point function in the twisted configuration. We carried out this calculation in subsections \ref{sec:4.2} and \ref{sec:5.2} in order to recover the chiral algebra correlators \eqref{6d-answer} and \eqref{4d-answer} respectively.

In both cases, it was enough to extract a finite number of singularities in the cross-ratio $\chi$ and discard the rest of the integral. As is familiar in 2d CFT, the infinitely many regular terms from one channel are instead captured by singular terms in the other channels. However, one might like to check that the full integral can indeed be evaluated and does not lead to surprises. We will show that this is the case for the $\left < \mathcal{O}_2\mathcal{O}_2\mathcal{O}_k\mathcal{O}_k \right >$ correlators in this appendix. While our original derivation appears to be more powerful, a brute force approach along these lines could be helpful for studying more general observables that do not obey crossing.

\subsection{Four dimensions}
We are interested in reproducing the following correlator.
\begin{align}
& \left < J^{(2)}(z_1, y_1) J^{(2)}(z_2, y_2) J^{(k)}(z_3, y_3) J^{(k)}(z_4, y_4) \right > = \left ( \frac{y_{12}}{z_{12}} \right )^2 \left ( \frac{y_{34}}{z_{34}} \right )^k \mathcal{F}_{22kk}(\chi; \alpha) \label{4d-22kk} \\
& \hspace{1.5cm} \mathcal{F}_{22kk}(\chi; \alpha) = 1 + \frac{k}{k_{2d}} \left [ \frac{\chi}{1 - \chi} - \frac{\chi(2 - k\chi)}{1 - \chi} \alpha + \frac{(k - 1)\chi^2}{1 - \chi} \alpha^2 \right ] \nonumber
\end{align}
This is a special case of \eqref{4d-answer}, which we have expressed in terms of $k_{2d} = -\frac{N^2 - 1}{2}$. Notice that \eqref{4d-22kk} is also $\frac{1}{k_{2d}}$ times the canonically normalized $\left < J J \mathcal{O} \mathcal{O} \right >$ which is often computed using the standard Ward identity
\begin{equation}
J_{a_1 a_2}(z) \mathcal{O}_{b_1 \dots b_k}(w) = \frac{\epsilon_{a_1 b_1} \mathcal{O}_{a_2 \dots b_k}(w) + \dots + \epsilon_{a_1 b_k} \mathcal{O}_{b_1 \dots a_2}(w) + (a_1 \leftrightarrow a_2)}{-2 (z - w)} + \dots \label{j-ward}
\end{equation}
for an affine $\mathfrak{su}(2)$ current.\footnote{If one prefers to work with adjoint indices, these can be re-introduced by taking
\begin{equation}
J_{ab} = \epsilon_{ac} (T_A)^c_{\;\;b} J^A = \frac{1}{2} \left [ \epsilon_{ac} (T_A)^c_{\;\;b} + \epsilon_{bc} (T_A)^c_{\;\;a} \right ] J^A \label{adjoint-form}
\end{equation}
where $T_A = \frac{1}{\sqrt{2}} \sigma_A$ are the generators, defined to have structure constants $f_{ABC} = \sqrt{2}\epsilon_{ABC}$.} This makes it easy to see that \eqref{4d-22kk} is more than just a tree-level correlator in the chiral algebra. It is exact to all orders since there is no way for loop corrections to modify the universal singular term in \eqref{j-ward}.

Let us now write down the associated Mellin amplitude in the parent theory. From \eqref{residueR}, we find
\begin{align}
\mathcal{M}_{22kk} & (s, t; \sigma, \tau) = -\frac{k}{N^2 (k - 2)!} \label{4d-22kk-mellin} \\
\times &\left [ \frac{(t - k - 2)(u - k - 2) + (s + 2)(t - k - 2)\sigma + (s + 2)(u - k - 2)\tau}{s - 2} \right . \nonumber \\
&\left. + \frac{(s - 2k)(u - k - 2)\tau^2 + (s - 2k)(t + k)\sigma\tau + (t + k)(u - k - 2)\tau}{t - k} \right. \nonumber \\
&\left. + \frac{(s - 2k)(t - k - 2)\sigma^2 + (t - k - 2)(u + k)\sigma + (s - 2k)(u + k)\sigma\tau}{u - k} \right ]. \nonumber
\end{align}
As in subsection \ref{sec:5.2}, we will implement the superconformal twist by taking $\alpha^{\prime -1} = \chi^\prime \rightarrow 0$, which is an MRV configuration for the $s$-channel. Going through the three lines of \eqref{4d-22kk-mellin}, we can now list the poles in $s$ which contribute in this limit.
\begin{itemize}
\item For the first line, there is an increasing sequence of poles starting at $s = 2$. Only the first (the single-particle one) contributes. This is because $\alpha^\prime$ enters linearly which means that $\chi^{\prime \frac{s - 2}{2}}$ is the smallest power of $\chi^\prime$ that we get.
\item For the other two lines, we only have the sequence of double-particle poles which starts at $s = 4$. Again, only the first contributes. This is because $\alpha^\prime$ enters quadratically which leads to $\chi^{\prime \frac{s - 4}{2}}$ as the lowest power.
\end{itemize}
After picking up these residues, the remaining integral to compute is
\begin{align}
\mathcal{F}_{22kk}(\chi; \alpha) = \frac{k}{N^2} & \int_{-i\infty}^{i\infty} \frac{\textup{d}t}{2\pi i} (1 - \chi)^{\frac{t - k - 2}{2}} \Gamma \left [ \frac{k + 2 - t}{2} \right ]^2 \label{4d-pre-trick} \\
& \left [ \chi\alpha (k + 2 - t) \Gamma \left [ \frac{t - k}{2} \right ]^2 + \chi(1 + \alpha)(k - t) \Gamma \left [ \frac{t - k}{2} \right ]^2 \right. \nonumber \\
& + \chi^2(1 - \alpha)^2 \frac{t - k + 2}{t - k} \Gamma \left [ \frac{t - k + 2}{2} \right ]^2 + \chi^2\alpha(1 - \alpha) \frac{t + k}{t - k} \Gamma \left [ \frac{t - k + 2}{2} \right ]^2 \nonumber \\
&\left. + \chi^2\alpha^2 \frac{t - k - 2}{t - k} \Gamma \left [ \frac{t - k + 2}{2} \right ]^2 + \chi^2\alpha(1 - \alpha) \frac{t - 3k}{t - k} \Gamma \left [ \frac{t - k + 2}{2} \right ]^2 \right ]. \nonumber
\end{align}
To simplify the integrand, an obvious approach is to absorb the explicit factors of $(t - k)^{-1}$ into the gamma functions. Even though there are two choices for how to do this, only one respects our prescription of encircling single-trace poles in the $t$-channel but not the $u$-channel. If the $t = k$ pole is kept (not kept), it should be absorbed into the gamma function that has $t$ appearing negatively (positively) so that the result can be treated with a standard Mellin-Barnes contour. In other words, the last line of \eqref{4d-pre-trick} goes with the last line of
\begin{align}\label{absorb-trick}
\begin{split}
\frac{2}{k - t} \Gamma \left ( \frac{k + 2 - t}{2} \right ) &= \Gamma \left ( \frac{k - t}{2} \right ) \\
\frac{2}{t - k} \Gamma \left ( \frac{t - k + 2}{2} \right ) &= \Gamma \left ( \frac{t - k}{2} \right )
\end{split}
\end{align}
and \textit{vice versa}. In view of \eqref{master1}, the integral becomes a sum of Gaussian hypergeometric functions after we make this simplification. All of their parameters can be seen to take the values $1$, $2$, $3$ or $4$. Analytic expressions for them follow from standard contiguous relations after using
\begin{equation}
{}_2F_1 \left ( 1, 1; 1; \frac{\chi}{\chi - 1} \right ) = 1 - \chi, \;\; {}_2F_1 \left ( 1, 1; 2; \frac{\chi}{\chi - 1} \right ) = \frac{\chi - 1}{\chi} \log(1 - \chi) \label{2f1-trick}
\end{equation}
as a seed. The cancellation of all logarithmic terms generated in this way is a non-trivial check of our calculation. By adding up all of the non-logarithmic terms, we find
\begin{equation}
\mathcal{F}_{22kk}(\chi; \alpha) = \frac{2k}{N^2} \frac{\chi}{\chi - 1} \left [ 1 + \alpha(k\chi - 2) - \alpha^2 \chi (k - 1) \right ]
\end{equation}
which is nothing but the tree-level piece of \eqref{4d-22kk}.

\subsection{Six dimensions}
The correlator to verify from Mellin space is now
\begin{align}
& \left < W^{(2)}(z_1) W^{(2)}(z_2) W^{(k)}(z_3) W^{(k)}(z_4) \right > = \frac{\mathcal{F}_{22kk}(\chi)}{z_{12}^4 z_{34}^{2k}} \label{6d-22kk} \\
& \hspace{1cm} \mathcal{F}_{22kk}(\chi) = 1 + \frac{2k}{c_{2d}} \frac{\chi^2 (k\chi^2 - 2\chi + 2)}{(1 - \chi)^2}. \nonumber
\end{align}
Once again, this is very special. Up to a $\frac{2}{c_{2d}}$ factor, it takes the form $\left < TT\mathcal{O}\mathcal{O} \right >$ which is easily computed using the Ward identity for the stress tensor.
\begin{equation}
T(z) \mathcal{O}(w) = \frac{h \mathcal{O}(w)}{(z - w)^2} + \frac{\partial \mathcal{O}(w)}{z - w} + \dots \label{t-ward}
\end{equation}
Clearly, this means that \eqref{6d-22kk} is $1 / c_{2d}$-exact. It is only the more singular Ward identities of \eqref{6d_ope} where there is room to have loop corrections in the form of normal ordered products.

The right Mellin amplitude to use is again obtained from \eqref{residueR}. To perform the superconformal twist and go to position space, we take $\alpha^{-1} = \alpha^{\prime -1} = \chi^\prime \rightarrow 0$ which localizes \eqref{inverse-mellin} onto a finite number of poles in $s$. We can again list them in two steps.
\begin{itemize}
\item The part for the $s$-channel has poles starting at $s = 4$. Only the first one (the first of two single-particle poles) contributes because of $\sigma$ and $\tau$ entering linearly. This leads to $\chi^{\prime \frac{s - 4}{2}}$ and higher powers of $\chi^\prime$.
\item The part for the other two channels has poles starting at $s = 8$ all of which are double-particle. Only the first contributes because these channels have $\sigma$ and $\tau$ appearing with total degree $2$. Hence, $\chi^{\prime \frac{s - 8}{2}}$ is the smallest power of $\chi^\prime$ produced.
\end{itemize}
This time, we will only write the part of the Mellin amplitude that survives after the unimportant terms are removed.
\begin{align}
\mathcal{M}_{22kk}(s, t; \chi^\prime) &\sim \frac{k \chi^{\prime -2}}{2N^3 (2k - 3)!} \frac{s(s + 2)}{s - 4} \label{6d-22kk-mellin} \\
&+ \frac{k^2 \chi^{\prime -4} (s - 4k)(s - 4k - 2)}{4N^3 (2k - 3)!} \left [ \frac{1}{t - 2k} + \frac{1}{t - 2k - 2} + \frac{1}{u - 2k} + \frac{1}{u - 2k - 2} \right ] \nonumber
\end{align}
After picking up the residues just described, we are left with
\begin{align}
\mathcal{F}_{22kk}(\chi) = \frac{k}{4N^3} & \int_{-i\infty}^{i\infty} \frac{\textup{d}t}{2\pi i} (1 - \chi)^{\frac{t}{2} - k - 2} \Gamma \left [ \frac{2k + 4 - t}{2} \right ]^2 \\
& \left [ 12\chi^2 \Gamma \left [ \frac{t - 2k}{2} \right ]^2 - \frac{2k\chi^4}{t - 2k} \Gamma \left [ \frac{t - 2k + 4}{2} \right ]^2 - \frac{\chi^4}{t - 2k - 2} \Gamma \left [ \frac{t - 2k + 4}{2} \right ]^2 \right. \nonumber \\
&\left. + \frac{2k\chi^4}{t - 2k} \Gamma \left [ \frac{t - 2k + 4}{2} \right ]^2 + \frac{\chi^4}{t - 2k + 2} \Gamma \left [ \frac{t - 2k + 4}{2} \right ]^2 \right ]. \nonumber
\end{align}
Two of the terms look like they cancel but they do not. They localize to a single $t = 2k$ residue because the $t = 2k$ pole is kept by one contour but not the other. The other terms can also be evaluated by paying attention to the contour. This leads to
\begin{align}
\mathcal{F}_{22kk}(\chi) &= \frac{k}{2N^3} \left ( \frac{\chi}{1 - \chi} \right )^2 \left [ k\chi^2 + 2{}_2F_1 \left ( 2, 2; 4; \frac{\chi}{\chi - 1} \right ) \right ] \nonumber \\
&+ \frac{k}{20N^3} \left ( \frac{\chi}{1 - \chi} \right )^4 \left [ 2{}_2F_1 \left ( 3, 4; 7; \frac{\chi}{\chi - 1} \right ) + 2{}_2F_1 \left ( 4, 4; 7; \frac{\chi}{\chi - 1} \right ) \right ]
\end{align}
which, upon using \eqref{2f1-trick}, matches \eqref{6d-22kk}.

\section{The infinite sum over descendants}
\label{sec:appb}

In a myriad of applications, one starts with \eqref{inverse-mellin} and extracts a pole in $s$ corresponding to a double-particle operator. Even though the remaining $t$ integral can always be evaluated exactly by \eqref{master3} and \eqref{master4}, these do not tell us what to do with the sum over $m$ which labels the level of the descendant being exchanged. Recalling the explicit form of the Mellin amplitude \eqref{residueR}, this sum is finite for $AdS_5 \times S^5$ and $AdS_7 \times S^4$. However, for the case of $AdS_4 \times S^7$, the sum can be either finite or infinite depending on the external weights. It is therefore necessary to develop some technology for the calculation being done in section \ref{sec:ope-coeffs}.

It is explained in footnote \ref{only-tu} that the $B[0]_{j_{\rm I}}^{[0, j_{\rm I} - j_{\rm II}, 2 j_{\rm II}, 0]}$ multiplets with maximal $j_{\rm I}$ are associated with the residue of an $\mathcal{M}^{(t)}$ or $\mathcal{M}^{(u)}$ Mellin amplitude at $s = s_{\mathrm{min}} \equiv \mathrm{min}(\Delta_1 + \Delta_2, \Delta_3 + \Delta_4)$.\footnote{It is useful to notice from selection rules that $\Delta_1 + \Delta_2$ and $\Delta_3 + \Delta_4$ are either both integers or both half-integers.} We therefore have two channels to analyze as far as the infinite sum is concerned. To start, we will take a closer look at the truncation conditions in each channel to see what they tell us about the parameters in our hypergeometric functions. After this, we will explain the appearance of \eqref{bailey-integral} in the protected OPE coefficients.

\subsection{Integers vs half-integers}
While it is clear that the parameters in \eqref{master3} and \eqref{master4} are either integers or half-integers, we are going to need a slightly more detailed statement. It is useful to define two constants that have a discrete choice available.
\begin{equation}
a \in \left \{ \frac{\Delta_1 + \Delta_4}{2}, \frac{\Delta_2 + \Delta_3}{2} \right \}, \;\;\;\; b \in \left \{ \frac{s_{\mathrm{min}} - \Delta_1 - \Delta_3}{2}, \frac{s_{\mathrm{min}} - \Delta_2 - \Delta_4}{2} \right \}. \label{ab-choices}
\end{equation}

In the $t$-channel, the condition for the sum \eqref{s-channel} to be infinite is
\begin{equation}
\Delta_1 + \Delta_4 - \frac{p}{2}, \;\; \Delta_2 + \Delta_3 - \frac{p}{2} \in 2\mathbb{Z} + 1. \label{t-truncation}
\end{equation}
The fact that $p \equiv |k_{14}|, |k_{23}| \; (\mathrm{mod} \; 2)$ only tells us that the above combinations are integers. Next, the hypergeometric function in \eqref{master3} is
\begin{equation}
T(\Delta_i; m, p) = {}_3F_2 \left [ \begin{tabular}{c} \begin{tabular}{ccc} $1$, & $\frac{1}{2}(\Delta_1 + \Delta_4) + b$, & $\frac{1}{2}(\Delta_2 + \Delta_3) + b$ \end{tabular} \\ \begin{tabular}{cc} $s_{\mathrm{min}}$, & $1 + b + m + \frac{p}{4}$ \end{tabular} \end{tabular} \right ]. \label{t-hyper}
\end{equation}
We can now quote a basic property of \eqref{t-hyper} which will be important in the next subsection. \textit{The denominator parameter that depends on $m$ differs from the non-trivial numerator parameters by a half-integer}. This follows immediately from \eqref{t-truncation} regardless of the value of $b$.

Proceeding identically in the $u$-channel, having an infinite sum in \eqref{s-channel} requires
\begin{equation}
\Delta_1 + \Delta_3 - \frac{p}{2}, \;\; \Delta_2 + \Delta_4 - \frac{p}{2} \in 2\mathbb{Z} + 1. \label{u-truncation}
\end{equation}
This is a refinement of $p \equiv |k_{13}|, |k_{24}| \; (\mathrm{mod} \; 2)$. From \eqref{master4}, we now have a hypergeometric function that looks like
\begin{equation}
U(\Delta_i; m, p) = {}_3F_2 \left [ \begin{tabular}{c} \begin{tabular}{ccc} $1$, & $a + \frac{1}{2}(s_{\mathrm{min}} - \Delta_1 - \Delta_3)$, & $a + \frac{1}{2}(s_{\mathrm{min}} - \Delta_2 - \Delta_4)$ \end{tabular} \\ \begin{tabular}{cc} $s_{\mathrm{min}}$, & $1 + a + m - \frac{1}{2}s_{\mathrm{max}} + \frac{p}{4}$ \end{tabular} \end{tabular} \right ] \label{u-hyper}
\end{equation}
with the obvious definition of $s_{\mathrm{max}}$. We observe that the previously discussed property also holds for \eqref{u-hyper}. \textit{The denominator parameter that depends on $m$ differs from the non-trivial numerator parameters by a half-integer}. This follows immediately from \eqref{u-truncation} regardless of the value of $a$.

\subsection{Translation to a finite sum}
Let us now consider the sum
\begin{equation}
I_k = \sum_{m = 0}^\infty \binom{m}{k} \frac{(a_1)_m (a_2)_m}{m! (b_1)_m (c + m - 1)} {}_3F_2 \left [ \begin{tabular}{c} \begin{tabular}{ccc} $1$, & $a_3$, & $a_4$ \end{tabular} \\ \begin{tabular}{cc} $b_2$, & $c + m$ \end{tabular} \end{tabular} \right ] \label{hard-sum}
\end{equation}
subject to the properties discussed above. Only $I_0$ will concern us since the other cases can all be obtained by shifting the parameters. The most useful way to proceed is to write the hypergeometric function as an Euler integral, essentially undoing the last step in the derivation of \eqref{master3} and \eqref{master4}. This leads to
\begin{align}
I_0 &= \sum_{m = 0}^\infty \frac{(a_1)_m (a_2)_m}{m! (b_1)_m} \int_0^1 {}_2F_1(a_3, a_4; b_2; z) (1 - z)^{c + m - 2} \textup{d}z \nonumber \\
&= \int_0^1 {}_2F_1(a_1, a_2; b_1; 1 - z) {}_2F_1(a_3, a_4; b_2; z) (1 - z)^{c - 2} \textup{d}z \label{kernel1}
\end{align}
which can be computed with the methods of \cite{gs18}. Performing two Pfaff transformations and passing to the Mellin-Barnes representation, we arrive at
\begin{align}
I_0 = & \frac{\Gamma(b_1)\Gamma(b_2)}{\Gamma(a_1)\Gamma(b_1 - a_2)\Gamma(a_3)\Gamma(b_2 - a_4)} \int_{-i\infty}^{i\infty} \frac{\textup{d}s \textup{d}t}{(2\pi i)^2} \int_0^\infty y^{t - s - a_1} (1 + y)^{a_1 + a_3 - c} \textup{d}y \nonumber \\
& \frac{\Gamma(-s)\Gamma(a_1 + s)\Gamma(b_1 - a_2 + s)}{\Gamma(b_1 + s)} \frac{\Gamma(-t)\Gamma(a_3 + t)\Gamma(b_2 - a_4 + t)}{\Gamma(b_2 + t)}. \label{kernel2}
\end{align}
This is where our proof that it is always possible to make $a_3 - c$ a half-integer will pay off. Since $a_1$ is also a half-integer, we get to use the binomial theorem with finitely many terms. Following \cite{gs18} again,
\begin{align}
I_0 = & \frac{(a_1)_{b_1 - a_1}(a_3)_{b_2 - a_3}}{\Gamma(b_1 - a_2)\Gamma(b_2 - a_4)} \sum_{n = 0}^{a_1 + a_3 - c} \binom{a_1 + a_3 - c}{n} \int_{-i\infty}^{i\infty} \frac{\textup{d}t}{2\pi i} \frac{\Gamma(-t)\Gamma(a_3 + t)\Gamma(b_2 - a_4 + t)}{\Gamma(b_2 + t)} \nonumber \\
& \frac{\Gamma(a_1 - n - 1 - t)\Gamma(1 + n + t)\Gamma(1 + n + b_1 - a_1 - a_2 + t)}{\Gamma(1 + n + b_1 - a_1 + t)}. \label{kernel3}
\end{align}

The Mellin-Barnes integral derived here bears an unmistakable resemblance to the crossing kernel for collinear blocks. As emphasized in \cite{gs18}, the crossing kernel has the special property that it is very well poised -- the gamma function arguments in the numerator and denominator have the same sum. This will only be true for \eqref{kernel3} if
\begin{equation}
a_1 - a_2 + a_3 - a_4 = 0. \label{very-well-poised}
\end{equation}
Remarkably, the amplitudes in \cite{az20} are such that
\begin{equation}
a_1 - a_2 = \begin{cases}
-\frac{1}{2}(\Delta_1 - \Delta_2 - \Delta_3 + \Delta_4), & \mathcal{M} = \mathcal{M}^{(t)} \\
-\frac{1}{2}(\Delta_1 - \Delta_2 + \Delta_3 - \Delta_4), & \mathcal{M} = \mathcal{M}^{(u)}
\end{cases}
\end{equation}
which means that $|a_1 - a_2| = |a_3 - a_4|$ after we compare to \eqref{t-hyper} and \eqref{u-hyper}. Since the original sum \eqref{hard-sum} is symmetric under $a_1 \leftrightarrow a_2$ and $a_3 \leftrightarrow a_4$ separately, we are free to choose the signs of the parameter differences and thereby ensure that \eqref{very-well-poised} holds.

\providecommand{\href}[2]{#2}\begingroup\raggedright\endgroup

\end{document}